\documentclass[aps,prc,superscriptaddress,showpacs,floatfix,twocolumn]{revtex4}

\usepackage{amsmath}
\usepackage{xspace}	
\usepackage{dcolumn}
\usepackage[dvips]{graphicx}
\usepackage{color}

\def\v2{\mbox{$v_2$}}

\newcommand \gevc{GeV/$c$\xspace}
\newcommand \pt{\mbox{$p_T$}\xspace}

\newcommand \Np{\mbox{$N_{\rm part}$}\xspace}

\newcommand \sqsn{\mbox{$\sqrt{s_{_{NN}}}$}\xspace}
\newcommand \KET{\mbox{${\rm KE}_T$}\xspace}
\newcommand \RP{\mbox{$\Psi_{\rm{RP}}$}\xspace}
\newcommand \Fig{Fig.\xspace}
\newcommand \mean[1]{\left\langle #1 \right\rangle} 
\newcommand \anpart{\mbox{$\langle N_{\rm part} \rangle$}}
\newcommand \Au{{Au+Au}\xspace}

\begin{document}

\hyphenation{author another created financial paper re-commend-ed Post-Script}

\title{Systematic Studies of Elliptic Flow Measurements \\
 in Au+Au Collisions at $\sqrt{s_{_{NN}}}$ = 200 GeV
}  

\newcommand{\abilene}{Abilene Christian University, Abilene, TX 79699, U.S.}
\newcommand{\banaras}{Department of Physics, Banaras Hindu University, Varanasi 221005, India}
\newcommand{\bnlphys}{Brookhaven National Laboratory, Upton, NY 11973-5000, U.S.}
\newcommand{\caucr}{University of California - Riverside, Riverside, CA 92521, U.S.}
\newcommand{\cns}{Center for Nuclear Study, Graduate School of Science, University of Tokyo, 7-3-1 Hongo, Bunkyo, Tokyo 113-0033, Japan}
\newcommand{\colorado}{University of Colorado, Boulder, CO 80309, U.S.}
\newcommand{\columbia}{Columbia University, New York, NY 10027 and Nevis Laboratories, Irvington, NY 10533, U.S.}
\newcommand{\dapnia}{Dapnia, CEA Saclay, F-91191, Gif-sur-Yvette, France}
\newcommand{\debrecen}{Debrecen University, H-4010 Debrecen, Egyetem t{\'e}r 1, Hungary}
\newcommand{\elte}{ELTE, E{\"o}tv{\"o}s Lor{\'a}nd University, H - 1117 Budapest, P{\'a}zm{\'a}ny P. s. 1/A, Hungary}
\newcommand{\fsu}{Florida State University, Tallahassee, FL 32306, U.S.}
\newcommand{\gsu}{Georgia State University, Atlanta, GA 30303, U.S.}
\newcommand{\hiroshima}{Hiroshima University, Kagamiyama, Higashi-Hiroshima 739-8526, Japan}
\newcommand{\ihepprot}{IHEP Protvino, State Research Center of Russian Federation, Institute for High Energy Physics, Protvino, 142281, Russia}
\newcommand{\illuiuc}{University of Illinois at Urbana-Champaign, Urbana, IL 61801, U.S.}
\newcommand{\isu}{Iowa State University, Ames, IA 50011, U.S.}
\newcommand{\jinrdubna}{Joint Institute for Nuclear Research, 141980 Dubna, Moscow Region, Russia}
\newcommand{\kaeri}{KAERI, Cyclotron Application Laboratory, Seoul, Korea}
\newcommand{\kek}{KEK, High Energy Accelerator Research Organization, Tsukuba, Ibaraki 305-0801, Japan}
\newcommand{\kfki}{KFKI Research Institute for Particle and Nuclear Physics of the Hungarian Academy of Sciences (MTA KFKI RMKI), H-1525 Budapest 114, POBox 49, Budapest, Hungary}
\newcommand{\korea}{Korea University, Seoul, 136-701, Korea}
\newcommand{\kurchatov}{Russian Research Center ``Kurchatov Institute", Moscow, Russia}
\newcommand{\kyoto}{Kyoto University, Kyoto 606-8502, Japan}
\newcommand{\labllr}{Laboratoire Leprince-Ringuet, Ecole Polytechnique, CNRS-IN2P3, Route de Saclay, F-91128, Palaiseau, France}
\newcommand{\lawllnl}{Lawrence Livermore National Laboratory, Livermore, CA 94550, U.S.}
\newcommand{\losalamos}{Los Alamos National Laboratory, Los Alamos, NM 87545, U.S.}
\newcommand{\lpc}{LPC, Universit{\'e} Blaise Pascal, CNRS-IN2P3, Clermont-Fd, 63177 Aubiere Cedex, France}
\newcommand{\lund}{Department of Physics, Lund University, Box 118, SE-221 00 Lund, Sweden}
\newcommand{\maryland}{University of Maryland, College Park, MD 20742, U.S.}
\newcommand{\muenster}{Institut f\"ur Kernphysik, University of Muenster, D-48149 Muenster, Germany}
\newcommand{\myongji}{Myongji University, Yongin, Kyonggido 449-728, Korea}
\newcommand{\nagasaki}{Nagasaki Institute of Applied Science, Nagasaki-shi, Nagasaki 851-0193, Japan}
\newcommand{\newmex}{University of New Mexico, Albuquerque, NM 87131, U.S. }
\newcommand{\nmsu}{New Mexico State University, Las Cruces, NM 88003, U.S.}
\newcommand{\ornl}{Oak Ridge National Laboratory, Oak Ridge, TN 37831, U.S.}
\newcommand{\orsay}{IPN-Orsay, Universite Paris Sud, CNRS-IN2P3, BP1, F-91406, Orsay, France}
\newcommand{\pnpi}{PNPI, Petersburg Nuclear Physics Institute, Gatchina, Leningrad region, 188300, Russia}
\newcommand{\riken}{RIKEN Nishina Center for Accelerator-Based Science, Wako, Saitama 351-0198, JAPAN}
\newcommand{\rikjrbrc}{RIKEN BNL Research Center, Brookhaven National Laboratory, Upton, NY 11973-5000, U.S.}
\newcommand{\rikkyo}{Physics Department, Rikkyo University, 3-34-1 Nishi-Ikebukuro, Toshima, Tokyo 171-8501, Japan}
\newcommand{\saispbstu}{Saint Petersburg State Polytechnic University, St. Petersburg, Russia}
\newcommand{\saopaulo}{Universidade de S{\~a}o Paulo, Instituto de F\'{\i}sica, Caixa Postal 66318, S{\~a}o Paulo CEP05315-970, Brazil}
\newcommand{\seoulnat}{System Electronics Laboratory, Seoul National University, Seoul, Korea}
\newcommand{\stonybrkc}{Chemistry Department, Stony Brook University, Stony Brook, SUNY, NY 11794-3400, U.S.}
\newcommand{\stonycrkp}{Department of Physics and Astronomy, Stony Brook University, SUNY, Stony Brook, NY 11794, U.S.}
\newcommand{\subatech}{SUBATECH (Ecole des Mines de Nantes, CNRS-IN2P3, Universit{\'e} de Nantes) BP 20722 - 44307, Nantes, France}
\newcommand{\tenn}{University of Tennessee, Knoxville, TN 37996, U.S.}
\newcommand{\titech}{Department of Physics, Tokyo Institute of Technology, Oh-okayama, Meguro, Tokyo 152-8551, Japan}
\newcommand{\tsukuba}{Institute of Physics, University of Tsukuba, Tsukuba, Ibaraki 305, Japan}
\newcommand{\vandy}{Vanderbilt University, Nashville, TN 37235, U.S.}
\newcommand{\waseda}{Waseda University, Advanced Research Institute for Science and Engineering, 17 Kikui-cho, Shinjuku-ku, Tokyo 162-0044, Japan}
\newcommand{\weizmann}{Weizmann Institute, Rehovot 76100, Israel}
\newcommand{\yonsei}{Yonsei University, IPAP, Seoul 120-749, Korea}
\affiliation{\abilene}
\affiliation{\banaras}
\affiliation{\bnlphys}
\affiliation{\caucr}
\affiliation{\cns}
\affiliation{\colorado}
\affiliation{\columbia}
\affiliation{\dapnia}
\affiliation{\debrecen}
\affiliation{\elte}
\affiliation{\fsu}
\affiliation{\gsu}
\affiliation{\hiroshima}
\affiliation{\ihepprot}
\affiliation{\illuiuc}
\affiliation{\isu}
\affiliation{\jinrdubna}
\affiliation{\kaeri}
\affiliation{\kek}
\affiliation{\kfki}
\affiliation{\korea}
\affiliation{\kurchatov}
\affiliation{\kyoto}
\affiliation{\labllr}
\affiliation{\lawllnl}
\affiliation{\losalamos}
\affiliation{\lpc}
\affiliation{\lund}
\affiliation{\maryland}
\affiliation{\muenster}
\affiliation{\myongji}
\affiliation{\nagasaki}
\affiliation{\newmex}
\affiliation{\nmsu}
\affiliation{\ornl}
\affiliation{\orsay}
\affiliation{\pnpi}
\affiliation{\riken}
\affiliation{\rikjrbrc}
\affiliation{\rikkyo}
\affiliation{\saispbstu}
\affiliation{\saopaulo}
\affiliation{\seoulnat}
\affiliation{\stonybrkc}
\affiliation{\stonycrkp}
\affiliation{\subatech}
\affiliation{\tenn}
\affiliation{\titech}
\affiliation{\tsukuba}
\affiliation{\vandy}
\affiliation{\waseda}
\affiliation{\weizmann}
\affiliation{\yonsei}
\author{S.~Afanasiev} \affiliation{\jinrdubna}
\author{C.~Aidala} \affiliation{\columbia}
\author{N.N.~Ajitanand} \affiliation{\stonybrkc}
\author{Y.~Akiba} \affiliation{\riken} \affiliation{\rikjrbrc}
\author{J.~Alexander} \affiliation{\stonybrkc}
\author{A.~Al-Jamel} \affiliation{\nmsu}
\author{K.~Aoki} \affiliation{\kyoto} \affiliation{\riken}
\author{L.~Aphecetche} \affiliation{\subatech}
\author{R.~Armendariz} \affiliation{\nmsu}
\author{S.H.~Aronson} \affiliation{\bnlphys}
% ...
\collaboration{PHENIX Collaboration} \noaffiliation

\begin{abstract}

We present inclusive charged hadron elliptic flow ($v_2$) measured 
over the pseudorapidity range $|\eta| < $ 0.35 in \Au collisions at 
\sqsn = 200 GeV. Results for $v_2$ are presented over a broad range 
of transverse momentum (\pt = 0.2--8.0 GeV/$c$) and centrality 
(0--60\%). In order to study non-flow effects that are not correlated 
with the reaction plane, as well as the fluctuations of $v_2$, we 
compare two different analysis methods: 
(1) event plane method from two independent sub-detectors at forward 
($|\eta|$ = 3.1--3.9) and beam ($|\eta| > 6.5$) pseudorapidities 
and 
(2) two-particle cumulant method extracted using correlations 
between particles detected at midrapidity.  The two event-plane 
results are consistent within systematic uncertainties over the 
measured \pt and in centrality 0--40\%. There is at most 20\% 
difference of the $v_2$ between the two event plane methods in 
peripheral (40--60\%) collisions.  The comparisons between the 
two-particle cumulant results and the standard event plane 
measurements are discussed.

\end{abstract}
\maketitle


\section{Introduction \label{sec:introduction}}

Collisions of \Au nuclei at the Relativistic Heavy Ion Collider 
(RHIC) produce matter at very high energy 
density~\cite{Arsene:2004fa,Adcox:2004mh,Back:2004je,Adams:2005dq}. 
The dynamical evolution of this hot and dense medium reflects 
its state and the degrees of freedom that govern the different 
stages it 
undergoes~\cite{Gyulassy:2004zy,Muller:2004kk,Shuryak:2004cy}. 
Azimuthal anisotropy measurements serve as a probe of the degree 
of thermalization, transport coefficients and the equation of 
state (EOS)~\cite{Ollitrault:1992bk,Kolb:2001qz,Hirano:2004en} 
of the produced medium.

Azimuthal correlation measurements in \Au collisions at RHIC 
have been shown to consist of a mixture of jet and harmonic 
contributions~\cite{Ajitanand:2002qd,Chiu:2002ma,Adler:2002tq,Adler:2004zd}. 
Jet contributions are found to be relatively small for \pt 
$\lesssim$ 2.0 \gevc, with away-side jet yields strongly 
suppressed~\cite{Adler:2002tq}. Significant modifications to the 
away-side jet topology have also been 
reported~\cite{Adams:2005ph,Adler:2005ee,Adare:2008cqb}. The 
harmonic contributions are typically characterized by the 
Fourier coefficients,
\begin{equation}
v_n = \left<\cos{(n[\phi-\RP])}\right> \quad (n = 1, 2, ...),
\end{equation}
where $\phi$ represents the azimuthal emission angle of a 
charged hadron and \RP is the azimuth of the reaction plane 
defined as containing both the direction of the impact parameter 
vector and the beam axis. The brackets denote statistical 
averaging over particles and events. The first two harmonics 
$v_1$ and $v_2$ are referred to as directed and elliptic flow, 
respectively.

It has been found that at low \pt 
(\pt~$\mathbin{\lower.3ex\hbox{$\buildrel<\over 
{\smash{\scriptstyle\sim}\vphantom{_x}}$}}~2.0$ \gevc) the 
magnitude and trends of \v2 are under-predicted by hadronic 
cascade models supplemented with string 
dynamics~\cite{Bleicher:2000sx}, but are well reproduced by 
models which either incorporate hydrodynamic 
flow~\cite{Shuryak:2004cy,Kolb:2001qz} with a first order phase 
transition and rapid thermalization, $\tau \sim 1$ fm/c 
\cite{Adler:2003kt}, or use a quasi-particle ansatz but include 
more than just 2-to-2 interactions~\cite{Xu:2008dv}.

The mass dependence of $v_2$ as a function of \pt has been 
studied using identified baryons and 
mesons~\cite{Adler:2003kt,Adams:2003am} and empirical scaling of 
elliptic flow per constituent quark was observed when the signal 
and the \pt of the hadron were divided by the number of 
constituent quarks $n_q$ ($n_q$ = 2 for mesons, 3 for baryons). 
This scaling is most clearly observed by plotting the data as a 
function of transverse kinetic energy 
\KET~$\equiv~m_T-m = \sqrt{\pt^2 + m^2}-m$~\cite{Adare:2006ti}, 
where $m_T$ and $m$ denote the transverse mass and mass of the 
particle, respectively. A recent study~\cite{Huang:2008vd} finds 
that the constituent quark scaling holds up to 
\KET~$\approx~1$~GeV. This indicates partonic, rather than 
hadronic flow, and suggests that the bulk matter collectivity 
develops before hadronization takes 
place~\cite{PhysRevLett.91.092301,PhysRevC.68.044902,PhysRevC.68.034904}. 
Results for the $v_2$ of the $\phi$ meson further validate the 
observation of partonic collectivity. The $\phi$ is not expected 
to be affected by hadronic interactions in the late stages of 
the medium evolution, due to its small interaction cross section 
with non-strange hadrons ~\cite{Shor:1984ui}.

All of the $v_2$ measurements referenced above were performed 
using the event plane method \cite{Poskanzer:1998yz}. In PHENIX 
the event plane was determined at forward and backward 
pseudorapidities ($|\eta|$ = 3.1--3.9) with the assumption that 
correlations induced by elliptic flow dominate over all other 
non-flow correlations \cite{Adler:2003kt}. Non-flow correlations 
are those which are not correlated with the reaction plane. 
Common sources of non-flow correlations include jets, the 
near-side ridge, quantum correlations and resonance decays. 
Simulation studies~\cite{Adler:2003kt,Jia:2006sb} have shown 
that the correlations from jets and dijets become negligible 
when the rapidity separation between the particles and the event 
plane is greater than three units. Thus we expect that the event 
plane at forward pseudorapidities $|\eta|$ = 3.1--3.9 in the 
PHENIX experiment would not have significant jet-correlation 
with particles measured within the PHENIX central arm 
spectrometer covering the pseudorapidity window $|\eta| < 0.35$. 
PHOBOS has observed that in central Au+Au collisions there is a 
ridge of particles~\cite{Alver:2009id} that are correlated in 
azimuthal angle with a high-$\pt$ particle and that this ridge 
of particles extends to $|\eta| < 4$ (for mid-rapidity 
triggers). The ridge could produce a non-flow correlation that 
we can provide information by using our $v_2$ measurements that 
are made with different techniques and at different rapidities.

Event-by-event flow fluctuations can also affect the magnitude 
of the extracted flow signal~\cite{Sorensen:2008zk}. This occurs 
because the event plane at forward pseudorapidities is 
reconstructed using particles from participant nucleons whose 
positions fluctuate event-by-event. Assuming that $v_2$ 
fluctuates according to a Gaussian distribution, the $v_2$ 
fluctuation is proportional to the fluctuation of the initial 
geometry. This effect scales as $1/\Np$, where \Np denotes the 
number of participant nucleons. The difference between $v_2$ 
values obtained from different methods can be quantitatively 
understood in terms of non-flow and fluctuation 
effects~\cite{Ollitrault:2009ie}.

Hence in this paper we will compare the $v_2$ results from the 
event plane determined at two different pseudorapidities with 
the goal to investigate the sensitivity of $v_2$ to non-flow and 
fluctuation effects. Additionally, we extract the elliptic flow 
with the two-particle cumulant method, which is expected to have 
higher sensitivity to non-flow contributions to $v_2$.

In this paper, we describe the PHENIX measurements of elliptic 
flow ($v_2$) at midrapidity ($|\eta| < 0.35$) in \Au collisions 
at \sqsn = 200 GeV obtained from a cumulant analysis of 
two-particle azimuthal correlations and the event plane method 
over a broad range of \pt (\pt = 0.2--8 \gevc) and centrality 
(0--60\%). The paper is organized as follows: 
Section~\ref{sec:experiment} describes the PHENIX apparatus, 
with an emphasis on the detectors relevant to the presented 
results, as well as the track selections used in the analysis. 
Section~\ref{sec:methods} gives details of the event-plane and 
the cumulant methods as applied in PHENIX. 
Section~\ref{sec:systematic_uncertainties} discusses the 
systematic uncertainties of the event-plane and cumulant 
methods. The results from the two methods are reported in 
Section~\ref{sec:results}. Section~\ref{sec:discussions} 
presents a comparison of $v_2$ results across different 
experiments and discussion.  The $v_2$ values obtained from the 
different methods are tabulated in the Appendix.

\section{Experimental Analysis \label{sec:experiment}}
\subsection{The PHENIX detector}
\begin{figure}[htbp]
\includegraphics[width=1.0\linewidth]{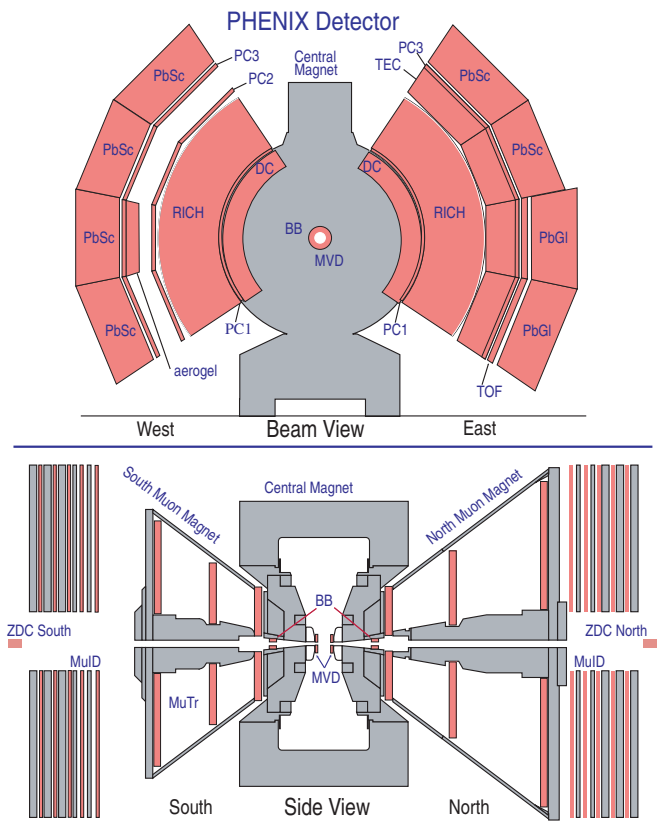}
\caption{PHENIX experimental layout in 2004. The top panel shows 
the PHENIX central arm spectrometers viewed along the beam axis. 
The bottom panel shows a side view of the PHENIX muon arm 
spectrometers.}
\label{fig:phenix}
\end{figure}

The PHENIX detector consists of two central spectrometer arms at 
midrapidity that are designated East and West for their location 
relative to the interaction region, and two muon spectrometers 
at forward rapidity, similarly called North and South. A 
detailed description of the PHENIX detector can be found in 
Ref.~\cite{Adcox:2003zm}. The layout of the PHENIX detector 
during data taking in 2004 is shown in Fig.~\ref{fig:phenix}. 
Each central spectrometer arm covers a pseudorapidity range of 
$|\eta| < 0.35$ subtending $90$ degrees in azimuth and is 
designed to detect electrons, photons and charged hadrons.  
Charged particles are tracked by drift chambers (DC) positioned 
between 2.0 m and 2.4 m radially outward from the beam axis and 
layers of multi-wire proportional chambers with pad readout (two 
in the east arm and three in the west arm) PC1, PC2 and PC3 
located at a radial distance of 2.4 m, 4.2 m and 5 m, 
respectively. Particle identification is provided by Ring 
Imaging $\check{\rm C}$erenkov counters (RICH), a time-of-flight 
scintillator wall (TOF), and two types of electromagnetic 
calorimeters (EMCAL), the lead scintillator (PbSc) and lead 
glass (PbGl).

The detectors used to characterize each event are the beam-beam 
counters (BBCs)~\cite{Allen:2003zt} and the zero-degree 
calorimeters (ZDCs)~\cite{Adler:2003sp}. These detectors are 
used to determine the time of the collision, the position of the 
collision vertex along the beam axis and the collision 
centrality and also provide the event trigger. In this analysis 
the BBCs are also used to determine the event plane. Each BBC is 
composed of 64 elements and a single BBC element consists of a 
one-inch diameter mesh dynode photomultiplier tube (PMT) mounted 
on a 3 cm long quartz radiator. The BBCs are installed on the 
north and south sides of the collision point along the beam axis 
at a distance of 144 cm from the center of the interaction 
region and surround the beam pipe. The BBC acceptance covers the 
pseudorapidity range $3.1 < |\eta | < 3.9 $ and the full range 
of azimuthal angles.

The ZDCs are hadronic calorimeters located on both sides of the 
PHENIX detector. Each ZDC is mechanically subdivided into 3 
identical modules of two interaction lengths. They cover a 
pseudorapidity range of $|\eta| > 6.5$ and measure the energy of 
the spectator neutrons with a 20 GeV energy 
resolution~\cite{Adler:2003sp}. The shower maximum detectors 
(ZDC-SMDs) are scintillator strip hodoscopes between the first 
and second ZDC modules. This location approximately corresponds 
to the maximum of the hadronic shower. The horizontal coordinate 
is sampled by 7 scintillator strips of 15 mm width, while the 
vertical coordinate is sampled by 8 strips of 20 mm width. The 
active area of a ZDC-SMD is 105 mm $\times$ 110 mm (horizontal 
$\times$ vertical dimension). Scintillation light is delivered 
to a multichannel PMT M16 by wavelength-shifter fibers. The 
ZDC-SMD position resolution depends on the energy deposited in 
the scintillator. It varies from $<$ 3 mm when the number of 
particles exceeds 100, to 10 mm for a smaller number of 
particles.

\subsection{Event selection}
\begin{figure}[htbp]
\includegraphics[width=1.0\linewidth]{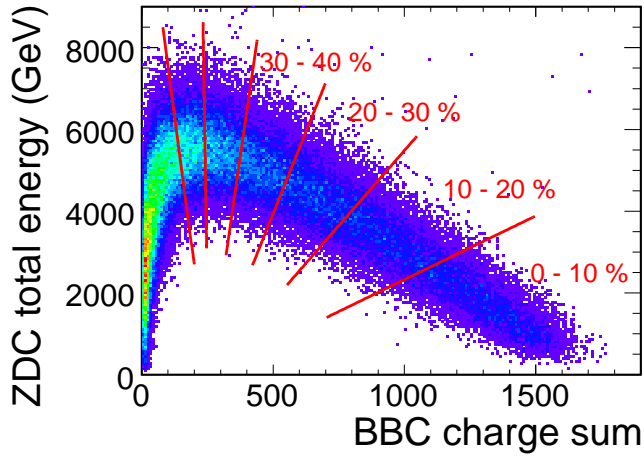}
\caption{ \label{bbc-zdc}
The correlation between ZDC energy and BBC charge sum for Au + Au collisions at \sqsn = 200 GeV. 
Solid lines represent the corresponding centrality boundaries up to 60\% centrality bin.
}
\end{figure}
For the analyses presented here we used approximately 850 
$\times$ 10$^6$ minimum-bias triggered events. The minimum-bias 
trigger was defined by a coincidence between North and South BBC 
signals and an energy threshold of one neutron in the ZDCs. The 
events are selected offline to be within a $z$-vertex of less 
than 30 cm from the nominal center of the PHENIX spectrometer. 
This selection corresponds to $92.2^{+2.5}_{-3.0}$\% of the 6.9 
barn \Au inelastic cross section at \sqsn = 200~GeV 
\cite{Miller:2007ri}. The event centrality was determined by 
correlating the charge detected in the BBCs with the energy 
measured in the ZDCs, as shown in Fig.~\ref{bbc-zdc}.

A Glauber model Monte-Carlo 
simulation~\cite{Glauber:1970jm,Adcox:2000sp,Adler:2003au} that 
includes the responses of BBC and ZDC gives an estimate of the 
average number of participating nucleons $\anpart$ for each 
centrality class. The simulation did not include fluctuations in 
the positions of the nucleons which give rise to eccentricity 
fluctuations. Table~\ref{tab:sys0} lists the calculated values 
of $\anpart$ for each centrality class.

\begin{table}[htbp]
\caption{\label{tab:sys0}
Centrality classes and average number of participant nucleons $\anpart$  obtained from a Glauber Monte-Carlo 
simulation of the BBC and ZDC responses for \Au collision at \sqsn = 200 GeV.  Each centrality class is expressed as 
a percentage of $\sigma_{\rm{AuAu}}$ = 6.9 b inelastic cross section. Errors denote systematic uncertainties from the 
Glauber MC simulation.
}
\begin{ruledtabular}\begin{tabular}{cccc}
&  Centrality     &       $\left<\Np\right>$ & \\ \hline
&   0--10\%        &       $325.2 \pm 3.3$    & \\  
&  10--20\%        &       $234.6 \pm 4.7$    & \\  
&  20--30\%        &       $166.6 \pm 5.4$    & \\ 
&  30--40\%        &       $114.2 \pm 4.4$    & \\  
&  40--50\%        &       $74.4  \pm 3.8$    & \\  
&  50--60\%        &       $45.5  \pm 3.3$    & \\  
\end{tabular}\end{ruledtabular}
\end{table}
\subsection{Track selection \label{subsec:track_selections}}

Charged particle tracks are measured using information from the 
DC, PC1 and PC3 detectors and the $z$-vertex from the BBC. The 
DC has 12 wire planes which are spaced at 0.6~cm intervals along 
the radial direction from the beam axis. Each wire provides a 
track position measurement, with better than 150 $\mu$m spatial 
resolution in the azimuthal ($\phi$) direction. The PC1 provides 
a space point in the $\phi$ and beam directions, albeit with 
lower resolution. This space point and the vertex position help 
determine the three-dimensional momentum vector by providing the 
polar angle for charged tracks at the exit of the DC. 
Trajectories are confirmed by requiring matching hits at PC3 to 
reduce secondary background. Tracks are then projected back to 
the collision vertex through the magnetic field to determine the 
momentum $\vec{p}$~\cite{Mitchell:2002wu}. The momentum 
resolution is $\delta p/p \simeq 0.7\% \oplus 1.0\%\times p$ 
(\gevc). The momentum scale is known to 0.7\%, as determined 
from the reconstructed proton mass using the TOF detector. 
Further details on track reconstruction and momentum 
determination can be found in 
Refs.~\cite{Mitchell:2002wu,Adler:2003au}.

The tracks reconstructed by the DC which do not originate from 
the event vertex have been investigated as potential background 
to the charged particle measurement. The main background sources 
include secondary particles from decays and $e^{+}e^{-}$ pairs 
from the conversion of photons in the material between the 
vertex and the DC~\cite{Adler:2003au}. Tracks are required to 
have a hit in the PC3, as well as in the EMCAL, within at most 2 
$\sigma$ of the expected hit location in both the azimuthal and 
beam directions. This cut reduces the background not originating 
in the direction of the vertex. In order to reduce the 
conversion background we further require tracks to have $E/\pt > 
0.2$, where $E$ denotes the energy deposited in the EMCAL and 
\pt is the transverse momentum of particles measured in the DC. 
Since most of the electrons from photon conversion are genuine 
low \pt particles that were reconstructed as high \pt particles, 
requiring a large deposit of energy in the EMCAL suppresses the 
electron background~\cite{Adler:2005ad}. We also require that 
there are no associated hits in the RICH. The RICH is filled 
with CO$_2$ gas at atmospheric pressure and has a charged 
particle threshold $\gamma_{\rm{th}}=35$ to emit 
$\check{\rm C}$erenkov photons.

\section{Methods of azimuthal anisotropy measurement\label{sec:methods}}

In this section we introduce the methods for azimuthal anisotropy measurements as used in 
the PHENIX experiment. Section III.A describes the event plane method using the BBCs and ZDC-SMDs 
detectors and Sec. III.B describes the two-particle cumulant method.

\subsection{Event plane method \label{subsec:eventplanemethod}}

The event plane method~\cite{Poskanzer:1998yz} uses the azimuthal anisotropy signal to estimate the angle
of the reaction plane. The estimated reaction plane is called 
the ``event plane" and is determined for each 
harmonic of the Fourier expansion of the azimuthal distribution. The event flow vector $\vec{Q}_n = (Q_x, Q_y)$ 
and azimuth of the event plane $\Psi_n$ for $n$-th harmonic of the azimuthal anisotropy can be expressed as
\begin{eqnarray}  
Q_x & \equiv & |\vec{Q}_n| \cos{(n\Psi_n)} = \sum_i^M w_i \cos{(n\phi_i)}, \label{eq:flowvector_x} \\
Q_y & \equiv & |\vec{Q}_n| \sin{(n\Psi_n)} = \sum_i^M w_i \sin{(n\phi_i)}, \label{eq:flowvector_y} \\
\Psi_n & = & \frac{1}{n} \tan^{-1}\left(\frac{Q_y}{Q_x}\right), \label{eq:eventplane_definition}
\end{eqnarray}
where $M$ denotes the number of particles used to determine the event plane, $\phi_i$ is the azimuthal 
angle of each particle, and $w_i$ is the weight chosen to optimize the event plane resolution. Once 
the event plane is determined, the elliptic flow $v_2$ can be extracted by correlating the azimuthal 
angle of emitted particles $\phi$ with the event plane 
\begin{eqnarray}
v_2 = \frac{v_2^{obs}}{\text{Res}\{\Psi_{n}\}} 
= \frac{\left<\cos{(2[\phi-\Psi_n])}\right>}{\left<\cos{(2[\Psi_{n}-\RP])}\right>},
\end{eqnarray}
where $\phi$ is the azimuthal angle of tracks in the laboratory frame, $\Psi_{n}$ is the $n$-th 
order event plane and the brackets denote an average over all charged tracks and events. The 
denominator Res\{$\Psi_n$\} is the event plane resolution that corrects for the difference 
between the estimated event plane $\Psi_n$ and true reaction plane \RP.

In this paper the second-harmonic event planes were independently determined with two BBCs located at 
forward (BBC South, referred to as BBCS) and backward (BBC North, referred to as BBCN) pseudorapidities 
$|\eta|$ = 3.1--3.9~\cite{Adler:2003kt}. The difference 
between the two independent event planes was 
used to estimate the event plane resolution. The planes were also combined to determine the event plane 
for the full event. A large pseudorapidity gap between the charged particles detected in the central 
arms and the event plane at the BBCs reduces the effect of possible non-flow contributions, especially 
those from dijets~\cite{Jia:2006sb}. The measured $v_2$ of hadrons in the central arms with respect to 
the combined second-harmonic BBC event plane will be denoted throughout this paper as $v_2$\{BBC\}.

Two first-harmonic event planes were also determined using spectator neutrons at the two shower maximum 
detectors (ZDC-SMDs) that are sandwiched between the first and second modules of the ZDCs. Forward 
(ZDCS) and backward (ZDCN) SMDs which cover pseudorapidity $|\eta| > $ 6.5 were used. The measured 
$v_2$ of hadrons in the central arms determined with respect to the first-harmonic ZDC-SMD event plane 
will be denoted as $v_2$\{ZDC-SMD\}.

The pseudorapidity gap between the hadrons measured in the central arms and the ZDC-SMDs is larger 
than that for the BBCs which could cause a further reduction of non-flow contributions on $v_2$\{ZDC-SMD\}. 
Since the ZDC-SMD measures spectator neutrons, the ZDC-SMD event plane should be insensitive to 
fluctuations in the participant event plane. Hence fluctuations in $v_2$\{ZDC-SMD\} should be suppressed 
up to fluctuations in the spectator positions.

For completeness, two further event planes are defined 1) a combined event plane defined by the weighted 
average of event planes at the forward and backward pseudorapidities for both BBCs and ZDC-SMDs, and 2) 
an event plane found using tracks in the central arm. The event plane at the central arms (CNT) is only 
used to estimate the resolution of BBC and ZDC-SMD event planes by using three subevents combination 
of the ZDC-SMD, BBC and CNT.

\subsubsection{Event plane determination}
\label{subsubsec:eventplane_determination}

To determine an event plane the contribution at each azimuthal angle needs to be appropriately weighted 
depending on the detector used. For the BBC we chose the weights to be the number of particles detected 
in each phototube, while for the ZDC-SMD the weights were based on the energy deposited in each of the 
SMD strips. For the CNT event plane the weight was taken to be proportional to \pt up to 2~\gevc and constant 
for \pt $>$ 2~\gevc. For the CNT event plane we also adopted a unit weight ($w_i = 1$)  and found that the 
resulting CNT event plane resolution extracted by comparing the CNT event plane with the BBC and ZDC-SMD 
planes was nearly identical when using the $\pt$-dependent or unit weights.

Corrections were performed to remove possible biases from the finite acceptance of the BBC and ZDC-SMD. 
In this analysis, we applied two corrections called the re-centering and shift methods. In the re-centering 
method, event flow vectors are shifted and normalized to a Gaussian distribution by using the mean $\left<Q\right>$ 
and width $\sigma$ of flow vectors;
\begin{eqnarray}
 Q_x' = \frac{Q_x - \left<Q_x\right>}{\sigma_x},\quad
 Q_y' = \frac{Q_y - \left<Q_y\right>}{\sigma_y}.
\end{eqnarray}
This correction reduces the dependence of the event plane resolution on the laboratory angle. Most acceptance 
effects were removed by the application of the re-centering method. However, remaining small corrections were 
applied after re-centering using the shift method~\cite{Poskanzer:1998yz}, in which the reaction plane is 
shifted by $\Delta\Psi_n$ defined by   
\begin{eqnarray}
n\Delta\Psi_n (\Psi_n) & = & \sum_{k=1}^{k_{\rm max}} \frac{2}{k} [ - \left<\sin{(kn\Psi_n)}\right>\cos{(kn\Psi_n)} \nonumber \\
& & ~ + \left<\cos{(kn\Psi_n)}\right>\sin{(kn\Psi_n)} ],
\label{eq:shift_correction}
\end{eqnarray}
where $k_{\rm max}$ = 8 in this analysis. The shift ensures that $dN/d\Psi_n$ is isotropic.
When $k_{\rm max}$ was reduced to $k_{\rm max} = 4$, the difference in the extracted $v_2$ was 
negligible and thus we include no systematic uncertainty due to the choice of $k_{\rm max}$ in our 
$v_2$ results.

Independent corrections were applied to each centrality selection in 5\% increments and in 20 cm steps 
in $z$-vertex in order to optimize the event plane resolution. The corrections were also done for each 
experimental run (the duration of a run is typically 1-3 hours) to minimize the possible time-dependent 
response of detectors.

\begin{figure}[htbp]
\includegraphics[width=1.0\linewidth]{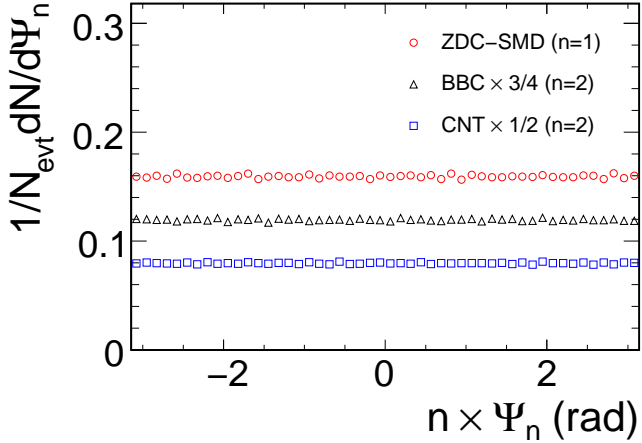}
\caption{\label{fig:eventplane_distribution_zdc-smd_bbc_cnt}
Event plane distributions after applying all corrections for the ZDC-SMD (circles), BBC (triangles) 
and CNT (squares). The statistical error bars are smaller than the symbols. The distributions for the 
BBC and CNT event planes are scaled by 3/4 and 1/2 to improve visibility. 
}
\end{figure}
Figure~\ref{fig:eventplane_distribution_zdc-smd_bbc_cnt} shows event plane distributions for a 
sub-sample of the entire data set. After all corrections are applied the event plane distributions 
are isotropic.

\subsubsection{Event plane resolution}
\label{subsubsec:eventplane_resolution}

The event plane resolution  for $v_2$ was evaluated by both the two-subevents and three-subevents methods. 
In the two-subevents method the event plane resolution~\cite{Poskanzer:1998yz} is expressed as
\begin{eqnarray}
& & \left<\cos{(kn[\Psi_n - \RP])}\right> 
= \frac{\sqrt{\pi}}{2\sqrt{2}}\chi_n e^{-\chi_n^2/4} \nonumber \\
& & \qquad \times \left[ I_{(k-1)/2}\left(\frac{\chi_n^2}{4}\right) + I_{(k+1)/2}\left(\frac{\chi_n^2}{4}\right) \right], 
\label{eq:resolution_formula}
\end{eqnarray}
where $\chi_n = v_n\sqrt{2M}$, $M$ is the number of particles used to determine the event 
plane $\Psi_n$, $I_k$ is the modified Bessel function of the first kind and $k$ = 1 for the 
second harmonic BBC event plane. For the ZDC-SMD event plane the resolution is estimated with 
both $k$ = 1 or 2 in Eq.~(\ref{eq:resolution_formula}). We will discuss the difference between 
these estimates in Sec.~\ref{subsec:syserror_eventplanemethod}.

To determine the event plane resolution we need to determine $\chi_n$. Since the North and 
South BBCs have approximately the same $\eta$ coverage, the event plane resolution of each 
sub-detector is expected to be the same. The same is true for the North and South ZDC-SMDs.
Thus, the subevent resolution for South and North event planes can be expressed as
\begin{eqnarray}
\left<\cos{(2[\Psi_n^{\rm S(N)} - \RP])}\right>
 = \sqrt{ \left<\cos{(2[\Psi_n^{\rm S} - \Psi_n^{\rm N}])}\right> },
 \label{eq:eventplane_resolution_twosubevents}
\end{eqnarray}
where $\Psi_n^{\rm S(N)}$ denotes the event plane determined by the South (North) BBC or ZDC-SMD. 
Once the subevent resolution is obtained from Eq.~(\ref{eq:eventplane_resolution_twosubevents}), 
one can calculate $\chi_n^{\rm sub}$ using Eq.~(\ref{eq:resolution_formula}). The $\chi_n$ for the 
full event can then be estimated by $\chi_n = \sqrt{2}\chi_n^{\rm sub}$. This is then substituted 
into Eq.~(\ref{eq:resolution_formula}) to give the full event resolution. Since the multiplicity 
of the full event is twice as large as that of the subevent, $\chi_n$ is proportional to $\sqrt{M}$.

In the three-subevents method the resolution of each subevent is calculated by adding a reference 
event plane $\Psi_n^{\rm C}$ in Eq.~(\ref{eq:eventplane_resolution_twosubevents}):  
\begin{eqnarray}
 & & \text{Res}\{\Psi_l^{\rm A}\} 
 = \sqrt{ \left< \cos{(2[\Psi_l^{\rm A} - \Psi_m^{\rm B}])} \right> } \nonumber \\
 & & \times \sqrt{ \frac{ \left< \cos{(2[\Psi_n^{\rm C} - \Psi_l^{\rm A}])} \right> }
 { \left< \cos{(2[\Psi_m^{\rm B} - \Psi_n^{\rm C}])} \right> } 
 },
 \label{eq:eventplane_resolution_threesubevents}
\end{eqnarray}
where $l, m, n$ are the harmonics of the event plane for subevent A, B and C, respectively. 
The multiplicity of each subevent is not necessarily the same in Eq.~(\ref{eq:eventplane_resolution_threesubevents}).

The resolution of each sub-detector for the BBC and ZDC-SMD can be evaluated with the 
three-subevents method. For the BBC event plane the reference event plane is chosen to be 
the ZDC-SMD event plane and vice versa. We found that the agreement of the event plane 
resolutions for BBCS and BBCN is much better than 1\%, while the ZDCS and ZDCN resolutions 
are comparable with each other within 2\%.

\begin{figure}[t]
\includegraphics[width=1.0\linewidth]{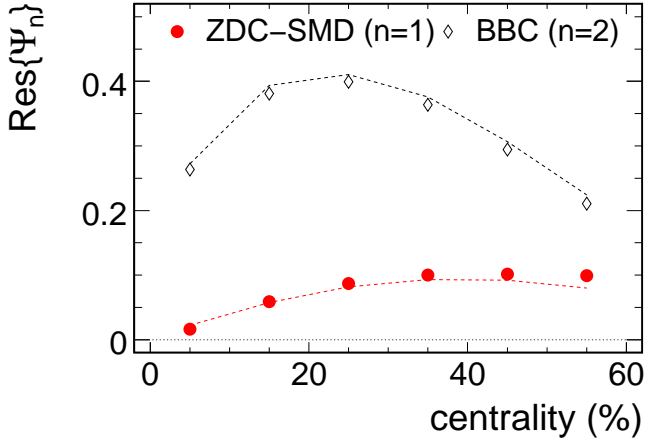}
\caption{\label{fig:eventplane_resolution_twosubevents}
Full-event resolutions for the ZDC-SMD (filled circles) and BBC (open diamonds) from 
the two-subevents method, Eq.~(\ref{eq:resolution_formula}), as a function of centrality 
in \Au at \sqsn = 200~GeV. The dashed lines represent resolutions from the three-subevents 
method with the CNT event plane as a reference. Statistical errors are smaller than the symbols.
}
\end{figure}
Figure~\ref{fig:eventplane_resolution_twosubevents} shows the full-event resolution as a function 
of centrality. The resolution of ZDC-SMD is much smaller than that of BBC because the resolution 
of the first-harmonic event plane is proportional to $(\chi_1)^2$. The dashed lines are the resolutions 
obtained from the three-subevents method with the CNT event plane as the reference plane. For 
example, the BBC event plane resolution is estimated by substituting 
$\Psi_l^{\rm A} \rightarrow \Psi_2^{\rm BBC}$, $\Psi_m^{\rm B} \rightarrow \Psi_2^{\rm CNT}$, and
$\Psi_n^{\rm C} \rightarrow \Psi_1^{\rm ZDC-SMD}$ in Eq.~(\ref{eq:eventplane_resolution_threesubevents}). 
By including the CNT event plane, the BBC resolution increases by about 3\% compared to that of the 
two-subevents method. For the ZDC-SMD we observe the opposite effect, namely the resolution 
decreases by about 8\%. In Sec.~\ref{sec:discussions} the resulting $v_2$\{BBC\} and $v_2$\{ZDC-SMD\}, 
corrected by the resolution obtained using the ZDC-BBC-CNT combination, will be compared to those 
with the resolution determined from South-North subevents.
Table~\ref{tab:summary_eventplane_resolution}
summarizes the event plane resolutions.

\begin{table}[htbp]
\caption{\label{tab:summary_eventplane_resolution}
Event plane resolutions for centrality 0--60\% at \sqsn = 200~GeV. 
S-N denotes the resolutions estimated from South and North 
correlation of BBC and ZDC-SMD using 
Eq.~(\ref{eq:resolution_formula}) 
and~(\ref{eq:eventplane_resolution_twosubevents}), and 
resolutions for ZDC-BBC-CNT are estimated from 
Eq.~(\ref{eq:eventplane_resolution_threesubevents}). The errors 
are statistical only.
}
\begin{ruledtabular} \begin{tabular}{ccc}
\multicolumn{3}{c}{Res\{$\Psi_2^{\rm BBC}$\}} \\  \hline
 Centrality   &  S-N & ZDC-BBC-CNT \\ \hline
  0--10\%  & 0.2637 $\pm$ 0.0003 & 0.272  $\pm$ 0.003  \\
 10--20\%  & 0.3809 $\pm$ 0.0002 & 0.394  $\pm$ 0.001  \\
 20--30\%  & 0.3990 $\pm$ 0.0002 & 0.4106 $\pm$ 0.0008 \\
 30--40\%  & 0.3634 $\pm$ 0.0002 & 0.3759 $\pm$ 0.0007 \\
 40--50\%  & 0.2943 $\pm$ 0.0003 & 0.3067 $\pm$ 0.0007 \\
 50--60\%  & 0.2106 $\pm$ 0.0004 & 0.2240 $\pm$ 0.0009 \\
\hline
 & & \\
\multicolumn{3}{c}{Res\{$\Psi_1^{\rm ZDC-SMD}$\}} \\ \hline
 Centrality   &  S-N & ZDC-BBC-CNT \\ \hline
  0--10\%  & 0.02  $\pm$ 0.01  & 0.0223 $\pm$ 0.0003  \\
 10--20\%  & 0.059 $\pm$ 0.003 & 0.0574 $\pm$ 0.0002  \\
 20--30\%  & 0.087 $\pm$ 0.002 & 0.0818 $\pm$ 0.0002  \\
 30--40\%  & 0.100 $\pm$ 0.002 & 0.0928 $\pm$ 0.0002  \\
 40--50\%  & 0.102 $\pm$ 0.002 & 0.0920 $\pm$ 0.0002  \\
 50--60\%  & 0.100 $\pm$ 0.002 & 0.0798 $\pm$ 0.0003  \\
\end{tabular}\end{ruledtabular}
\end{table}

\subsubsection{Correlation of event planes}
\label{subsubsec:correlation_eventplanes}
\begin{figure}[tbh]
\includegraphics[width=1.0\linewidth]{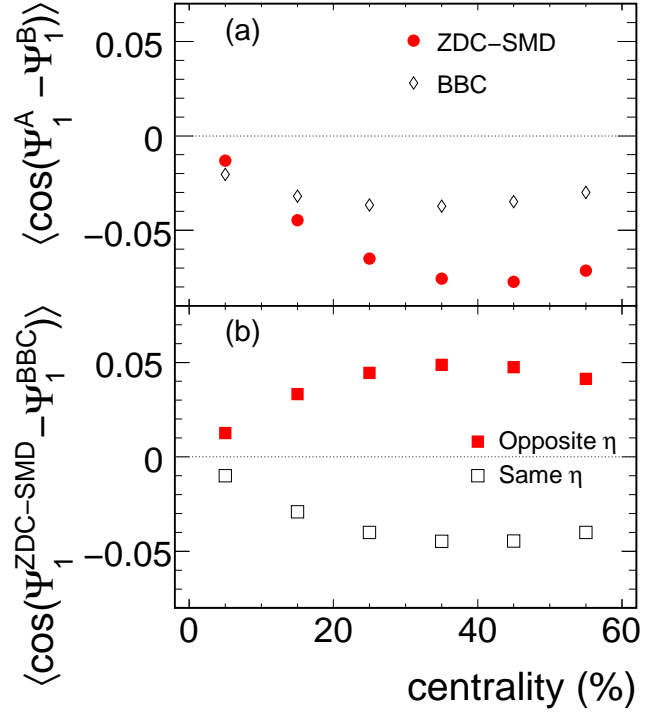}
\caption{\label{fig:correlation_zdc-smd_bbc_1storder}
(a) Correlation of first harmonic event planes between forward and backward 
ZDC-SMDs (filled circles) and BBCs (open diamonds) as a function of centrality.
(b) Correlation of first harmonic event planes between ZDC-SMDs and BBCs as a 
function of centrality, where filled (open) squares are the correlation for opposite side 
(same side) of $\eta$ subevents. Statistical errors are smaller than the symbols.
}
\end{figure}
Figure~\ref{fig:correlation_zdc-smd_bbc_1storder} shows the correlation of two different event 
planes as a function of centrality. The first harmonic event plane correlation for South-North 
detector combinations is negative both for the ZDC-SMDs and the BBCs over all centrality bins, 
as shown in \Fig~\ref{fig:correlation_zdc-smd_bbc_1storder}(a). 
This is due to the fact that $v_1$ 
is an odd function of $\eta$. The magnitude of the ZDC-SMDs correlation is about a factor of two 
larger than that of the BBCs for midcentral collisions.  This indicates that the magnitude of $v_1$ 
and/or the subevent multiplicity at higher pseudorapidities are larger compared to that at the BBC 
location, since the magnitude of the correlation is proportional to $v_1^2M$. 
Fig.~\ref{fig:correlation_zdc-smd_bbc_1storder}(b) shows the 
correlation of the first harmonic event 
planes between BBC and ZDC-SMD. The same-side $\eta$ correlation is negative while the 
opposite-side $\eta$ correlation is positive, which shows that the particles detected at the BBCs 
(dominantly charged pions emitted from participant nucleons) have the opposite sign of $v_1$ 
compared to the spectator neutrons detected at the ZDCs-SMDs. 


\begin{figure}[thb]
\includegraphics[width=1.0\linewidth]{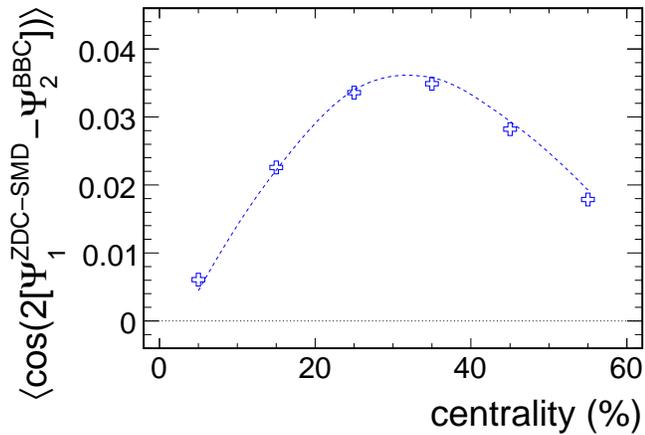}
\caption{\label{fig:correlation_zdc-smd_bbc_2ndorder}
The correlation between the first harmonic ZDC-SMD and the second harmonic BBC 
event planes as a function of centrality. The dashed line shows the result obtained using 
Eq.~(\ref{eq:mixed_harmonic_eventplane_correlation_approximation}).
Statistical errors are smaller than the data symbols.
}
\end{figure}
The correlation of the mixed harmonic event planes provides the sign of $v_2$  
since the correlation is given by the expression~\cite{Poskanzer:1998yz}
\begin{eqnarray}
& & \left< \cos{(2[\Psi_1^{\rm ZDC-SMD}-\Psi_2^{\rm BBC}])} \right> \nonumber \\
& & \qquad \approx \frac{2}{\pi} \left(\text{Res}\{\Psi_1^{\rm ZDC-SMD}\}\right)^2
          \text{Res}\{\Psi_2^{\rm BBC}\} \nonumber \\
& & \qquad = \pm 2\sqrt{2} \frac{2}{\pi} 
  \left< \cos{(\Psi_1^{\rm ZDCS}-\Psi_{1}^{\rm ZDCN})} \right> \nonumber \\
& & \qquad ~~ \times \sqrt{\left< \cos{(2[\Psi_2^{\rm BBCS}-\Psi_{2}^{\rm BBCN}])} \right>}.
\label{eq:mixed_harmonic_eventplane_correlation_approximation}
\end{eqnarray}
Three assumptions were made to obtain 
Eq.~(\ref{eq:mixed_harmonic_eventplane_correlation_approximation}):
(1) the BBC and ZDC-SMD are statistically independent, 
(2) the weak flow limit is applicable, and
(3) the subevent multiplicity $M$ is equal in the North-South direction for the same detector type.
Thus the sign of the correlation of the mixed harmonic event planes in 
Eq.~(\ref{eq:mixed_harmonic_eventplane_correlation_approximation}) is determined by the term 
Res$\{\Psi_2^{\rm BBC}\}$, which in turn determines the sign of $v_2$ measured at the BBC.

Figure~\ref{fig:correlation_zdc-smd_bbc_2ndorder} shows the mixed harmonic correlation of the 
ZDC-SMD and BBC event planes as a function of centrality. The approximations in 
Eq.~(\ref{eq:mixed_harmonic_eventplane_correlation_approximation}) provide a good description of 
the magnitude of the measured correlation as shown by the dashed line.  The correlation is positive 
over all centrality bins. This result indicates that the sign of $v_2$ at the BBC is positive.


\subsection{Cumulant method \label{subsec:cumulantmethod}}

In this section, we present the application of the cumulant method for azimuthal 
anisotropy measurements in PHENIX.  This method uses cumulants of multiparticle 
correlations~\cite{Borghini:2001vi,Borghini:2001zr} to extract the azimuthal anisotropy.  
The cumulant method has been successfully 
applied in several heavy-ion experiments utilizing detectors with full azimuthal coverage 
(NA49, STAR)~\cite{Alt:2003ab,Adler:2002pu}. Here, we describe the first application of the method for a 
detector with only partial azimuthal coverage. The cumulant method does not require the measurement of the 
reaction plane, instead the cumulants of multi-particle azimuthal correlations are related to the flow 
harmonics $v_n$, where $n$ is the harmonic being evaluated. The cumulants can be constructed in increasing 
order according to the number of particles that are correlated with each other. Since PHENIX has partial 
azimuthal coverage, reliable extraction of azimuthal anisotropy requires the choice of a fixed number of 
particles from each event in order to avoid additional numerical errors~\cite{Borghini:2001vi}. 

Particles in an event are selected over a fixed \pt range where there is sufficient multiplicity.
These particles (called ``integral particles" hereafter) are 
used to determine integrated flow, that is
flow measured over a large (\pt,$\eta$) bin. For differential flow measurement, we select particles 
(called ``differential" particles) over small (\pt,$\eta$) bins, 
from which the integral particles are excluded 
so as to avoid autocorrelations. For each event a fixed number $M$ of particles, chosen at random among 
the integral particles in the event, are used to reconstruct the integrated flow through 
the generating function $G_2(z)$ defined by:

\begin{eqnarray}
\label{genfunc}
G_2(z) & = & \prod_{j=1}^M  \left[ 1+\frac{w_j}{M}( z^* e^{2i\phi_j} 
+ z e^{-2i\phi_j})\right],
\end{eqnarray}
where $w_j $ is the weight, chosen to be equal to 1 in our analysis, $\phi_j$ is the azimuth of the detected 
particles, and $M$ is the multiplicity chosen for the integrated flow reconstruction. $G_2(z)$ is a function 
of the complex variable $z$. The average of $G_2(z)$ over events is then expanded in a power series to 
generate multi-particle azimuthal correlations. The generating function of the cumulants, defined by
\begin{equation}
\label{defc0}
{\cal C}_n(z)\equiv
M\left(\mean{G_n(z)}^{1/M}-1\right).
\end{equation}
generates cumulants of azimuthal correlations to all orders, the lowest being the second order, as detailed in 
Section~II.B of Ref.~\cite{Borghini:2001vi}. The formulas used to compute the cumulants from 
which the $v_2$ is computed 
are given in Appendix B of Ref.~\cite{Borghini:2001vi}. In the case of a perfect acceptance 
the relations between 
the anisotropy parameter $v_2$ and the lowest order cumulants are 
\begin{eqnarray}
\label{v2int}
v_2\{2\}^2 & = & c_2\{2\}, \\
v_2\{4\}^4 & = & -c_2\{4\}, 
\end{eqnarray}
for the integrated anisotropy. Here $v_2\{2\}$ and $v_2\{4\}$ 
are the second and fourth order $v_2$, respectively; whereas, 
$c_2\{2\}$ and $c_2\{4\}$ are the second and fourth order 
cumulants.  Because the typical multiplicity of charged hadrons 
in PHENIX did not allow a reliable calculation of $v_2$\{4\}, we 
report here only the $v_2$\{2\} results.

The remaining differential particles in the same event are selected in different (\pt, $\eta$) bins and 
the differential cumulants are calculated from the generating function
\begin{eqnarray}
\label{defcm}
{\cal D}_{2/2}(z)
\equiv\frac{\mean{e^{2i\psi}G_2(z)}}{\mean{G_2(z)}},
\end{eqnarray}
where $\mean{G(z)}$ denotes an average over all events, and $\psi$ is the azimuth of each differential particle.
$D_{2/2}$ denotes the second order differential cumulant computed with respect to the second order integral
cumulant.

The differential $v_{2/2}\{2\}(\pt,\eta)$, the second order differential \v2 with respect to the second order 
integrated \v2, is calculated from the relation
\begin{eqnarray}
\label{v2diff}
v_{2/2}\{2\}(\pt,\eta) & = &  \frac{d_{2/2}\{2\}(\pt,\eta)}{v_{2}\{2\}}, 
\end{eqnarray}
where $d_{2/2}\{2\}(\pt,\eta)$ is the second order differential cumulant. These relations have to be modified 
through acceptance corrections which are detailed below.

\subsubsection{Acceptance/efficiency corrections}
The central arms detectors in PHENIX have only partial azimuthal coverage and the implementation of the cumulant 
method requires an additional acceptance correction. In order to correct for the influence of the detector acceptance 
on the raw anisotropy values, we apply a correction factor using the prescription described in Ref.~\cite{Borghini:2001vi}. 
The acceptance and efficiency of the detector is characterized by a function $A(\phi, \pt, \eta)$ which is expressed in 
terms of the Fourier series

\begin{equation}\label{Aphi}
A(\phi,\pt,\eta) = \sum_{p=-\infty}^{+\infty} a_p(\pt,\eta)\, e^{ip\phi}.
\end{equation}
The Fourier coefficients $a_p(\pt,\eta)$ for the detector acceptance were 
extracted from the fit of the respective azimuthal distributions of 
integral and differential particles. The coefficients resulting from such 
fits were then used to calculate the correction factor for the raw values 
of the $v_2$ following the procedure detailed in Appendix C of 
Ref.~\cite{Borghini:2001vi}.

\begin{figure}[hbt]
\includegraphics[width=1.0\linewidth]{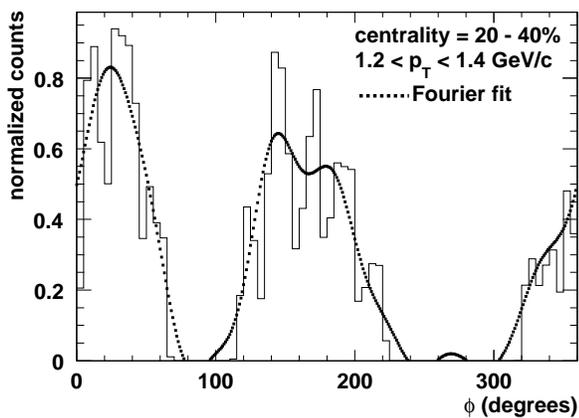}
\caption{\label{phiandfour} 
Azimuthal angular distribution and corresponding Fourier fit
for centrality 20--40\% and \pt = 1.2--1.4 \gevc.}
\end{figure}

\begin{figure}[hbt]
\includegraphics[width=1.0\linewidth]{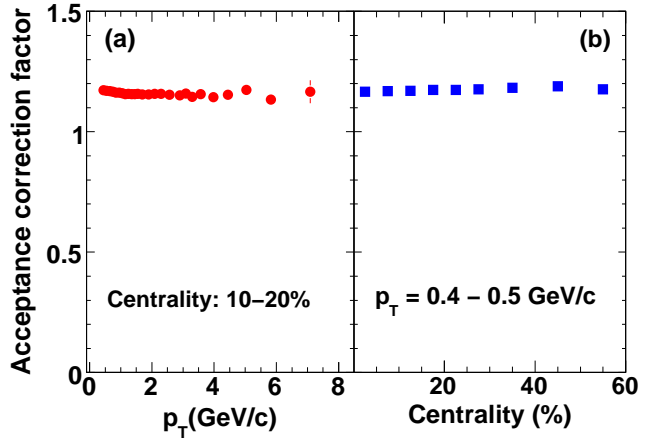}
\caption{\label{corrfac} (a) Acceptance correction factor for 
differential $v_2\{2\}$ as a function of \pt for centrality 
10--20\% (b) Acceptance correction factor as a function of 
centrality for \pt range 0.4--0.5 GeV/$c$ in \Au collisions at 
\sqsn = 200~GeV.}
\end{figure}

Figure~\ref{phiandfour} shows a typical azimuthal angular distribution of 
differential particles detected in the PHENIX central arms and the 
corresponding Fourier fit used to correct for acceptance inhomogeneities. 
The Fourier fit reproduces well the overall features of the acceptance 
profile. This produces typical correction factors that are in the range 
1.1--1.2 for the differential flow and depend very little on centrality 
and \pt, as shown in Fig.~\ref{corrfac}.

\subsubsection{Simulations}

While \Fig~\ref{phiandfour} shows that the uneven detector acceptance is 
reproduced by the Fourier fit, a better test of the cumulant method is to 
use Monte-Carlo simulations, as was done in Ref.~\cite{Borghini:2001vi}.  
For these tests events were generated with particles having a distribution 
of the form $1 + 2v_1\cos{\phi} + 2v_2\cos{2\phi}$, with known integrated 
and differential azimuthal anisotropies. The anisotropy was introduced 
into the events by way of a Fourier weighted selection of the azimuthal 
angles followed by a random event rotation designed to simulate the random 
orientation of the reaction plane. The multiplicity of these events was 
chosen to reflect the typical multiplicity measured with the PHENIX 
detector and the $\phi$ angles were chosen from a filter that is 
representative of the PHENIX acceptance. We extracted Fourier components 
from these simulated results and applied these to extract corrected 
elliptic flow values.

\begin{figure}[hbt]
\includegraphics[width=1.0\linewidth]{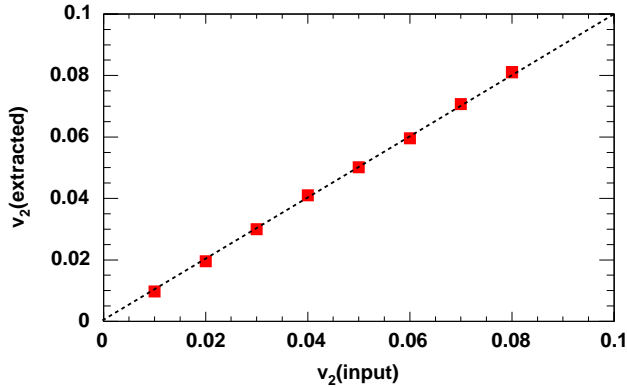}
\caption{\label{v2outvsv2in}
Comparison of input and extracted differential $v_{2}$ values for a fixed 
integral $v_{2}$ of 8$\%$. The dotted line indicates the expectation if 
input and reconstructed values are the same.}
\end{figure}

Figure~\ref{v2outvsv2in} shows selected results from these 
simulations.  Corrected differential anisotropy values are 
compared for various input differential $v_2$ values, with the 
integral $v_2$ kept fixed. The dotted line shows the trend 
expected if the extracted $v_2$ is identical to the input value 
used to generate the events. The good agreement between the 
input and extracted $v_2$ attests to the reliability of the 
analysis method within the acceptance of the PHENIX central 
arms.

\section{Systematic Uncertainties\label{sec:systematic_uncertainties}}

In this section, we present the systematic uncertainties on the $v_2$ from the event plane method 
(Sec.~\ref{subsec:syserror_eventplanemethod}) and the two-particle cumulant method 
(Sec.~\ref{subsec:syserror_cumulantmethod}). 
Table~\ref{tab:summary_systematicerror_eventplane} lists
the different sources of systematic errors for each method.
The errors in Tab.~\ref{tab:summary_systematicerror_eventplane}
are categorized by type:
  \begin{itemize}
    \item[A] point-to-point error uncorrelated between \pt bins,
    \item[B] \pt correlated, all points move in the same direction
but not by the same factor,
    \item[C] an overall normalization error in which all points
move by the same factor independent of \pt.
  \end{itemize}

\begin{table}[htbp]
\caption{\label{tab:summary_systematicerror_eventplane} 
List of systematic uncertainties given in percent on the $v_2$\{ZDC-SMD\}, $v_2$\{BBC\} and $v_2$\{2\} 
measurements. The ranges correspond to different systematic errors for different centrality bins.
}
\begin{ruledtabular} \begin{tabular}{lccc}
Error source              & \multicolumn{2}{c}{Percentage error}          &Type\\ \hline
                          & $v_2$\{BBC\} \  &  \ $v_2$\{ZDC-SMD\}         &    \\ \hline
Background contribution   & \multicolumn{2}{c}{$<$ 5\% in \pt $<$ 4 \gevc}& B  \\
	                          & \multicolumn{2}{c}{5--30\% in \pt $>$ 4 \gevc} & B  \\
Event plane calibration   & \multicolumn{2}{c}{1--5\%}                     & C  \\
Event plane determination &  1--4\%        & 1--16\%                        & C  \\
Acceptance effect         &  1\%          & 1--25\%                        & C  \\ 
on event planes           &               &                               &    \\ \hline
                          & \multicolumn{2}{c}{$v_2$\{2\}}                &    \\ \hline
Fixed multiplicity        & \multicolumn{2}{c}{ 5\%}                      & B  \\
Integrated \pt range      & \multicolumn{2}{c}{ 3--8\%}                    & B  \\
Background correction     & \multicolumn{2}{c}{ 6--10\%}                   & B  \\
\end{tabular} \end{ruledtabular}
\end{table}

\subsection{Event plane method \label{subsec:syserror_eventplanemethod}}

\subsubsection{Background contributions}

In order to study the influence of background on our results, we varied one of the track 
selections while keeping other cuts fixed and investigated the effect on $v_2$ in the following
two cases:
(i) the PC3 and EMCAL matching cuts, $\pm$ 1.5 and $\pm$ 2.5$\sigma$ matching cuts and
(ii) $E > 0.15 \pt$ and $E > 0.25 \pt$.
For both conditions, we found that the difference of the $v_2$ is 1--2\% for \pt $<$ 4 \gevc,
and 5--20\% for \pt $>$ 4 \gevc depending on \pt and centrality.

\begin{figure}[t]
\includegraphics[width=1.0\linewidth]{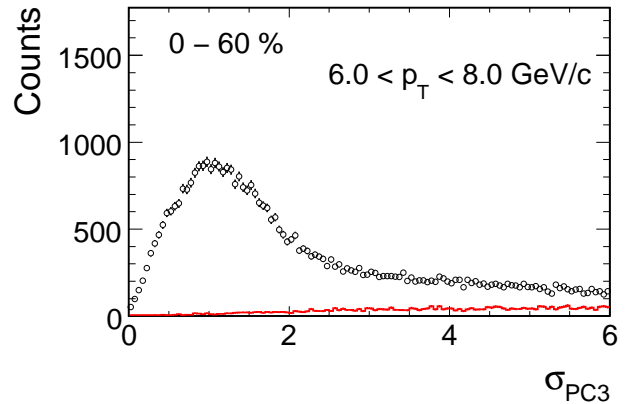}
\caption{\label{fig:pc3matching_real_and_random}
The radial PC3 matching distribution for real (open circles) and random tracks (solid lines)
for 6 $<$ \pt $<$ 8 \gevc in centrality 0--60\%.
}
\end{figure}

The effect of the RICH veto cut has also been studied. Since the 
contribution of charged $\pi$ increases without the RICH veto 
cut, the $p/\pi$ ratio decreases at high \pt. Thus, the $v_2$ 
for charged hadrons could be modified due to the difference of 
$v_2$ between protons and $\pi$ in the range 4 $<$ \pt $<$ 8 
\gevc. We found that $v_2$ is 10--20\% different without the 
RICH veto cut for \pt $>$ 4--5 \gevc, where the charged $\pi$ starts 
firing the RICH.

One of the remaining sources of background contribution comes 
from the random tracks that are accidentally associated with the 
tracks in PC3. These random tracks have been estimated by 
swapping the z-coordinate of the PC3 hits and then by 
associating those hits with the real tracks. 
Figure~\ref{fig:pc3matching_real_and_random} shows the 
comparison of the radial PC3 matching distribution between the 
real and random tracks for 6 $<$ \pt $<$ 8~\gevc.  The signal to 
background ratio $S/B$ is evaluated in the $\sigma_{\rm PC3} < 2$ 
window, and is $\sim$ 52 for 6 $<$ \pt $<$ 8 \gevc in 
centrality 0--60\%.

\begin{figure}[t]
\includegraphics[width=1.0\linewidth]{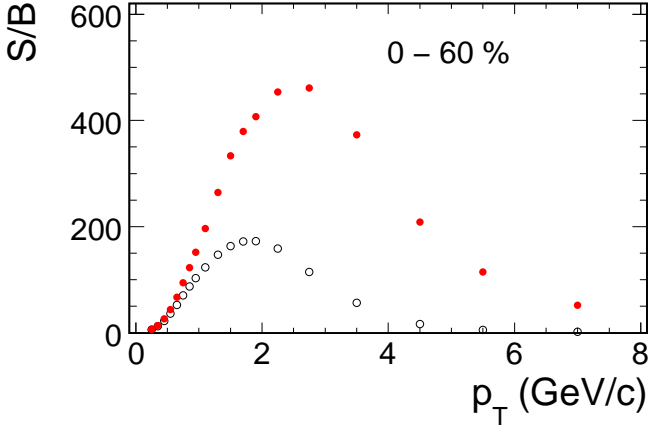}
\caption{\label{fig:sb_vs_pt_real_and_random}
The ratio of real $S$ to random tracks $B$ as a function of \pt 
in centrality 0--60\%. Solid and open circles show the $S/B$ 
ratio with and without $E/\pt > 0.2$, respectively.
}
\end{figure}

The ratio of real and random tracks with and without the 
$E/\pt > 0.2$ cut is shown as a function of \pt for centrality 
0--60\% in Fig.~\ref{fig:sb_vs_pt_real_and_random}. 
The $E/\pt > 0.2$ cut reduces the random tracks and improves the 
$S/B$ ratio by a factor of $\approx$ 10--24 for \pt $>$ 4~\gevc. 
Since random tracks are not expected to be correlated with the 
event plane, we assume that their $v_2 = 0$ and evaluate the 
systematic uncertainty on $v_2$ to be less than 2\% for \pt $>$ 
4~\gevc, increasing to 5\% for \pt $<$ 0.5~\gevc.

\begin{figure}[b]
\includegraphics[width=1.0\linewidth]{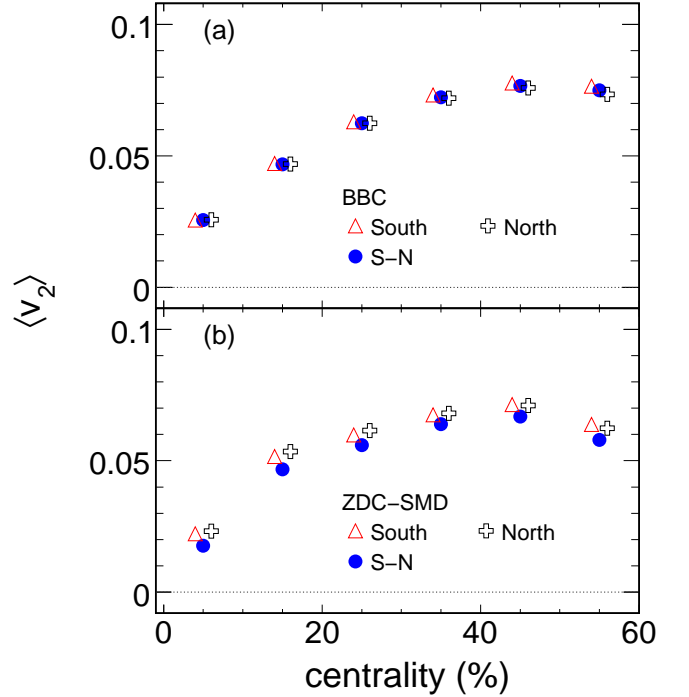}
\caption{\label{fig:comparison_v2cent_bbc_and_zdc-smd_differentep} 
(a) Comparison of $\left<v_2\right>$ averaged over 0.2 $<$ \pt 
$<$ 8 GeV/$c$ as a function of centrality for the BBC event 
planes. Open triangles and crosses represent the $v_2$ with 
respect to the event planes from South and North sub-detectors 
and filled circles show the $v_2$ from combined South-North 
event planes. Results from South and North event planes are 
shifted in the x-direction to improve visibility. 
(b) The same comparison for the ZDC-SMD event planes. Only 
statistical errors are shown and they are smaller than the 
symbols.
}
\end{figure}
 
There is a finite residual background contribution even after 
the $E/\pt > 0.2$ has been applied, as observed in 
Fig.~\ref{fig:pc3matching_real_and_random}. The residual 
backgrounds have been estimated by fitting the $\sigma_{\rm 
PC3}$ with a double Gaussian while requiring that the signal and 
residual background $\sigma_{\rm PC3}$ distribution have the 
same mean. For the highest \pt bin, we found that the signal to 
background ratio is $\sim$ 5 for $\sigma_{\rm PC3} < 2$. The 
systematic error on $v_2$ is evaluated by comparing the measured 
$v_2$ with that of signal
\begin{eqnarray}
v_2^{S} = \left( 1+\frac{B}{S} \right)v_2 - \frac{B}{S}v_2^{B},
\end{eqnarray}
where $v_2^S$, $v_2^B$ and $v_2$ are respectively $v_2$ of 
signal, background estimated for $\sigma_{\rm PC3} > 3$, and 
measured within the 2$\sigma$ matching window. The systematic 
uncertainties are less than 5\% for \pt $<$ 4~\gevc, and $\sim$ 
5--10\% for higher \pt. All the above systematic errors are added 
in quadrature and the overall systematic error from the 
background contribution is estimated to vary from $<$ 5\% for 
\pt $<$ 4~\gevc to $\sim$30\% for higher \pt.

\subsubsection{Event plane calibrations}
\label{subsubsec:syserror_eventplane_calibrations}

The procedures used in the determination and calibration of 
event planes are the dominant sources of systematic errors on 
$v_2$ and are discussed in the following sections.


Different calibration procedures of the BBC event plane were 
extensively studied for previous Au + Au data 
sets~\cite{Adler:2003kt}. We followed the same procedure to 
study the systematic errors on the BBC and ZDC-SMD event planes. 
Systematic uncertainties from the shift methods on $v_2$\{BBC\} 
are $\sim$ 1-5\% depending on the centrality. The systematic 
errors on the $v_2$\{ZDC-SMD\} are 1-2\% larger than those on 
$v_2$\{BBC\} for centrality 10--30\% and 50--60\%, although 
those 
are still less than 5\%.

\subsubsection{Event plane determination}
\label{subsubsec:syserror_eventplane_determination}

Figure~\ref{fig:comparison_v2cent_bbc_and_zdc-smd_differentep} shows the comparison of $\left<v_2\right>$ 
for different sub-detectors with respect to the BBC and ZDC-SMD event planes as a function of 
centrality. Systematic errors are estimated by taking the maximum difference of the $v_2$ from the
South and North event planes to that from the combined South-North event plane scaled by $2/\sqrt{12}$ 
for each centrality. Systematic errors range from 1-4\% for the BBC, and 1-16\% for the ZDC-SMD 
event planes depending on the centrality bins.
 
\subsubsection{Effect of non-uniform acceptance on $v_2$}
\label{subsubsec:syserror_nonuniform_acceptance}

In this subsection we discuss the effect of non-uniform acceptance on the measured $v_2$. 
In practice, the imperfect azimuthal acceptance of the BBC or ZDC-SMD or the central arms
could induce an azimuthal-dependent event plane resolution and/or smear the magnitude of 
$v_2$. In order to study the possible effect of non-uniform acceptance,
the measured $v_2$ is decomposed into X and Y components \cite{Selyuzhenkov:2007zi}:
 \begin{eqnarray}
v_2^{\rm X} & = & \frac{\sqrt{2}}{a^+_4}\frac{\left<\cos{(2\phi)}\cos{(2\Psi_n^{\rm A})}\right>}{{\rm Res}\{\Psi_n^{\rm A};{\rm X}\}}, \nonumber \\ 
v_2^{\rm Y} & = & \frac{\sqrt{2}}{a^-_4}\frac{\left<\sin{(2\phi)}\sin{(2\Psi_n^{\rm A})}\right>}{{\rm Res}\{\Psi_n^{\rm A};{\rm Y}\}},
\label{eq:v2xy}
\end{eqnarray}
where $\phi$ denotes the azimuthal angle of hadrons measured in the central arms 
and $a^{\pm}_4 = 1 \pm \left<\cos{(4\phi)}\right>$ are the acceptance correction factors of the measured 
$v_2$ in the central arms. The coefficient $a_4^{\pm}$ should be unity in the case of perfect azimuthal 
acceptance. Res$\{\Psi_n^{\rm A};{\rm X}\}$ and Res$\{\Psi_n^{\rm A};{\rm Y}\}$ denote the event plane 
resolution for $v_2^{\rm X}$ and $v_2^{\rm Y}$ respectively and are expressed as
\begin{eqnarray}
& & {\rm Res}\{\Psi_l^{\rm A};{\rm X}\} = \sqrt{ \left<\cos{(2\Psi_l^{\rm A})}\cos{(2\Psi_m^{\rm B}))}\right> } \nonumber \\
& \times & \sqrt{\frac{\left<\cos{(2\Psi_n^{\rm C})}\cos{(2\Psi_l^{\rm A})}\right>}{\left<\cos{(2\Psi_m^{\rm B})}\cos{(2\Psi_n^{\rm C})}\right>}}, \nonumber \\
& & {\rm Res}\{\Psi_l^{\rm A};{\rm Y}\} = \sqrt{ \left<\sin{(2\Psi_l^{\rm A})}\sin{(2\Psi_m^{\rm B}))}\right> } \nonumber \\
& \times & \sqrt{\frac{\left<\sin{(2\Psi_n^{\rm C})}\sin{(2\Psi_l^{\rm A})}\right>}{\left<\sin{(2\Psi_m^{\rm B})}\sin{(2\Psi_n^{\rm C})}\right>}},
\label{eq:resolution_xy}
\end{eqnarray}
where $l, m, n$ are the harmonics of event planes for subevents 
A, B, and C, respectively.   Another acceptance effect from the 
difference between Res$\{\Psi_n^{\rm A};{\rm X}\}$ and 
Res$\{\Psi_n^{\rm A};{\rm Y}\}$ is discussed below.

\begin{figure}[htbp]
\includegraphics[width=1.0\linewidth]{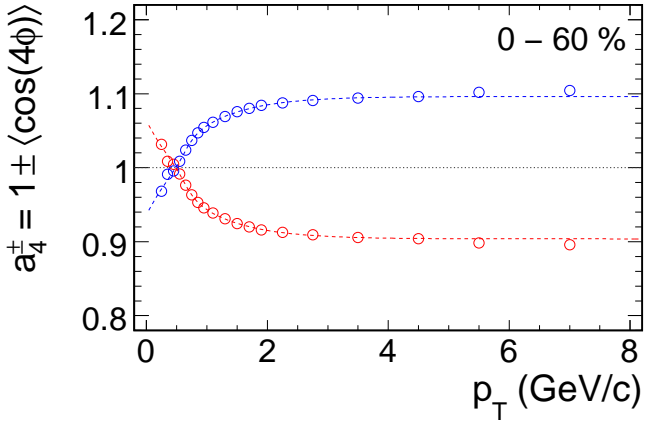}
\caption{\label{fig:acceptancecorrection_ptdep_cent0-60}
Acceptance correction factors $a^{\pm}_4$ in the central arms as 
a function of \pt for centrality 0--60\%. Correction factors 
become unity for a perfect azimuthal acceptance. Statistical 
errors are smaller than the symbols.
}
\end{figure}
Figure~\ref{fig:acceptancecorrection_ptdep_cent0-60} shows the 
acceptance correction factor $a^{\pm}_4$ as a function of \pt in 
the central arms for centrality 0--60\%. The \pt dependence is 
parameterized by
\begin{eqnarray}
a^{\pm}_4(\pt) = 1 \mp \left( p_0e^{-p_1p_T} + \frac{p_2}{1 + e^{(p_T-p_3)/p_4}} + p_5 \right), 
\label{eq:acceptance_correction_v2ep}
\end{eqnarray}
where $p_n$ ($n$ = 0,1,...,5) are free parameters. From the fit, we get
$p_0 = 0.131$,
$p_1 = 1.203$,
$p_2 = 0.029$,
$p_3 = 0.640$,
$p_4 = 0.096$ and
$p_5 = -0.097$.
There is no centrality dependence of the acceptance corrections in the measured
centrality range and these same correction factors are applied for all centrality bins.

\begin{figure}[htbp]
\includegraphics[width=1.0\linewidth]{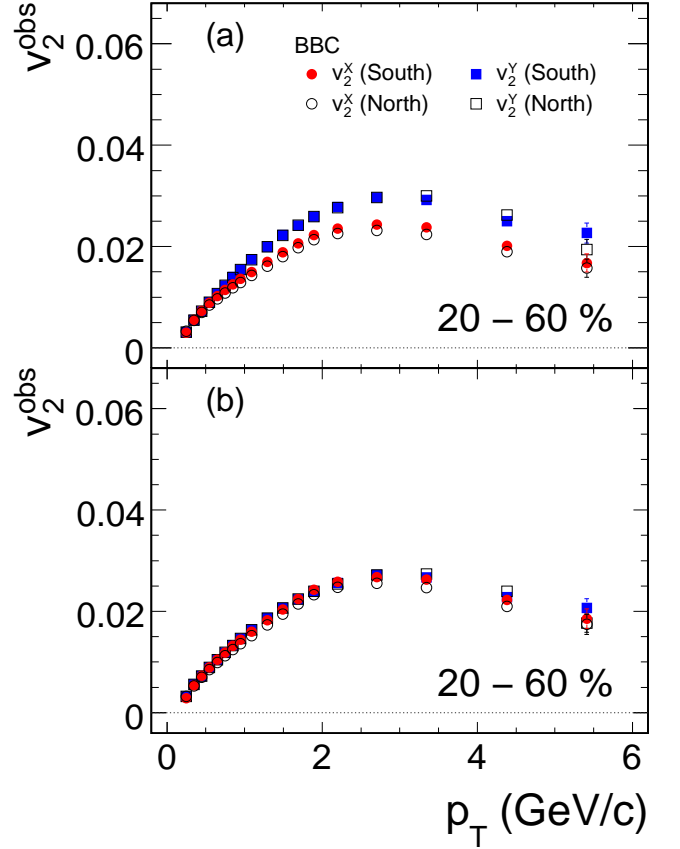}
\caption{\label{fig:v2pt_xy_before_and_after_correction_cent20-60} 
(a) Raw $v_2$\{BBC\} without the acceptance correction as a 
function of \pt in centrality 20--60\% for $v_2^{\rm X}$ (filled 
circles), $v_2^{\rm Y}$ (filled squares) with the South BBC 
event plane and or $v_2^{\rm X}$ (open circles), $v_2^{\rm Y}$ 
(open squares) with the North BBC event plane.
(b) The same comparison with the acceptance correction.
}
\end{figure}

Figure~\ref{fig:v2pt_xy_before_and_after_correction_cent20-60} shows the 
raw $v_2$\{BBC\} as a function of \pt in 20--60\% centrality bin. $v_2^{\rm 
Y}$ is systematically higher than $v_2^{\rm X}$ for \pt $>$ 1 \gevc as 
shown in Fig.~\ref{fig:v2pt_xy_before_and_after_correction_cent20-60}(a). 
Figure~\ref{fig:v2pt_xy_before_and_after_correction_cent20-60}(b) shows 
that $v_2^{\rm X}$ and $v_2^{\rm Y}$ agree with each other after dividing 
$v_2^{obs}$ by $a^{\pm}_{4}$, the remaining difference between them being 
accounted for as a systematic error. For the ZDC-SMD event plane we 
observed a similar trend for $v_2^{\rm X}$ and $v_2^{\rm Y}$.

A possible non-uniform acceptance of the BBC and ZDC-SMD could lead to the difference between 
Res$\{\Psi_n;{\rm X}\}$ and Res$\{\Psi_n;{\rm Y}\}$. If the azimuthal coverage of both detectors 
is perfect, Res$\{\Psi_n;{\rm X}\}$ and Res$\{\Psi_n;{\rm Y}\}$ should be identical. Therefore, 
the effect of the acceptance of the detector on the event plane resolution can be assessed by  
comparing Res$\{\Psi_n;{\rm X}\}$ and Res$\{\Psi_n;{\rm Y}\}$.

\begin{figure}[thb]
\includegraphics[width=1.0\linewidth]{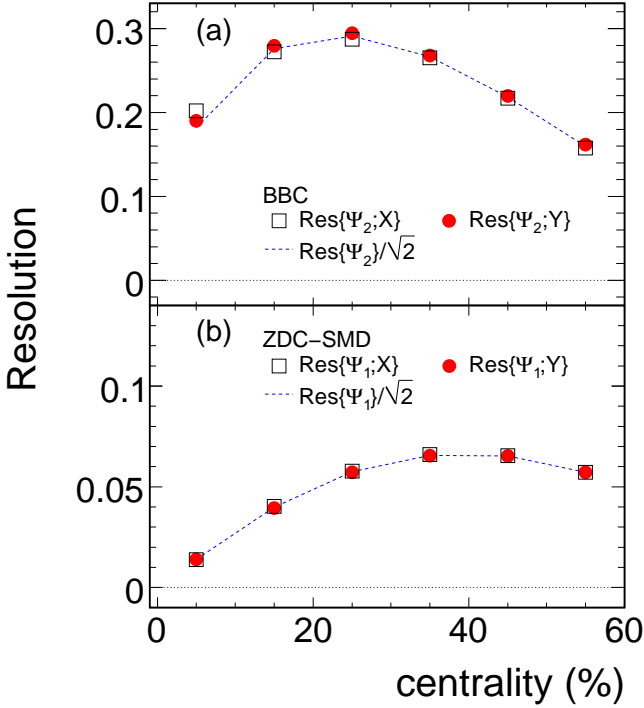}

\caption{\label{fig:resolution_xy_vs_centrality_BBC_and_ZDC-SMD}
(a) Comparison of Res$\{\Psi_n;{\rm X}\}$ (open squares) and 
Res$\{\Psi_n;{\rm Y}\}$ (filled circles) with Res$\{\Psi_n\}$ (dashed 
lines) for the BBC event plane ($n$ = 2) as a function of centrality. The 
resolutions are calculated by using Eq.~(\ref{eq:resolution_xy}) with the 
ZDC-SMD, BBC and CNT event planes. Res$\{\Psi_n\}$ is divided by $\sqrt{2}$ 
in order to compare Res$\{\Psi_n;{\rm X}\}$ and Res$\{\Psi_n;{\rm Y}\}$.
(b) The same comparison for the ZDC-SMD event plane ($n$ = 1).
Only statistical errors are shown and are smaller than symbols.
}
\end{figure}

Figure~\ref{fig:resolution_xy_vs_centrality_BBC_and_ZDC-SMD} shows 
Res$\{\Psi_n;{\rm X}\}$ and Res$\{\Psi_n;{\rm Y}\}$ of the BBC and ZDC-SMD 
as a function of centrality. The resolutions are calculated by using 
Eq.~(\ref{eq:resolution_xy}) with the ZDC-SMD, BBC and CNT event planes. 
Res$\{\Psi_n;{\rm X}\}$ was comparable with Res$\{\Psi_n;{\rm Y}\}$ for 
both the BBC and ZDC-SMD event planes. They also agreed, within statistical 
errors, with the expected resolution, namely the full event resolution 
scaled by 1/$\sqrt{2}$. We also evaluated Res$\{\Psi_n;{\rm X}\}$ and 
Res$\{\Psi_n;{\rm Y}\}$ of BBC and ZDC-SMD for the two-subevents method. 
Res$\{\Psi_2^{\rm BBC};{\rm X}\}$ was consistent with Res$\{\Psi_2^{\rm 
BBC};{\rm Y}\}$. However, for the ZDC-SMD event plane, Res$\{\Psi_1^{\rm 
ZDC-SMD};{\rm Y}\}$ (Res$\{\Psi_1^{\rm ZDC-SMD};{\rm X}\}$) was 
systematically higher (lower) by about 30\% than the expected resolution 
when the resolutions were calculated with $k$ = 1 in 
Eq.~(\ref{eq:resolution_formula}). The difference between Res$\{\Psi_1^{\rm 
ZDC-SMD};{\rm X}\}$ and Res$\{\Psi_1^{\rm ZDC-SMD};{\rm Y}\}$ for the 
two-subevents method is attributed to the non-uniform acceptance between 
horizontal (x) and vertical (y) directions of the ZDC-SMD. Those 
resolutions of the ZDC-SMD were consistent with each other using $k$ = 2. 
For $k$ = 2, the non-uniform acceptance in the azimuthal directions cancels 
out since Res$\{\Psi_1^{\rm ZDC-SMD};{\rm X,Y}\}$ contain both 
$\left<\cos{(\Psi)}\right>$ and $\left<\sin{(\Psi)}\right>$ terms. Thus, 
Res$\{\Psi_1^{\rm ZDC-SMD};{\rm X,Y}\}$ should be the same and consistent 
with that from the expected resolution.

\begin{figure}[thb]
\includegraphics[width=1.0\linewidth]{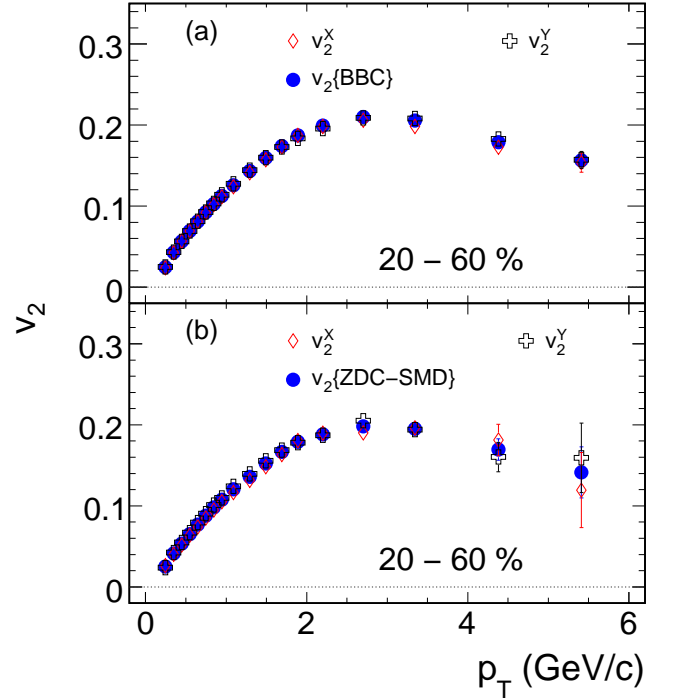}
\caption{\label{fig:comparison_v2xy_and_v2_cent20-60_bbc_and_zdc-smd}
(a) Comparison of $v_2^{\rm X}$ (open diamonds) and $v_2^{\rm Y}$ (open 
crosses) with the total $v_2$ (filled circles) for the BBC event plane as a 
function of \pt for the centrality bin 20--60\%. Res$\{\Psi_n;{\rm X}\}$ 
and Res$\{\Psi_n;{\rm Y}\}$ are calculated by the combination of the 
ZDC-SMD, BBC and CNT event planes. Acceptance corrections are included into 
$v_2^{\rm X}$ and $v_2^{\rm Y}$. Error bars denote statistical errors.
(b) The same comparison for the ZDC-SMD event plane.
}
\end{figure}

The comparison of $v_2^{\rm X}$ and $v_2^{\rm Y}$ with $v_2$ 
with respect to the BBC and ZDC-SMD event planes is shown in 
Fig.~\ref{fig:comparison_v2xy_and_v2_cent20-60_bbc_and_zdc-smd}. 
The maximum difference of $v_2^{\rm X}$ and $v_2^{\rm Y}$ 
relative to $v_2$\{BBC\} is about 2\% for the centrality range 
20--60\% and is independent of centrality. Systematic 
uncertainties are evaluated by scaling the maximum difference by 
$2/\sqrt{12}$. The same comparison is also made for 
$v_2$\{ZDC-SMD\} as shown in the bottom panel in 
\Fig~\ref{fig:comparison_v2xy_and_v2_cent20-60_bbc_and_zdc-smd}. 
The systematic errors range from 1--25\% and in this case 
strongly depend on the centrality, as well as on the corrections 
by the different event plane resolutions. $v_2^{\rm X}$ and 
$v_2^{\rm Y}$ are $\sim$ 10--25\% different from $v_2$\{ZDC-SMD\} 
in the 0--20\% centrality bin due to the very low resolution. 
This systematic uncertainty is denoted as ``Acceptance effect on 
event planes" in Table 
\ref{tab:summary_systematicerror_eventplane}.

\subsection{Cumulant method \label{subsec:syserror_cumulantmethod}}
The potential sources of systematic errors on the cumulant 
measurements are detailed below.

\subsubsection{Fixed multiplicity cut}
Following Ref.~\cite{Borghini:2001vi} a fixed multiplicity is 
used to reconstruct the integrated flow to avoid introducing 
additional errors arising from a fluctuating multiplicity. In 
our analysis the systematic errors were estimated by varying the 
fixed multiplicity cut used for the reconstruction of the 
integrated flow and studying its effect on the differential flow 
values.

\begin{figure}[hbt]
\begin{center}
\includegraphics[width=1.0\linewidth]{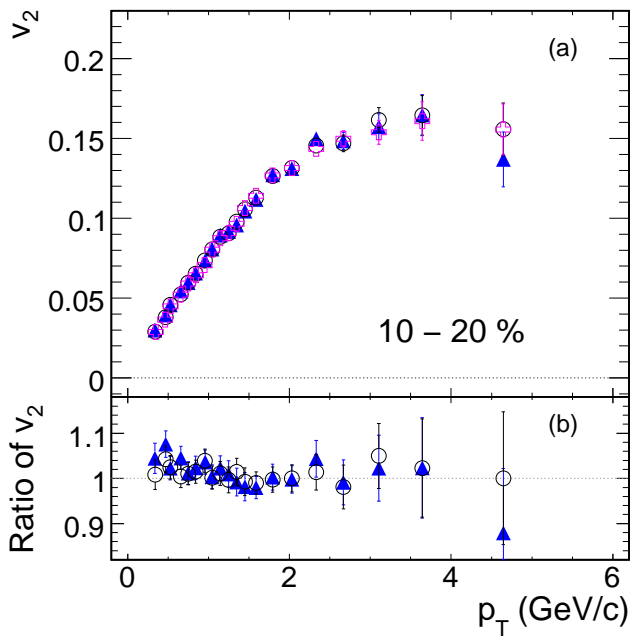}
\caption{\label{v2multcuts}
(a)  $v_{2}$\{2\} as a function of \pt for centrality 10--20$\%$ in \Au 
collisions at \sqsn = 200 GeV for different fixed multiplicity cuts, 
corresponding to 60\% (filled triangles), 70\% (open circles) and 80\% 
(open crosses) of the mean multiplicity.
(b) The ratio of $v_2$(\pt) for the two lowest multiplicity cuts to 
$v_2$(\pt) for 80\% of the mean multiplicity.
}
\end{center}
\end{figure}

Figure~\ref{v2multcuts}(a) shows the variation of $v_2$ with \pt for 
integral multiplicity cuts equal to 60\%, 70\%, and 80\% of the mean 
multiplicity for the centrality bin 20--40\%. The ratio of the differential 
$v_2$ values, shown in Fig.~\ref{v2multcuts}(b), is used to estimate the 
systematic error on our measurements, which is $\sim$ 5\%.

\subsubsection{\pt range for integrated flow}

In order to assess the influence of the \pt range used to estimate the 
integrated flow on the differential flow, we chose different \pt ranges 
over which the integral particles were selected. Differential $v_2$ results 
were obtained for three \pt ranges: 0.25~-~2.0 \gevc, 0.25~-~1.5 \gevc and 
0.3~-~1.5 \gevc. The systematic error from this source is estimated to be 
3-8\% depending on centrality and \pt.

\subsubsection{Background contribution}

The procedures followed for studying the background contribution to 
$v_2$\{2\} were the same as for the event plane method. After background 
subtraction the systematic error is calculated by determining the 
difference between the $v_2$ obtained from using 2$\sigma$ and 3$\sigma$ 
association cuts. We determined that the overall systematic error due to 
these differences is 6--10\% depending on \pt and centrality.

\section{Results\label{sec:results}}

\subsection{\pt dependence of $v_2$}

\begin{figure*}[htb!]
\includegraphics[width=1.0\linewidth]{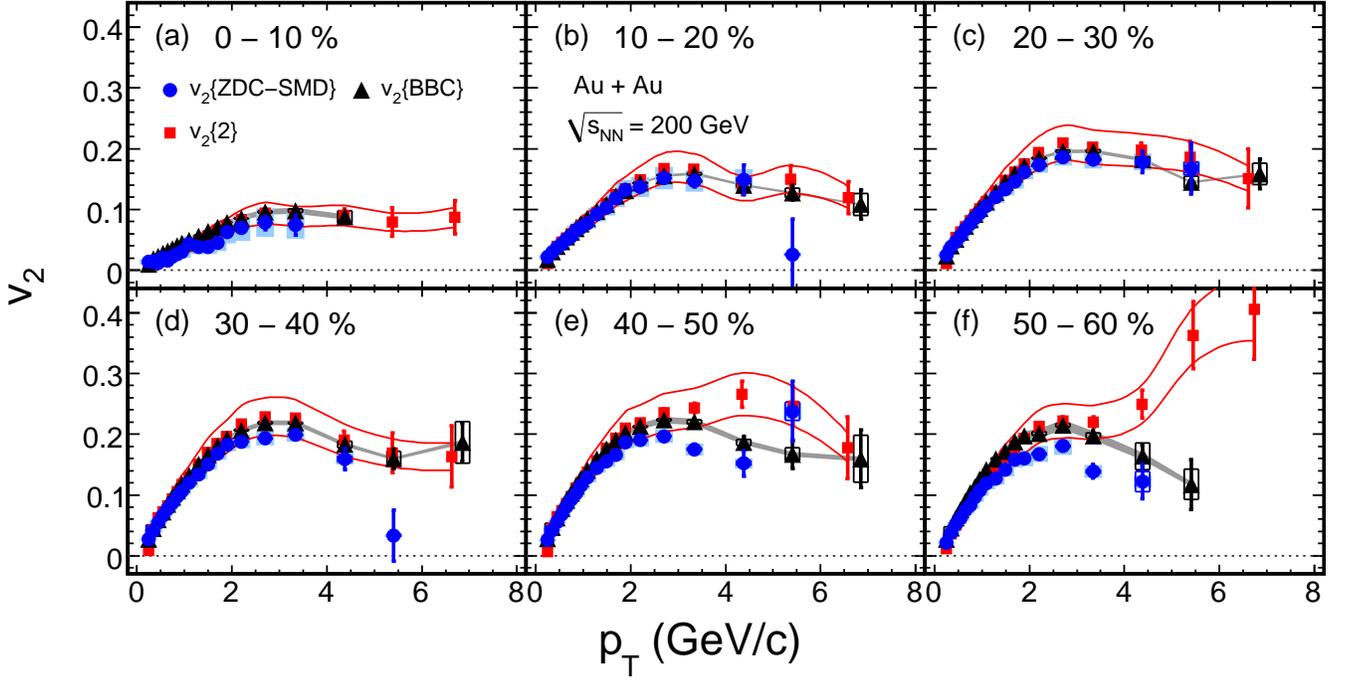}
\caption{\label{fig:v2pt_centdep_zdcsmd_bbc_twosubevents_wide} 
Charged hadron $v_2$(\pt) in \Au collisions at \sqsn = 200 GeV 
from the two-particle cumulant method (filled squares), the BBC 
event plane (filled triangles) and the ZDC-SMD event plane 
(filled circles) for centrality (a) 0--10\%, (b) 10--20\%, (c) 
20--30\%, (d) 30--40\%, (e) 40--50\%, and (f) 50--60\%. Error 
bars denote statistical errors. The type B systematic uncertainties 
are represented by the open boxes for the $v_2$\{BBC\} and 
$v_2$\{ZDC-SMD\}, and by the solid lines for the $v_2$\{2\}. The 
gray bands and blue boxes represent the type C systematic 
uncertainties on the $v_2$\{BBC\} and $v_2$\{ZDC-SMD\}, 
respectively.
}
\end{figure*}

\begin{figure*}[htb!]
\includegraphics[width=1.0\linewidth]{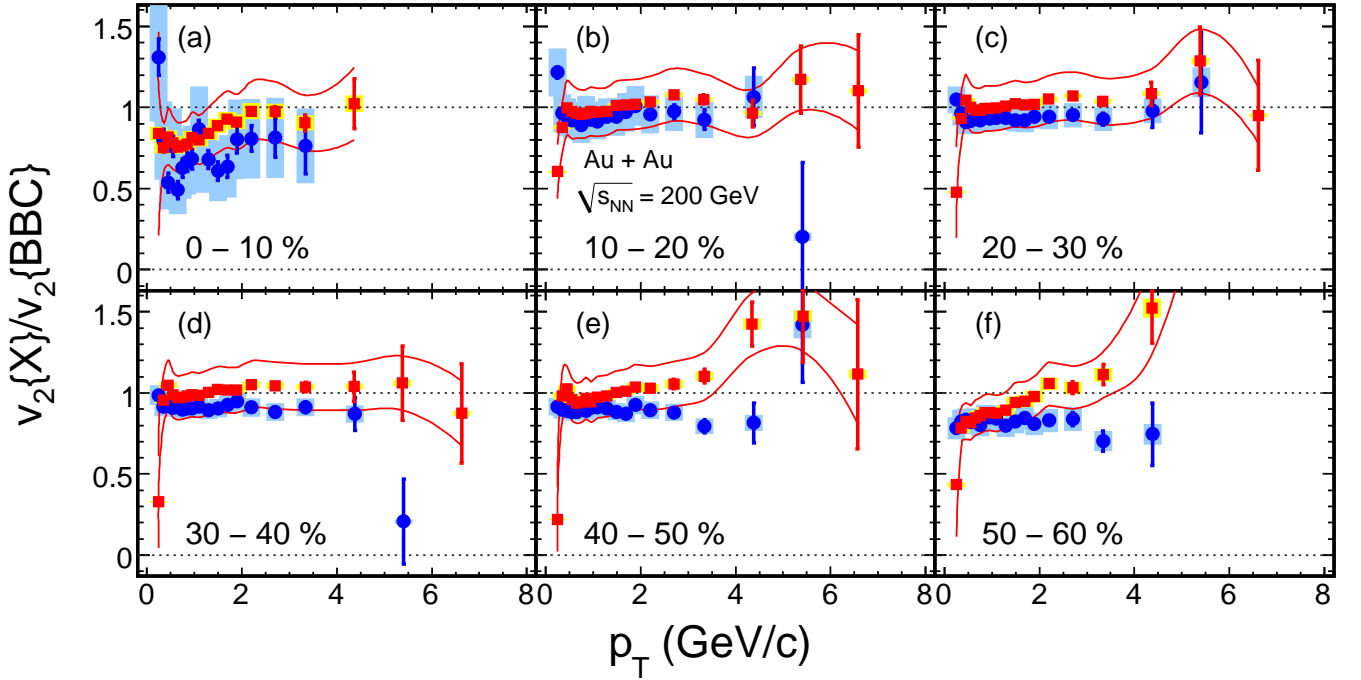}
\caption{\label{fig:ratio_v2pt_centdep_zdcsmd_bbc_twosubevents_wide}
The ratio of $v_2$ to $v_2$\{BBC\} as a function of \pt for six centrality bins over the 
range 0--60\% in \Au 
collisions at \sqsn = 200 GeV. Data symbols are the same as in the \Fig~\ref{fig:v2pt_centdep_zdcsmd_bbc_twosubevents_wide}.
Error bars denote statistical errors. The solid red lines represent the type B systematic errors on the 
$v_2$\{2\}. The blue and yellow bands represent type C systematic uncertainties 
on $v_2$\{ZDC-SMD\} and $v_2$\{2\}.
 }
\end{figure*}

\begin{figure*}[htb!]
\includegraphics[width=0.55\linewidth]{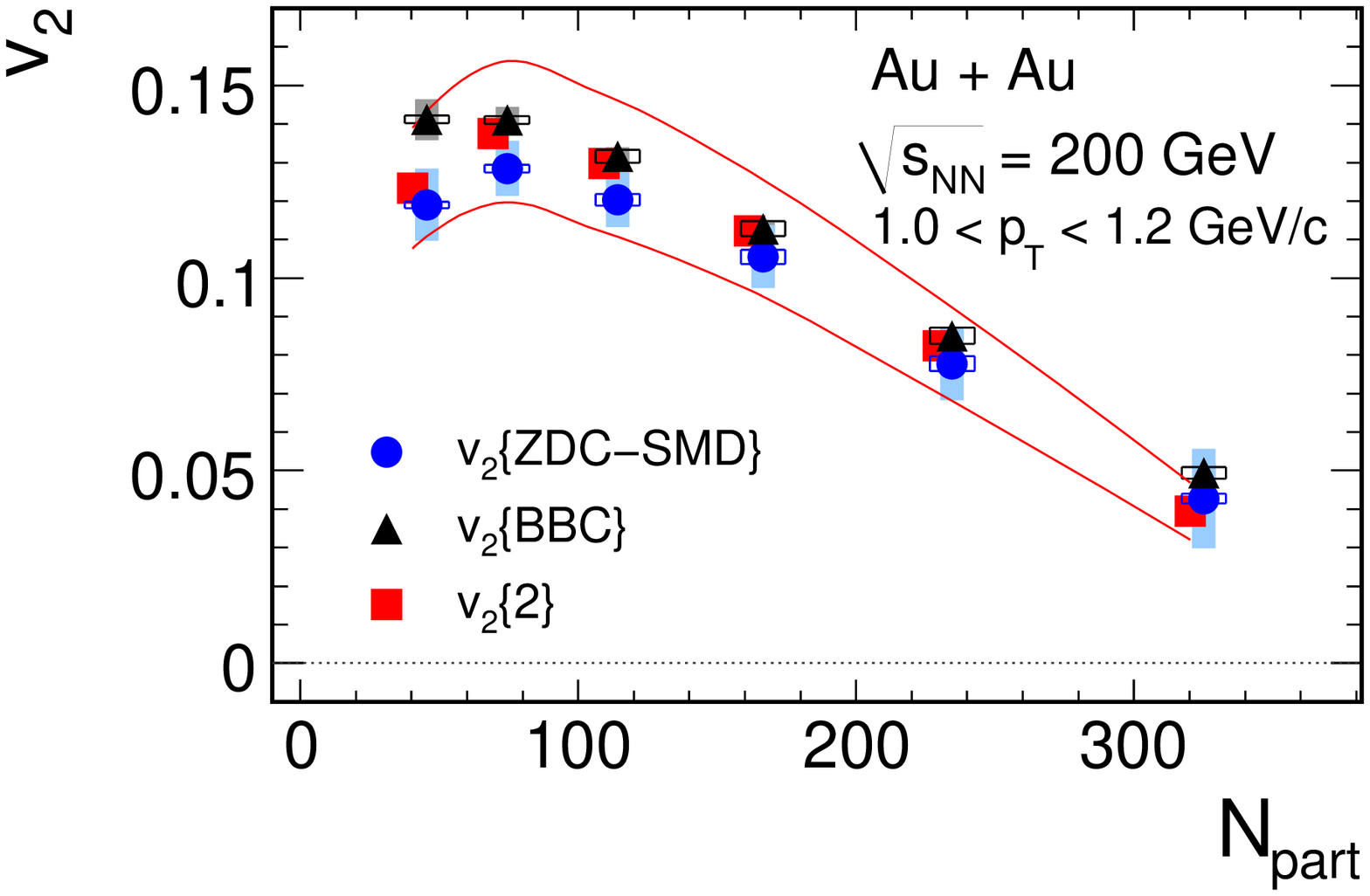}
\caption{\label{fig:v2npart_zdc-smd_bbc_cumulants.eps}
Comparison of charged hadron $v_2$ at 1 $<$ \pt $<$ 1.2 \gevc as a function of \Np for 
$v_2$\{BBC\} (filled triangles),  $v_2$\{ZDC-SMD\} (filled circles) and $v_2$\{2\} (filled squares) 
in \Au at \sqsn = 200 GeV. The error bars represent statistical errors. The open boxes represent 
type B systematic uncertainties on $v_2$\{BBC\} and $v_2$\{ZDC-SMD\}. Type B systematic 
uncertainties on $v_2$\{2\} are represented by solid red lines. The gray and blue bands represent 
type C systematic errors on $v_2$\{BBC\} and $v_2$\{ZDC-SMD\}, respectively.  $v_2$\{2\} values 
are shifted in the x-axis to improve the plot.
}
\end{figure*}

\begin{figure*}[hbt!]
\includegraphics[width=0.55\linewidth]{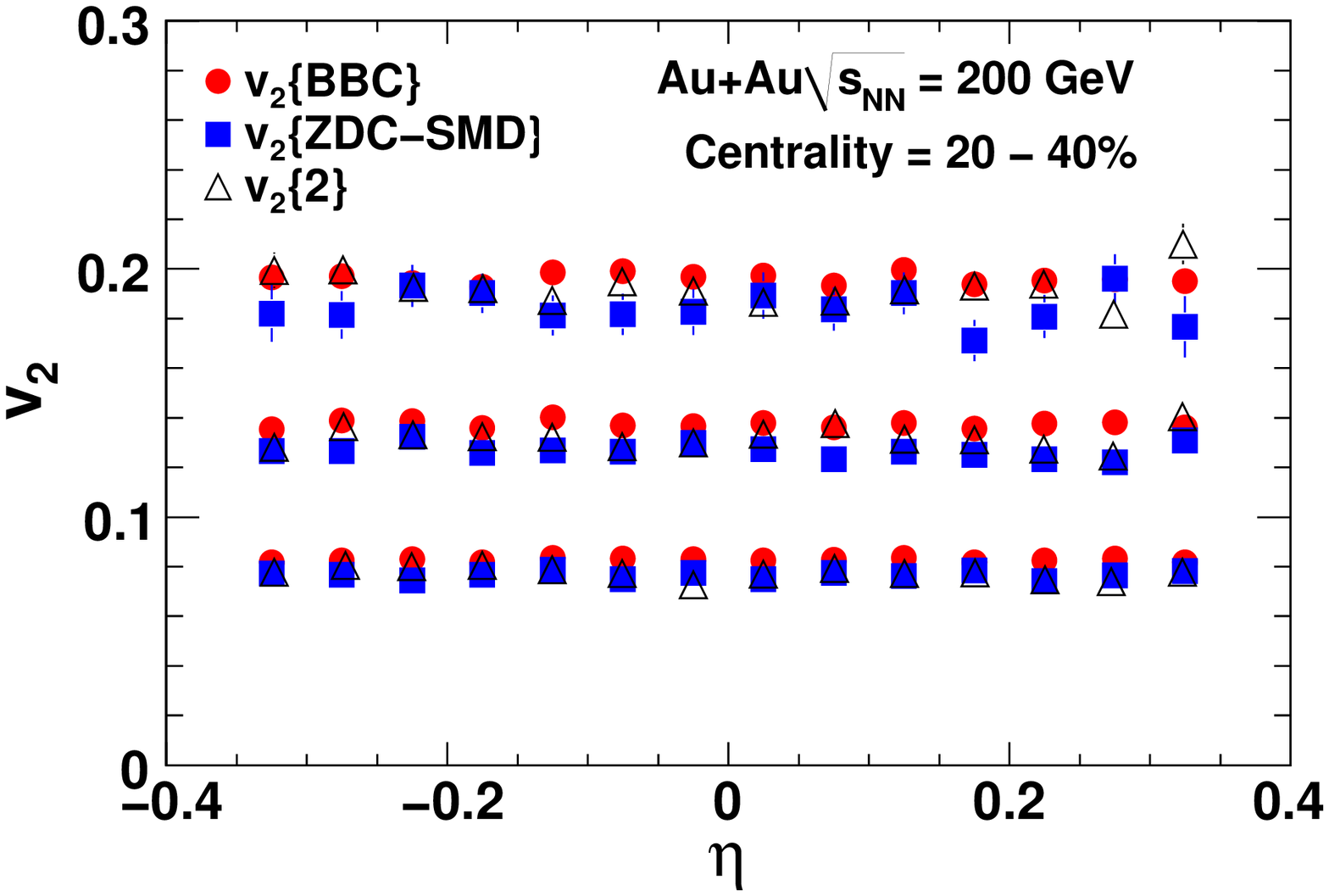}
\caption{\label{fig:v2vseta_20-40_3methods.eps} Anisotropy parameter $v_2$ 
as a function of pseudorapidity within the PHENIX central arms using event 
planes from the BBC (filled circles), ZDC-SMD (filled squares), and from 
the two-particle cumulant method (open triangles) for centrality 20--40\%. 
The results are shown for three \pt bins, which are from top to bottom: 
2.0--3.0, 1.2--1.4 and 0.6--0.8 \gevc.  Only statistical errors are shown.
}
\end{figure*}

The \pt dependence of $v_2$ has been instrumental in revealing 
the hydrodynamic properties of the matter formed at 
RHIC~\cite{Adler:2003kt,Adams:2003am}. In this context, it is 
important to compare the \pt dependence of $v_2$ from different 
methods to establish the robustness of our $v_2$ measurements. 
This comparison is displayed in 
Fig.~\ref{fig:v2pt_centdep_zdcsmd_bbc_twosubevents_wide} which 
shows the differential charged hadron $v_2$ as a function of \pt 
from the event plane and cumulant methods for different 
centrality bins in the range 0--60\% in \Au at \sqsn = 200 GeV. 
$v_2$\{2\} increases up to \pt $\approx$ 3~\gevc and saturates 
at \mbox{~0.1--0.25}, depending on centrality, for higher \pt. 
On the other hand, $v_2$\{BBC\} and $v_2$\{ZDC-SMD\} reach their 
maximum value at \pt $\approx$ 3~\gevc, and decrease for higher 
\pt.

The differences between $v_2$\{BBC\} and $v_2$\{ZDC-SMD\} are 
independent of \pt within systematic errors in the measured 
centrality range. $v_2$\{ZDC-SMD\} is consistent with 
$v_2$\{BBC\} within systematic errors in the 0--40\% centrality 
range, but is $\sim$ 10--20\% smaller than $v_2$\{BBC\} in the 
40--60\% centrality range. These results could indicate that the 
influence of non-flow effects on $v_2$\{BBC\} is small and 
within the systematic errors, because non-flow effects are not 
expected to influence $v_2$\{ZDC-SMD\}. The difference between 
$v_2$\{BBC\} and $v_2$\{ZDC-SMD\} in peripheral collisions could 
be attributed to non-flow contributions that might be 
proportionally larger in more peripheral collisions.

The cumulant and event plane $v_2$ agree well within systematic 
uncertainties in the centrality range 0--40\%. In more 
peripheral collisions, there may be some differences developing 
above $p_T$ $\simeq$ 4 \gevc. Correlations between particles 
from jets affect the cumulant results, but have less influence 
on $v_2$\{BBC\}, as explained in Ref.~\cite{Jia:2006sb}, where 
it was shown that the smaller the rapidity gap between the 
leading particle from a jet and the event plane, the greater the 
$v_2$ of the leading particle of the jet.

In order to illustrate more clearly the differences between the 
different methods, 
\Fig~\ref{fig:ratio_v2pt_centdep_zdcsmd_bbc_twosubevents_wide} 
shows the ratio of $v_2$\{ZDC-SMD\} and $v_2$\{2\} to 
$v_2$\{BBC\}. The results from the three methods are comparable 
in magnitude within systematic errors, except for the central 
and peripheral bins where the largest deviations occur. In 
addition, $v_2$\{2\} and $v_2$\{ZDC-SMD\} show different 
behaviors at \pt $>$ 3 \gevc, with $v_2$\{2\} being larger, and 
$v_2$\{ZDC-SMD\}, smaller than $v_2$\{BBC\}.

\begin{figure*}[htb!]
\includegraphics[width=0.85\linewidth]{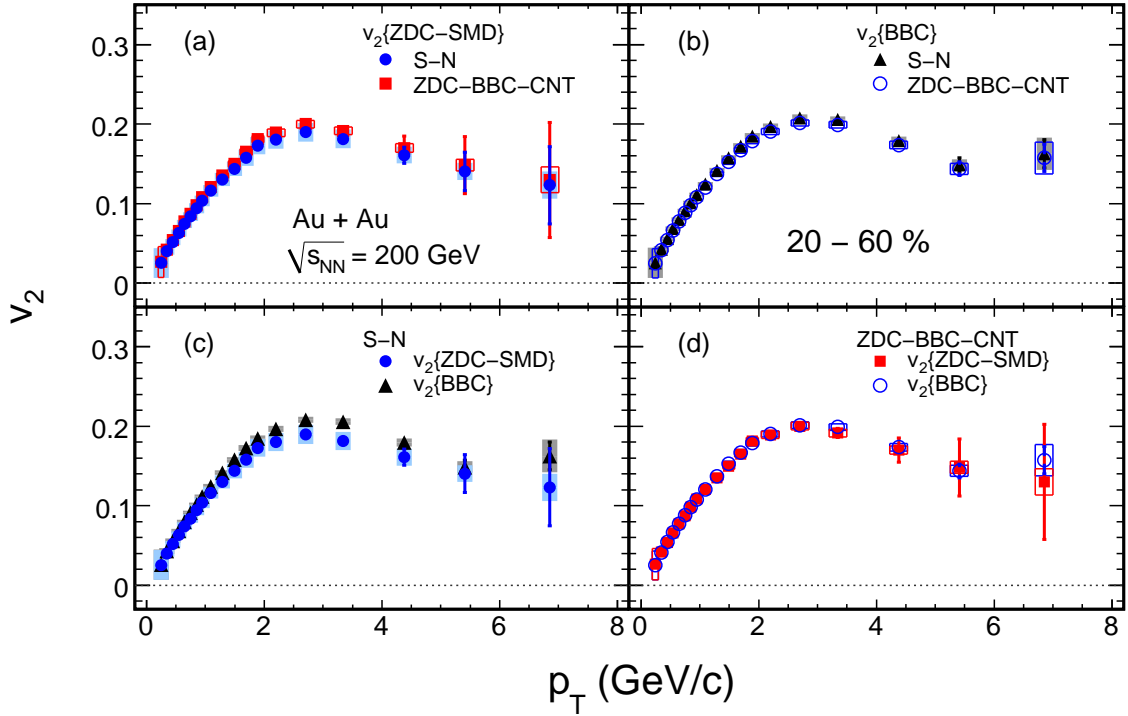}

\caption{\label{fig:comparison_v2pt_zdc-smd_bbc_withCNT_cent20-60}
(a) Comparison of the $v_2$\{ZDC-SMD\} from the S-N (filled circles) 
and ZDC-BBC-CNT subevents (filled squares) as a function of \pt 
in the 20--60\% centrality range.
(b) The same comparison as (a) for the $v_2$\{BBC\}, where 
filled triangles and open circles represent the $v_2$ from the 
S-N and ZDC-BBC-CNT subevents, respectively.
(c) Comparison of $v_2$ between BBC (filled triangles) and 
ZDC-SMD event planes (filled circles) from the S-N subevent as a 
function of \pt in centrality 20--60\%.
(d) The same comparison as (c) from the ZDC-BBC-CNT subevent, 
where filled squares and open circles represent the 
$v_2$\{ZDC-SMD\} and $v_2$\{BBC\}, respectively. Error bars 
denote statistical errors. Open boxes and shaded bands describe 
the quadratic sum of type B and C systematic uncertainties from 
the S-N and ZDC-BBC-CNT subevents, respectively.
}
\end{figure*}

\subsection{Centrality dependence of $v_2$}

Figure~\ref{fig:v2npart_zdc-smd_bbc_cumulants.eps} shows the \Np 
dependence of $v_2$ from different methods for charged hadrons 
in the range 1.0 $<$ \pt $<$ 1.2 \gevc. $v_2$ is observed to 
increase with decreasing \Np and then decrease slightly for \Np 
$\lesssim$ 75. Note that $v_2$ values obtained with 
the different methods agree well within systematic errors for 
all centralities. This is \pt dependent, as shown in 
\Fig~\ref{fig:v2pt_centdep_zdcsmd_bbc_twosubevents_wide}.

\subsection{Pseudorapidity dependence of v$_2$}

Figure~\ref{fig:v2vseta_20-40_3methods.eps} compares the 
pseudorapidity dependence of the $v_2$ of charged hadrons within 
the $\eta$ range ($\pm$ 0.35) of the PHENIX central arms for 
different \pt selections. It can be observed that $v_2$ is 
constant over the $\eta$ coverage of the PHENIX detector and the 
constancy does not depend on \pt.  This is not the case when the 
$v_2$ is measured far from midrapidity where the PHOBOS and STAR 
collaborations observe a drop in $v_2$ for $|\eta| > 
1.0$~\cite{Adams:2004bi,Alver:2006wh}.

\section{Discussion\label{sec:discussions}}

\subsection{Effect of CNT event plane resolution}

Figure~\ref{fig:comparison_v2pt_zdc-smd_bbc_withCNT_cent20-60} 
shows the comparison of $v_2$\{ZDC-SMD\} and $v_2$\{BBC\} as a 
function of \pt corrected either by the resolution from 
South-North correlations from the same detectors or by the 
resolution from ZDC-SMD-CNT correlations in the 20--60\% 
centrality bin. 
Figures~\ref{fig:comparison_v2pt_zdc-smd_bbc_withCNT_cent20-60}(a)~and~(b) 
compare the $v_2$ obtained by using two different 
corrections from the South-North and ZDC-BBC-CNT subevents for 
the BBC (a) and ZDC-SMD event planes (b). The $v_2$ from the 
South-North subevent is consistent with that from the 
ZDC-BBC-CNT subevent, within systematic uncertainties. The small 
difference between South-North and ZDC-BBC-CNT subevents is 
attributed to the difference between the event plane resolution, 
as shown in \Fig~\ref{fig:eventplane_resolution_twosubevents}.  
Figures~\ref{fig:comparison_v2pt_zdc-smd_bbc_withCNT_cent20-60}(c)~and~(d)
compare $v_2$\{ZDC-SMD\} with $v_2$\{BBC\} for the 
South-North (c) and ZDC-BBC-CNT subevent (d). The data 
points in 
Fig.~\ref{fig:comparison_v2pt_zdc-smd_bbc_withCNT_cent20-60}(c)~and~(d) 
are the same as in 
Fig.~\ref{fig:comparison_v2pt_zdc-smd_bbc_withCNT_cent20-60}(a)~and~(b).
Figure~\ref{fig:comparison_v2pt_zdc-smd_bbc_withCNT_cent20-60}(c) 
shows that $v_2$\{ZDC-SMD\} is about 10\% smaller than 
$v_2$\{BBC\} for the South-North subevent. The ratio of 
$v_2$\{ZDC-SMD\} to $v_2$\{BBC\} is found to be independent of 
\pt except for $6 < \pt < 8$ \gevc.  If jets are the dominant 
source of non-flow, one expects its contribution to $v_2$ to 
become larger at higher \pt.  The constant ratio suggests that 
the non-flow contribution from jets is small and $v_2$ 
fluctuations may affect $v_2$\{BBC\} below \pt $ \approx 6$ 
\gevc since the effect of fluctuations is expected to be 
independent of \pt. $v_2$\{ZDC-SMD\} agrees with $v_2$\{BBC\} 
within systematic uncertainties for the ZDC-BBC-CNT subevent as 
shown in 
Fig.~\ref{fig:comparison_v2pt_zdc-smd_bbc_withCNT_cent20-60}(d). 
The event plane resolution from the ZDC-BBC-CNT subevents 
includes the effect of non-flow contributions and $v_2$ 
fluctuations since the CNT and BBC event planes are sensitive to 
both effects, though non-flow effects especially from jets could 
be negligible in the BBC event plane, as discussed earlier. The 
consistency between $v_2$ from the ZDC-SMD and BBC event planes 
may suggest that $v_2$\{ZDC-SMD\} becomes sensitive to $v_2$ 
fluctuations by the inclusion of the BBC and CNT event planes to 
estimate the resolution.

\subsection{Comparison with other experiments}

\begin{figure}[htbp!]
\includegraphics[width=1.0\linewidth]{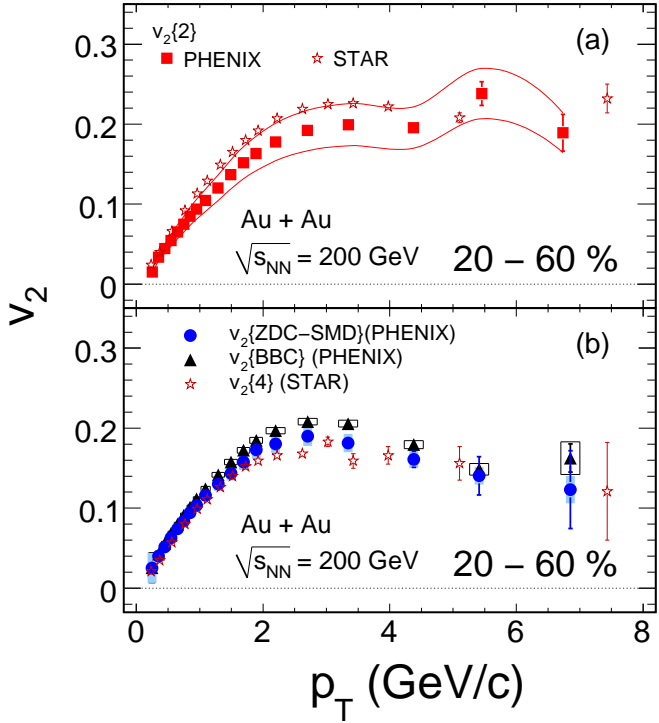}
\caption{\label{fig:comparison_v2pt_STAR_cent20-60}
(a) Comparison of charged hadron $v_2$\{2\} between PHENIX (filled squares) 
and STAR experiments (open stars) as a function of \pt in 
centrality 20--60\%.  Solid lines represent the quadratic sum of type B 
and C systematic errors on the PHENIX $v_2$\{2\}.
(b) Comparison of charged hadron $v_2$ from four-particle cumulant 
$v_2$\{4\} at STAR (open stars) with the PHENIX $v_2$\{BBC\} (filled 
triangles) and $v_2$\{ZDC-SMD\} (filled circles) as a function of \pt in 
centrality 20--60\%. Open boxes and shaded bands represent the quadratic 
sum of type B and C systematic errors on the $v_2$\{BBC\} and 
$v_2$\{ZDC-SMD\}, respectively. STAR results are taken from 
Ref.~\cite{Adams:2004wz}. Systematic errors on the STAR $v_2$ are not 
plotted, see text for more details.
}
\end{figure}
It is instructive to compare measurements made by different experiments at 
RHIC. Figure~\ref{fig:comparison_v2pt_STAR_cent20-60} shows a comparison 
of the \pt dependence of charged hadron $v_2$ in the 20--60\% centrality 
range between PHENIX and STAR experiments~\cite{Adams:2004wz}. The 
relative systematic errors on the STAR $v_2$\{2\} and $v_2$\{4\} 
measurements range up to 10\% for \pt $<$ 1 \gevc, with the lowest \pt bin 
having the largest error $\sim$ 10\%, while they are of the order of 1\% 
above 1 \gevc~\cite{Adams:2004wz}. The $v_2$\{2\} from PHENIX is lower 
than that from STAR, but they are comparable within systematic 
uncertainties, as shown in 
Fig.~\ref{fig:comparison_v2pt_STAR_cent20-60}(a). 
Figure~\ref{fig:comparison_v2pt_STAR_cent20-60}(b) compares $v_2$\{BBC\} 
and $v_2$\{ZDC-SMD\} with $v_2$\{4\}, obtained from four particle 
cumulants, as measured in STAR. For \pt $>$ 2 \gevc, the STAR $v_2$\{4\} 
is systematically smaller than the PHENIX event plane $v_2$, while 
$v_2$\{ZDC-SMD\} is lower than $v_2$\{BBC\}. However, the three set of 
measurements are consistent within systematic errors. The order of $v_2$, 
$v_2$\{BBC\} $>$ $v_2$\{ZDC-SMD\} $>$ $v_2$\{4\} could be explained by the 
effect of flow fluctuations~\cite{Bhalerao:2006tp,Miller:2003kd} if other 
non-flow contributions are small.

\begin{figure}[tbp!]
\includegraphics[width=1.0\linewidth]{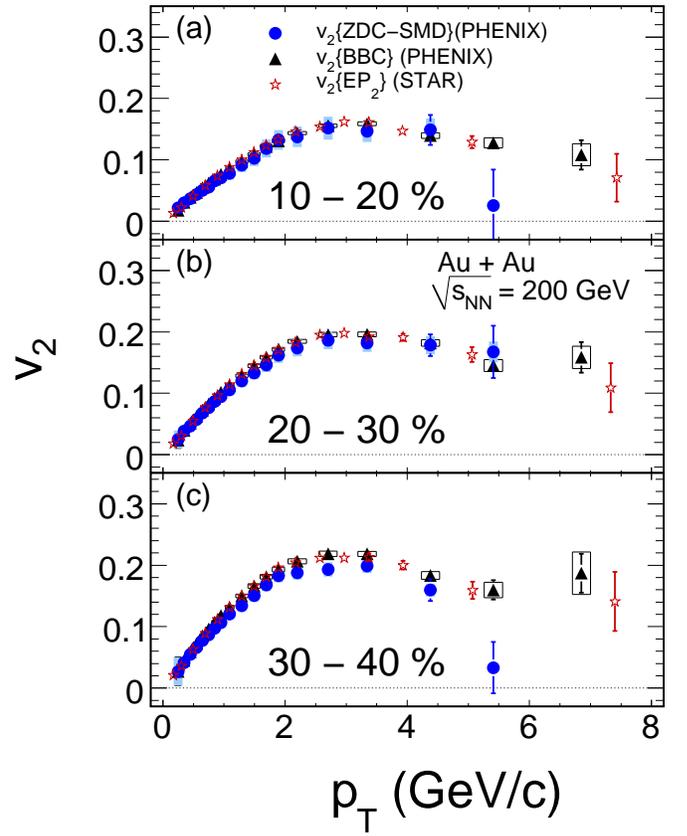}
\caption{\label{fig:comparison_v2pt_centdep_zdcsmd_STAR_twosubevents} 
Comparison of the PHENIX $v_2$\{BBC\} (filled triangles) and 
$v_2$\{ZDC-SMD\} (filled circles) with the STAR $v_2$ from 
modified event plane method (open stars) for charged 
hadrons~\cite{Adams:2004bi} as a function of \pt in centrality 
(a) 10--20\%, (b) 20--30\%, and (c) 30--40\%. Open boxes and 
shaded bands represent the quadratic sum of type B and C systematic 
errors on $v_2$\{BBC\} and $v_2$\{ZDC-SMD\}, respectively.
}
\end{figure}

Figure~\ref{fig:comparison_v2pt_centdep_zdcsmd_STAR_twosubevents} 
shows the comparison of our charged hadron $v_2$ from the BBC 
and ZDC-SMD event planes to $v_2$ from a modified event plane 
method~\cite{Adams:2004bi}, labelled $v_2$\{EP$_2$\}, from the 
STAR experiment for three centrality bins in the range 10--40\%. 
Particles within $|\Delta\eta| < 0.5$ around the highest \pt 
particle were excluded for the determination of the modified 
event plane in order to reduce some of the non-flow effects at 
high \pt. We find that $v_2$\{BBC\} agrees well with 
$v_2$\{EP$_2$\} over the measured \pt range, whereas 
$v_2$\{ZDC-SMD\} is generally slightly smaller than 
$v_2$\{EP$_2$\} .

\section{Summary\label{sec:conclusion}}

In summary we have presented PHENIX elliptic flow measurements for 
unidentified charged hadrons from the event plane and the 
two-particle cumulant methods as a function of \pt and centrality at 
midrapidity ($|\eta| < 0.35$) in \Au collisions at \sqsn = 200 GeV. 
The first harmonic ZDC-SMD event plane is used to measure $v_2$ and 
is compared to $v_2$ from the second harmonic BBC event plane in 
order to understand the possible non-flow contributions as well as 
the effect of $v_2$ fluctuations on $v_2$\{BBC\}.

The comparison between $v_2$ from two-particle cumulant and event 
plane methods shows that they agree within systematic errors. 
However, non-flow effects from jet correlations begin to contribute 
to the two-particle cumulant $v_2$, especially for peripheral 
collisions and at high \pt.

In contrast, non-flow effects on $v_2$\{BBC\} are very small. 
The measured $v_2$\{BBC\} values decrease by about 3\% when the 
central arm event plane is included in the estimate of the BBC 
reaction plane resolution.  This could be due to a partial 
compensation of the non-flow effects on the measured $v_2$, 
though the results of $v_2$\{BBC\} with and without the CNT 
event plane resolution are consistent within systematic errors. 
The strongest evidence that non-flow effects on $v_2$\{BBC\} are 
small comes from the observation that $v_2$\{ZDC-SMD\} is 
comparable with $v_2$\{BBC\} within systematic uncertainties in 
the 0--40\% centrality range, and are only $\sim$ 5--10\% 
smaller than $v_2$\{BBC\} for the 40--60\% centrality bin. The 
magnitude of this difference could indicate the level at which 
non-flow effects such as jets or the ridge could impact the 
measured flow. However, PHOBOS has observed the ridge to be 
strongest in central collisions\cite{Alver:2009id} where we 
observe that $v_2$\{ZDC-SMD\} is comparable with $v_2$\{BBC\}. 
For collisions that are more peripheral than 40\% centrality, 
PHOBOS observes no ridge\cite{Alver:2009id}, so it is unlikely 
that our observation that $v_2$\{ZDC-SMD\} is $\sim$ 5--10\% 
smaller than $v_2$\{BBC\} for the 40--60\% centrality bin is 
caused by the ridge. Moreover, the difference between 
$v_2$\{ZDC-SMD\} and $v_2$\{BBC\} is independent of \pt in the 
measured centrality range.

Due to the large pseudorapidity gap between the event plane and 
the particles detected in the central arms spectrometer, and the 
first harmonic event plane from directed flow by spectator 
neutrons, $v_2$\{ZDC-SMD\} is considered to provide an unbiased 
measure of the elliptic flow.  Within systematic uncertainties 
the measured $v_2$\{ZDC-SMD\} from PHENIX is consistent with 
$v_2$ from the four particle cumulant method measured by the 
STAR experiment in the 20--60\% centrality bin, and is also 
consistent with the STAR $v_2$ from a modified event plane 
method in 10--40\% centrality bins.  These comparisons (1) 
further demonstrate the validity of the $v_2$\{ZDC-SMD\}, 
because both STAR results aim to minimize the non-flow effects, 
(2) reinforce the robustness of the BBC event plane method at 
RHIC, and (3) confirm previous studies of the influence of jets 
on the measured $v_2$ for different rapidity gaps.  Hence, 
$v_2$\{BBC\} can be used to infer constraints on the 
hydrodynamic behavior of heavy-ion collisions at RHIC.

\section*{Acknowledgements}  


We thank the staff of the Collider-Accelerator and Physics
Departments at Brookhaven National Laboratory and the staff of
the other PHENIX participating institutions for their vital
contributions.  We acknowledge support from the 
Office of Nuclear Physics in the
Office of Science of the Department of Energy, the
National Science Foundation, Abilene Christian University
Research Council, Research Foundation of SUNY, and Dean of the
College of Arts and Sciences, Vanderbilt University (U.S.A),
Ministry of Education, Culture, Sports, Science, and Technology
and the Japan Society for the Promotion of Science (Japan),
Conselho Nacional de Desenvolvimento Cient\'{\i}fico e
Tecnol{\'o}gico and Funda\c c{\~a}o de Amparo {\`a} Pesquisa do
Estado de S{\~a}o Paulo (Brazil),
Natural Science Foundation of China (People's Republic of China),
Centre National de la Recherche Scientifique, Commissariat
{\`a} l'{\'E}nergie Atomique, and Institut National de Physique
Nucl{\'e}aire et de Physique des Particules (France),
Ministry of Industry, Science and Tekhnologies,
Bundesministerium f\"ur Bildung und Forschung, Deutscher
Akademischer Austausch Dienst, 
and Alexander von Humboldt Stiftung (Germany),
Hungarian National Science Fund, OTKA (Hungary), 
Department of Atomic Energy (India), 
Israel Science Foundation (Israel), 
Korea Research Foundation 
and Korea Science and Engineering Foundation (Korea),
Ministry of Education and Science, Rassia Academy of Sciences,
Federal Agency of Atomic Energy (Russia),
VR and the Wallenberg Foundation (Sweden), 
the U.S. Civilian Research and Development Foundation for the
Independent States of the Former Soviet Union, the US-Hungarian
NSF-OTKA-MTA, and the US-Israel Binational Science Foundation.

\appendix*

\section{Data tables of $v_2$}

Tables~\ref{tab:table_v2cumulant_cent0-30}--\ref{tab:table_v2zdc_smd_cent30-60} 
show numerical data in the same units as plotted in the figures: 
\pt (GeV/$c$), $v_2$, type A statistical error $\sigma_{\rm 
stat}$, type B systematic error $\sigma_{\rm syst}^B$ and type C 
systematic error $\sigma_{\rm syst}^C$.

\begin{table*}[htbp]
\caption{\label{tab:table_v2cumulant_cent0-30}
 $v_2$\{2\} as a function of \pt in centralities 
0--10\%, 10--20\%, 20--30\%, 30--40\%, 40--50\%, and 50--60\%.
}
\begingroup \squeezetable
\begin{ruledtabular} \begin{tabular}{c|ccccc||c|ccccc}
Centrality & & & & & & Centrality & & & & & \\
$v_2$\{\} & $\pt$ (\gevc) & $v_2$ & $\sigma_{\rm stat}$ & $\sigma_{\rm syst}^B$ & $\sigma_{\rm syst}^C$ &
$v_2$\{\} & $\pt$ (\gevc) & $v_2$ & $\sigma_{\rm stat}$ & $\sigma_{\rm syst}^B$ & $\sigma_{\rm syst}^C$ \\ 
\hline
 & 0.247 & 0.00859 & 0.00014 & 0.00001   &  0.00000 &  & 0.250 & 0.00898   & 0.00021 &  0.00001  & 0.00000 \\
 & 0.347 & 0.01406 & 0.00019 & 0.00004   &  0.00000 &  & 0.349 & 0.04323   & 0.00030 &  0.00026  & 0.00000 \\
 & 0.450 & 0.01882 & 0.00023 & 0.00007   &  0.00000 &  & 0.448 & 0.06214   & 0.00036 &  0.00053  & 0.00000 \\
 & 0.547 & 0.02140 & 0.00027 & 0.00009   &  0.00000 &  & 0.548 & 0.07193   & 0.00042 &  0.00071  & 0.00000 \\
 & 0.649 & 0.02395 & 0.00031 & 0.00011   &  0.00000 &  & 0.648 & 0.08243   & 0.00048 &  0.00093  & 0.00000 \\
 & 0.748 & 0.02718 & 0.00036 & 0.00014   &  0.00000 &  & 0.748 & 0.09401   & 0.00055 &  0.00121  & 0.00000 \\
 & 0.847 & 0.03087 & 0.00041 & 0.00018   &  0.00000 &  & 0.848 & 0.10533   & 0.00063 &  0.00153  & 0.00000 \\
 & 0.949 & 0.03605 & 0.00047 & 0.00024   &  0.00000 &  & 0.948 & 0.11678   & 0.00071 &  0.00187  & 0.00000 \\
 & 1.090 & 0.03950 & 0.00041 & 0.00029   &  0.00000 &  & 1.092 & 0.12972   & 0.00063 &  0.00231  & 0.00000 \\
0--10\% 
 & 1.291 & 0.04734 &  0.00053 & 0.00042  & 0.00000 &  30--40\%& 1.291 & 0.15059   & 0.00081 &  0.00312 & 0.00000 \\
$v_2$\{2\}  
& 1.490 &    0.05633   &  0.00070   & 0.00059 & 0.00000 &  $v_2$\{2\} & 1.489 & 0.16955   & 0.00107 &  0.00395 & 0.00000 \\
 & 1.689 &   0.06542   &  0.00095   & 0.00080  & 0.00000 &  	      & 1.689 & 0.18422   & 0.00147 &  0.00467 & 0.00000 \\
 & 1.890 &   0.07148   &  0.00128   & 0.00096  & 0.00000 &            & 1.891 & 0.19625   & 0.00198 &  0.00529 & 0.00000 \\
 & 2.194 &   0.08352   &  0.00128   & 0.00130  & 0.00000 &            & 2.197 & 0.21718   & 0.00196 &  0.00648 & 0.00000 \\
 & 2.698 &   0.09362   &  0.00249   & 0.00164  & 0.00000 &            & 2.702 & 0.22835   & 0.00369 &  0.00717 & 0.00000 \\
 & 3.329 &   0.08866   &  0.00421   & 0.00147  & 0.00000 &            & 3.338 & 0.22623   & 0.00556 &  0.00704 & 0.00000 \\
 & 4.365 &   0.08997   &  0.01134   & 0.00151  & 0.00000 &            & 4.360 & 0.19059   & 0.01496 &  0.00499 & 0.00000 \\
 & 5.376 &   0.07933   &  0.02365   & 0.00118  & 0.00000 &            & 5.379 & 0.16931   & 0.03256 &  0.00394 & 0.00000 \\
 & 6.695 &   0.08701   &  0.02720   & 0.00142  & 0.00000 &            & 6.628 & 0.16346   & 0.05010 &  0.00367 & 0.00000 \\
\hline 
 & 0.248 &  0.01089   & 0.00013  & 0.00002  & 0.00000 &  & 0.250 & 0.00625   &  0.00032 & 0.00001  & 0.00000 \\
 & 0.348 &  0.02714   & 0.00018  & 0.00011  & 0.00000 &  & 0.349 & 0.04611   &  0.00044 & 0.00028  & 0.00000 \\
 & 0.449 &  0.03914   & 0.00023  & 0.00023  & 0.00000 &  & 0.448 & 0.06387   &  0.00054 & 0.00054  & 0.00000 \\
 & 0.547 &  0.04592   & 0.00027  & 0.00032  & 0.00000 &  & 0.548 & 0.07455   &  0.00062 & 0.00073  & 0.00000 \\
 & 0.649 &  0.05281   & 0.00030  & 0.00042  & 0.00000 &  & 0.648 & 0.08575   &  0.00072 & 0.00097  & 0.00000 \\
 & 0.748 &  0.05977   & 0.00035  & 0.00054  & 0.00000 &  & 0.748 & 0.09774   &  0.00082 & 0.00126  & 0.00000 \\
 & 0.848 &  0.06637   & 0.00040  & 0.00066  & 0.00000 &  & 0.848 & 0.11126   &  0.00094 & 0.00163  & 0.00000 \\
 & 0.948 &  0.07459   & 0.00045  & 0.00083  & 0.00000 &  & 0.948 & 0.11974   &  0.00108 & 0.00189  & 0.00000 \\
 & 1.092 &  0.08249   & 0.00040  & 0.00102  & 0.00000 &  & 1.092 & 0.13745   &  0.00095 & 0.00249  & 0.00000 \\
10--20\% 
 & 1.291 & 0.09506   & 0.00051 &  0.00136  & 0.00000 &  40--50\% & 1.291 & 0.15672   &  0.00123 & 0.00324  & 0.00000 \\
$v_2$\{2\} 
 & 1.490 & 0.10997   & 0.00067   & 0.00181  & 0.00000 &  $v_2$\{2\} & 1.489 & 0.17633	&  0.00166 & 0.00410  & 0.00000 \\
 & 1.689 & 0.12394   & 0.00090   & 0.00230  & 0.00000 &             & 1.689 & 0.19315	&  0.00229 & 0.00492  & 0.00000 \\
 & 1.891 & 0.13378   & 0.00121   & 0.00268  & 0.00000 &             & 1.891 & 0.20965	&  0.00309 & 0.00580  & 0.00000 \\
 & 2.196 & 0.14881   & 0.00121   & 0.00332  & 0.00000 &             & 2.199 & 0.21909	&  0.00304 & 0.00633  & 0.00000 \\
 & 2.699 & 0.16781   & 0.00232   & 0.00422  & 0.00000 &             & 2.701 & 0.23572	&  0.00567 & 0.00733  & 0.00000 \\
 & 3.328 & 0.16669   & 0.00382   & 0.00417  & 0.00000 &             & 3.344 & 0.24331	&  0.00808 & 0.00781  & 0.00000 \\
 & 4.357 & 0.13468   & 0.01047   & 0.00272  & 0.00000 &             & 4.346 & 0.26575	&  0.02124 & 0.00932  & 0.00000 \\
 & 5.371 & 0.14951   & 0.02244   & 0.00335  & 0.00000 &             & 5.414 & 0.24613	&  0.03288 & 0.00799  & 0.00000 \\
 & 6.587 & 0.11931   & 0.02641   & 0.00214  & 0.00000 &             & 6.566 & 0.17786	&  0.05097 & 0.00417  & 0.00000 \\
\hline 
 & 0.249 & 0.01127   &  0.00015  &  0.00002 & 0.00000 &  & 0.251 & 0.01201   & 0.00052 & 0.00002  & 0.00000 \\
 & 0.349 & 0.03713   &  0.00022  &  0.00019 & 0.00000 &  & 0.349 & 0.03575   & 0.00056 & 0.00016  & 0.00000 \\
 & 0.448 & 0.05370   &  0.00028  &  0.00040 & 0.00000 &  & 0.448 & 0.05111   & 0.00063 & 0.00033  & 0.00000 \\
 & 0.548 & 0.06252   &  0.00032  &  0.00054 & 0.00000 &  & 0.548 & 0.06256   & 0.00071 & 0.00050  & 0.00000 \\
 & 0.648 & 0.07147   &  0.00036  &  0.00070 & 0.00000 &  & 0.648 & 0.07591   & 0.00080 & 0.00073  & 0.00000 \\
 & 0.748 & 0.08144   &  0.00041  &  0.00091 & 0.00000 &  & 0.748 & 0.08903   & 0.00091 & 0.00101  & 0.00000 \\
 & 0.848 & 0.09118   &  0.00047  &  0.00114 & 0.00000 &  & 0.848 & 0.09965   & 0.00103 & 0.00126  & 0.00000 \\
 & 0.948 & 0.10071   &  0.00053  &  0.00139 & 0.00000 &  & 0.948 & 0.11124   & 0.00118 & 0.00157  & 0.00000 \\
 & 1.092 & 0.11227   &  0.00047  &  0.00173 & 0.00000 &  & 1.091 & 0.12340   & 0.00103 & 0.00193  & 0.00000 \\
20--30\% 
 & 1.291 & 0.12982   &  0.00060  &  0.00232   & 0.00000 &  50--60\% & 1.290 & 0.14241   & 0.00133 & 0.00257  & 0.00000 \\
$v_2$\{2\} 
 & 1.489 & 0.14786   &  0.00079  &  0.00301  & 0.00000 &  $v_2$\{2\} & 1.489 & 0.16236   & 0.00178 & 0.00334  & 0.00000 \\
 & 1.689 & 0.16113   &  0.00107  &  0.00357  & 0.00000 &             & 1.689 & 0.17737   & 0.00248 & 0.00399  & 0.00000 \\
 & 1.891 & 0.17515   &  0.00145  &  0.00422  & 0.00000 &             & 1.890 & 0.19295   & 0.00337 & 0.00472  & 0.00000 \\
 & 2.196 & 0.19364   &  0.00143  &  0.00515  & 0.00000 &             & 2.198 & 0.21282   & 0.00330 & 0.00575  & 0.00000 \\
 & 2.699 & 0.20931   &  0.00271  &  0.00602  & 0.00000 &             & 2.700 & 0.22201   & 0.00623 & 0.00625  & 0.00000 \\
 & 3.333 & 0.20299   &  0.00430  &  0.00567  & 0.00000 &             & 3.348 & 0.21980   & 0.00917 & 0.00613  & 0.00000 \\
 & 4.356 & 0.19729   &  0.01175  &  0.00535  & 0.00000 &             & 4.373 & 0.24935   & 0.02292 & 0.00789  & 0.00000 \\
 & 5.383 & 0.18635   &  0.02567  &  0.00477  & 0.00000 &             & 5.452 & 0.36285   & 0.05515 & 0.01671  & 0.00000 \\
 & 6.611 & 0.15079   &  0.04839  &  0.00313  & 0.00000 &             & 6.734 & 0.40554   & 0.08167 & 0.02087  & 0.00000 \\
\end{tabular} \end{ruledtabular} \endgroup
\end{table*}

\begin{table*}[htbp]
\caption{\label{tab:table_v2cumulant_cent20-60}
 $v_2$\{2\} as a function of \pt in centrality 20--60\%.
}
\begingroup \squeezetable
\begin{ruledtabular} \begin{tabular}{c|cccccc|cccccc}
Centrality & & & & & & & & & & & \\
$v_2$\{\} & $\pt$ (\gevc) & $v_2$ & $\sigma_{\rm stat}$ & $\sigma_{\rm syst}^B$ & $\sigma_{\rm syst}^C$ 
 & & $\pt$ (\gevc) & $v_2$ & $\sigma_{\rm stat}$ & $\sigma_{\rm syst}^B$ & $\sigma_{\rm syst}^C$ \\ 
\hline
 & 0.251 & 0.00778   &  0.00011  & 0.00001 & 0.00000 && 1.489 & 0.14884   &  0.00058  & 0.00292 & 0.00000 \\ 
 & 0.349 & 0.03793   &  0.00016  & 0.00019 & 0.00000 && 1.689 & 0.16226   &  0.00080  & 0.00347 & 0.00000 \\
 & 0.448 & 0.05476   &  0.00020  & 0.00040 & 0.00000 && 1.890 & 0.17456   &  0.00108  & 0.00402 & 0.00000 \\
 & 0.548 & 0.06374   &  0.00023  & 0.00054 & 0.00000 && 2.198 & 0.19027   &  0.00106  & 0.00478 & 0.00000 \\
20--60\%
 & 0.648 & 0.07303   &  0.00026  & 0.00070 & 0.00000 && 2.700 & 0.20415   &  0.00201  & 0.00550 & 0.00000 \\
$v_2$\{2\}
 & 0.748 & 0.08283   &  0.00030  & 0.00091 & 0.00000 && 3.348 & 0.21363   &  0.00304  & 0.00602 & 0.00000 \\
 & 0.848 & 0.09301   &  0.00034  & 0.00114 & 0.00000 && 4.373 & 0.19568   &  0.00653  & 0.00505 & 0.00000 \\
 & 0.948 & 0.10247   &  0.00039  & 0.00139 & 0.00000 && 5.452 & 0.23823   &  0.01494  & 0.00749 & 0.00000 \\
 & 1.091 & 0.11444   &  0.00034  & 0.00173 & 0.00000 && 6.734 & 0.18915   &  0.02297  & 0.00472 & 0.00000 \\
 & 1.290 & 0.13201   &  0.00044  & 0.00230 & 0.00000 && &  &  &  &
\end{tabular} \end{ruledtabular} \endgroup
\end{table*}

\begin{table*}[htbp]
\caption{\label{tab:table_v2_cent20-60}
$v_2$\{BBC\} and $v_2$\{ZDC-SMD\}
from S-N and ZDC-BBC-CNT subevents as a function of \pt in centrality 
20--60\%.
}
\begingroup \squeezetable
\begin{ruledtabular} \begin{tabular}{c|cccccccccccc}
Centrality
 & & & \multicolumn{4}{c}{S-N subevents} & & \multicolumn{4}{c}{ZDC-BBC-CNT subevents} \\ 
$v_2$\{\}
 &$\pt$ (\gevc) & & $v_2$ & $\sigma_{\rm stat}$ & $\sigma_{\rm syst}^B$ & $\sigma_{\rm syst}^C$ 
& & $v_2$ & $\sigma_{\rm stat}$ & $\sigma_{\rm syst}^B$ & $\sigma_{\rm syst}^C$ \\ 
\hline
 & 0.247 && 0.02569 & 0.00009 & 0.00049 & 0.00001 & & 0.02486 & 0.00009 & 0.00045 & 0.00001 \\ 
 & 0.348 && 0.04271 & 0.00009 & 0.00016 & 0.00003 & & 0.04133 & 0.00010 & 0.00015 & 0.00003 \\ 
 & 0.448 && 0.05587 & 0.00010 & 0.00014 & 0.00006 & & 0.05407 & 0.00012 & 0.00013 & 0.00005 \\ 
 & 0.548 && 0.06846 & 0.00011 & 0.00015 & 0.00009 & & 0.06625 & 0.00013 & 0.00014 & 0.00008 \\ 
 & 0.648 && 0.08009 & 0.00013 & 0.00015 & 0.00012 & & 0.07751 & 0.00015 & 0.00014 & 0.00011 \\ 
 & 0.748 && 0.09123 & 0.00014 & 0.00016 & 0.00015 & & 0.08828 & 0.00017 & 0.00015 & 0.00014 \\ 
 & 0.848 && 0.10124 & 0.00016 & 0.00019 & 0.00019 & & 0.09798 & 0.00019 & 0.00018 & 0.00018 \\ 
 & 0.948 && 0.11159 & 0.00018 & 0.00017 & 0.00023 & & 0.10799 & 0.00021 & 0.00016 & 0.00021 \\ 
20--60\%
 & 1.092 && 0.12439 & 0.00016 & 0.00018 & 0.00029 & & 0.12038 & 0.00020 & 0.00017 & 0.00027 \\ 
$v_2$\{BBC\}
 & 1.292 && 0.14170 & 0.00020 & 0.00019 & 0.00037 & & 0.13713 & 0.00025 & 0.00018 & 0.00035 \\ 
 & 1.492 && 0.15770 & 0.00027 & 0.00027 & 0.00046 & & 0.15261 & 0.00031 & 0.00025 & 0.00043 \\ 
 & 1.692 && 0.17244 & 0.00037 & 0.00027 & 0.00055 & & 0.16688 & 0.00040 & 0.00026 & 0.00051 \\ 
 & 1.892 && 0.18481 & 0.00050 & 0.00030 & 0.00063 & & 0.17885 & 0.00052 & 0.00028 & 0.00059 \\ 
 & 2.200 && 0.19684 & 0.00049 & 0.00029 & 0.00071 & & 0.19049 & 0.00052 & 0.00027 & 0.00067 \\ 
 & 2.703 && 0.20803 & 0.00092 & 0.00025 & 0.00080 & & 0.20132 & 0.00092 & 0.00023 & 0.00075 \\ 
 & 3.343 && 0.20569 & 0.00141 & 0.00039 & 0.00078 & & 0.19905 & 0.00138 & 0.00037 & 0.00073 \\ 
 & 4.381 && 0.17942 & 0.00371 & 0.00066 & 0.00059 & & 0.17363 & 0.00360 & 0.00062 & 0.00056 \\ 
 & 5.410 && 0.14862 & 0.00877 & 0.00098 & 0.00041 & & 0.14382 & 0.00849 & 0.00092 & 0.00038 \\ 
 & 6.852 && 0.16262 & 0.01770 & 0.00328 & 0.00049 & & 0.15738 & 0.01713 & 0.00308 & 0.00046 \\ 
\hline
 & 0.247 && 0.02532 & 0.00025 & 0.00047 & 0.00004 & & 0.02661 & 0.00035 & 0.00052 & 0.00002 \\ 
 & 0.348 && 0.04002 & 0.00029 & 0.00014 & 0.00010 & & 0.04188 & 0.00037 & 0.00015 & 0.00004 \\ 
 & 0.448 && 0.05165 & 0.00032 & 0.00012 & 0.00017 & & 0.05395 & 0.00041 & 0.00013 & 0.00007 \\ 
 & 0.548 && 0.06296 & 0.00036 & 0.00013 & 0.00025 & & 0.06567 & 0.00046 & 0.00014 & 0.00010 \\ 
 & 0.648 && 0.07433 & 0.00041 & 0.00013 & 0.00035 & & 0.07746 & 0.00051 & 0.00014 & 0.00014 \\ 
 & 0.748 && 0.08377 & 0.00046 & 0.00013 & 0.00044 & & 0.08730 & 0.00057 & 0.00015 & 0.00017 \\ 
 & 0.848 && 0.09429 & 0.00052 & 0.00017 & 0.00056 & & 0.09827 & 0.00065 & 0.00018 & 0.00022 \\ 
 & 0.948 && 0.10365 & 0.00059 & 0.00015 & 0.00067 & & 0.10808 & 0.00074 & 0.00016 & 0.00027 \\ 
20--60\%
 & 1.092 && 0.11617 & 0.00053 & 0.00016 & 0.00085 & & 0.12065 & 0.00063 & 0.00017 & 0.00033 \\ 
$v_2$\{ZDC-SMD\}
 & 1.292 && 0.13006 & 0.00066 & 0.00016 & 0.00106 & & 0.13535 & 0.00081 & 0.00018 & 0.00042 \\ 
 & 1.492 && 0.14367 & 0.00086 & 0.00023 & 0.00129 & & 0.14994 & 0.00109 & 0.00024 & 0.00052 \\ 
 & 1.692 && 0.15763 & 0.00115 & 0.00023 & 0.00156 & & 0.16504 & 0.00150 & 0.00025 & 0.00062 \\ 
 & 1.892 && 0.17281 & 0.00151 & 0.00026 & 0.00187 & & 0.18136 & 0.00203 & 0.00029 & 0.00075 \\ 
 & 2.200 && 0.18031 & 0.00149 & 0.00024 & 0.00204 & & 0.18912 & 0.00200 & 0.00026 & 0.00082 \\ 
 & 2.703 && 0.18983 & 0.00263 & 0.00021 & 0.00226 & & 0.19998 & 0.00375 & 0.00023 & 0.00092 \\ 
 & 3.343 && 0.18147 & 0.00393 & 0.00030 & 0.00206 & & 0.19147 & 0.00576 & 0.00034 & 0.00084 \\ 
 & 4.381 && 0.16102 & 0.01018 & 0.00053 & 0.00162 & & 0.17005 & 0.01517 & 0.00059 & 0.00066 \\ 
 & 5.410 && 0.14043 & 0.02402 & 0.00088 & 0.00124 & & 0.14833 & 0.03585 & 0.00098 & 0.00050 \\ 
 & 6.852 && 0.12310 & 0.04849 & 0.00188 & 0.00095 & & 0.13003 & 0.07240 & 0.00210 & 0.00039 \\ 
\end{tabular} \end{ruledtabular} \endgroup
\end{table*}

\begin{table*}[htbp]
\caption{\label{tab:table_v2bbc_cent0-30}
$v_2$\{BBC\} from S-N and ZDC-BBC-CNT subevents as a function of \pt in centrality 
0--10\%, 10--20\%, and 20--30\%.
}
\begingroup \squeezetable
\begin{ruledtabular} \begin{tabular}{c|cccccccccccc}
Centrality
 & & & \multicolumn{4}{c}{S-N subevents} & & \multicolumn{4}{c}{ZDC-BBC-CNT subevents} \\ 
$v_2$\{\}
 &$\pt$ (\gevc) & & $v_2$ & $\sigma_{\rm stat}$ & $\sigma_{\rm syst}^B$ & $\sigma_{\rm syst}^C$ 
& & $v_2$ & $\sigma_{\rm stat}$ & $\sigma_{\rm syst}^B$ & $\sigma_{\rm syst}^C$ \\ 
\hline
 & 0.247 && 0.01025 & 0.00012 & 0.00008 & 0.00001 & & 0.00966 & 0.00016 & 0.00007 & 0.00000 \\ 
 & 0.348 && 0.01868 & 0.00014 & 0.00003 & 0.00002 & & 0.01762 & 0.00025 & 0.00003 & 0.00002 \\ 
 & 0.448 && 0.02300 & 0.00016 & 0.00005 & 0.00003 & & 0.02169 & 0.00030 & 0.00005 & 0.00002 \\ 
 & 0.548 && 0.02741 & 0.00018 & 0.00007 & 0.00004 & & 0.02586 & 0.00035 & 0.00006 & 0.00003 \\ 
 & 0.648 && 0.03174 & 0.00020 & 0.00007 & 0.00005 & & 0.02993 & 0.00041 & 0.00006 & 0.00005 \\ 
 & 0.748 && 0.03570 & 0.00023 & 0.00007 & 0.00006 & & 0.03367 & 0.00046 & 0.00006 & 0.00006 \\ 
 & 0.848 && 0.03990 & 0.00026 & 0.00007 & 0.00008 & & 0.03763 & 0.00051 & 0.00006 & 0.00007 \\ 
 & 0.948 && 0.04428 & 0.00029 & 0.00008 & 0.00010 & & 0.04176 & 0.00057 & 0.00007 & 0.00009 \\ 
 & 1.092 && 0.04941 & 0.00025 & 0.00008 & 0.00012 & & 0.04660 & 0.00061 & 0.00007 & 0.00011 \\ 
0--10\%
 & 1.292 && 0.05631 & 0.00032 & 0.00008 & 0.00016 & & 0.05310 & 0.00070 & 0.00007 & 0.00014 \\ 
$v_2$\{BBC\}
 & 1.492 && 0.06349 & 0.00042 & 0.00008 & 0.00020 & & 0.05988 & 0.00082 & 0.00007 & 0.00018 \\ 
 & 1.692 && 0.07065 & 0.00058 & 0.00012 & 0.00025 & & 0.06663 & 0.00096 & 0.00010 & 0.00022 \\ 
 & 1.892 && 0.07859 & 0.00078 & 0.00011 & 0.00031 & & 0.07412 & 0.00115 & 0.00010 & 0.00028 \\ 
 & 2.200 && 0.08557 & 0.00078 & 0.00009 & 0.00037 & & 0.08070 & 0.00121 & 0.00008 & 0.00033 \\ 
 & 2.703 && 0.09598 & 0.00151 & 0.00015 & 0.00046 & & 0.09052 & 0.00179 & 0.00014 & 0.00041 \\ 
 & 3.343 && 0.09806 & 0.00245 & 0.00031 & 0.00049 & & 0.09249 & 0.00257 & 0.00028 & 0.00043 \\ 
 & 4.381 && 0.08795 & 0.00699 & 0.00089 & 0.00039 & & 0.08295 & 0.00667 & 0.00079 & 0.00035 \\ 
 & 5.410 && -- & -- & -- & -- & & -- & -- & -- & -- \\ 
 & 6.852 && -- & -- & -- & -- & & -- & -- & -- & -- \\ 
\hline 
 & 0.247 && 0.01804 & 0.00010 & 0.00008 & 0.00000 & & 0.01754 & 0.00011 & 0.00007 & 0.00000 \\ 
 & 0.348 && 0.03095 & 0.00011 & 0.00008 & 0.00001 & & 0.03008 & 0.00015 & 0.00008 & 0.00001 \\ 
 & 0.448 && 0.03927 & 0.00012 & 0.00012 & 0.00002 & & 0.03816 & 0.00018 & 0.00011 & 0.00002 \\ 
 & 0.548 && 0.04714 & 0.00014 & 0.00018 & 0.00003 & & 0.04582 & 0.00020 & 0.00017 & 0.00003 \\ 
 & 0.648 && 0.05480 & 0.00015 & 0.00016 & 0.00004 & & 0.05326 & 0.00023 & 0.00015 & 0.00004 \\ 
 & 0.748 && 0.06236 & 0.00017 & 0.00016 & 0.00006 & & 0.06060 & 0.00026 & 0.00015 & 0.00005 \\ 
 & 0.848 && 0.06895 & 0.00019 & 0.00016 & 0.00007 & & 0.06701 & 0.00029 & 0.00015 & 0.00006 \\ 
 & 0.948 && 0.07647 & 0.00022 & 0.00017 & 0.00008 & & 0.07432 & 0.00033 & 0.00016 & 0.00008 \\ 
 & 1.092 && 0.08498 & 0.00019 & 0.00018 & 0.00010 & & 0.08259 & 0.00033 & 0.00017 & 0.00010 \\ 
10--20\%
 & 1.292 && 0.09731 & 0.00024 & 0.00018 & 0.00014 & & 0.09457 & 0.00040 & 0.00017 & 0.00013 \\ 
$v_2$\{BBC\}
 & 1.492 && 0.10883 & 0.00032 & 0.00022 & 0.00017 & & 0.10576 & 0.00047 & 0.00021 & 0.00016 \\ 
 & 1.692 && 0.12204 & 0.00044 & 0.00021 & 0.00021 & & 0.11860 & 0.00058 & 0.00020 & 0.00020 \\ 
 & 1.892 && 0.13129 & 0.00059 & 0.00029 & 0.00025 & & 0.12760 & 0.00072 & 0.00027 & 0.00023 \\ 
 & 2.200 && 0.14375 & 0.00058 & 0.00021 & 0.00030 & & 0.13970 & 0.00074 & 0.00020 & 0.00028 \\ 
 & 2.703 && 0.15569 & 0.00112 & 0.00023 & 0.00035 & & 0.15130 & 0.00120 & 0.00022 & 0.00033 \\ 
 & 3.343 && 0.15885 & 0.00177 & 0.00033 & 0.00037 & & 0.15437 & 0.00180 & 0.00031 & 0.00034 \\ 
 & 4.381 && 0.13970 & 0.00491 & 0.00056 & 0.00028 & & 0.13577 & 0.00480 & 0.00053 & 0.00027 \\ 
 & 5.410 && 0.12763 & 0.01194 & 0.00101 & 0.00024 & & 0.12403 & 0.01161 & 0.00095 & 0.00022 \\ 
 & 6.852 && 0.10820 & 0.02401 & 0.00193 & 0.00017 & & 0.10515 & 0.02334 & 0.00183 & 0.00016 \\ 
\hline
 & 0.247 && 0.02367 & 0.00011 & 0.00032 & 0.00001 & & 0.02303 & 0.00012 & 0.00030 & 0.00001 \\ 
 & 0.348 && 0.03981 & 0.00012 & 0.00014 & 0.00002 & & 0.03874 & 0.00015 & 0.00014 & 0.00002 \\ 
 & 0.448 && 0.05138 & 0.00014 & 0.00016 & 0.00004 & & 0.04999 & 0.00017 & 0.00015 & 0.00004 \\ 
 & 0.548 && 0.06250 & 0.00015 & 0.00019 & 0.00006 & & 0.06081 & 0.00020 & 0.00018 & 0.00005 \\ 
 & 0.648 && 0.07276 & 0.00017 & 0.00020 & 0.00008 & & 0.07080 & 0.00023 & 0.00019 & 0.00007 \\ 
 & 0.748 && 0.08298 & 0.00019 & 0.00018 & 0.00010 & & 0.08075 & 0.00026 & 0.00017 & 0.00010 \\ 
 & 0.848 && 0.09184 & 0.00022 & 0.00020 & 0.00012 & & 0.08937 & 0.00029 & 0.00019 & 0.00012 \\ 
 & 0.948 && 0.10139 & 0.00024 & 0.00020 & 0.00015 & & 0.09866 & 0.00032 & 0.00019 & 0.00014 \\ 
 & 1.092 && 0.11279 & 0.00021 & 0.00022 & 0.00019 & & 0.10976 & 0.00032 & 0.00021 & 0.00018 \\ 
20--30\%
 & 1.292 && 0.12862 & 0.00027 & 0.00023 & 0.00024 & & 0.12516 & 0.00038 & 0.00022 & 0.00023 \\ 
$v_2$\{BBC\}
 & 1.492 && 0.14459 & 0.00036 & 0.00029 & 0.00031 & & 0.14070 & 0.00046 & 0.00027 & 0.00029 \\ 
 & 1.692 && 0.15864 & 0.00049 & 0.00030 & 0.00037 & & 0.15437 & 0.00058 & 0.00029 & 0.00035 \\ 
 & 1.892 && 0.17169 & 0.00066 & 0.00032 & 0.00043 & & 0.16707 & 0.00074 & 0.00030 & 0.00041 \\ 
 & 2.200 && 0.18437 & 0.00065 & 0.00032 & 0.00050 & & 0.17941 & 0.00075 & 0.00030 & 0.00047 \\ 
 & 2.703 && 0.19554 & 0.00123 & 0.00042 & 0.00056 & & 0.19028 & 0.00127 & 0.00039 & 0.00053 \\ 
 & 3.343 && 0.19585 & 0.00192 & 0.00048 & 0.00056 & & 0.19058 & 0.00192 & 0.00046 & 0.00053 \\ 
 & 4.381 && 0.18189 & 0.00521 & 0.00088 & 0.00049 & & 0.17700 & 0.00509 & 0.00083 & 0.00046 \\ 
 & 5.410 && 0.14502 & 0.01244 & 0.00138 & 0.00031 & & 0.14112 & 0.01211 & 0.00131 & 0.00029 \\ 
 & 6.852 && 0.15856 & 0.02490 & 0.00286 & 0.00037 & & 0.15430 & 0.02423 & 0.00271 & 0.00035 \\ 
\end{tabular} \end{ruledtabular} \endgroup
\end{table*}

\begin{table*}[htbp]
\caption{\label{tab:table_v2bbc_cent30-60}
$v_2$\{BBC\} from S-N and ZDC-BBC-CNT subevents as a function of \pt in centrality 
30--40\%, 40--50\%, and 50--60\%.
}
\begingroup \squeezetable
\begin{ruledtabular} \begin{tabular}{c|ccccccccccc}
Centrality
 & & & \multicolumn{4}{c}{S-N subevents} & & \multicolumn{4}{c}{ZDC-BBC-CNT subevents} \\ 
$v_2$\{\}
 &$\pt$ (\gevc) & & $v_2$ & $\sigma_{\rm stat}$ & $\sigma_{\rm syst}^B$ & $\sigma_{\rm syst}^C$ 
& & $v_2$ & $\sigma_{\rm stat}$ & $\sigma_{\rm syst}^B$ & $\sigma_{\rm syst}^C$ \\ 
\hline
 & 0.247 && 0.02733 & 0.00015 & 0.00064 & 0.00001 & & 0.02643 & 0.00016 & 0.00059 & 0.00001 \\ 
 & 0.348 && 0.04523 & 0.00016 & 0.00017 & 0.00004 & & 0.04375 & 0.00018 & 0.00016 & 0.00003 \\ 
 & 0.448 && 0.05935 & 0.00018 & 0.00017 & 0.00006 & & 0.05740 & 0.00021 & 0.00016 & 0.00006 \\ 
 & 0.548 && 0.07263 & 0.00020 & 0.00016 & 0.00010 & & 0.07024 & 0.00024 & 0.00015 & 0.00009 \\ 
 & 0.648 && 0.08502 & 0.00023 & 0.00015 & 0.00013 & & 0.08223 & 0.00028 & 0.00014 & 0.00012 \\ 
 & 0.748 && 0.09651 & 0.00025 & 0.00020 & 0.00017 & & 0.09334 & 0.00031 & 0.00018 & 0.00016 \\ 
 & 0.848 && 0.10742 & 0.00029 & 0.00019 & 0.00021 & & 0.10390 & 0.00035 & 0.00018 & 0.00020 \\ 
 & 0.948 && 0.11793 & 0.00033 & 0.00018 & 0.00025 & & 0.11406 & 0.00039 & 0.00017 & 0.00024 \\ 
 & 1.092 && 0.13156 & 0.00028 & 0.00022 & 0.00032 & & 0.12724 & 0.00038 & 0.00020 & 0.00030 \\ 
30--40\%
 & 1.292 && 0.15004 & 0.00036 & 0.00019 & 0.00041 & & 0.14512 & 0.00046 & 0.00018 & 0.00038 \\ 
$v_2$\{BBC\}
 & 1.492 && 0.16604 & 0.00048 & 0.00030 & 0.00050 & & 0.16059 & 0.00057 & 0.00028 & 0.00047 \\ 
 & 1.692 && 0.18107 & 0.00066 & 0.00029 & 0.00060 & & 0.17513 & 0.00073 & 0.00027 & 0.00056 \\ 
 & 1.892 && 0.19290 & 0.00089 & 0.00034 & 0.00068 & & 0.18657 & 0.00094 & 0.00032 & 0.00063 \\ 
 & 2.200 && 0.20640 & 0.00088 & 0.00035 & 0.00078 & & 0.19962 & 0.00094 & 0.00032 & 0.00073 \\ 
 & 2.703 && 0.21859 & 0.00164 & 0.00042 & 0.00087 & & 0.21142 & 0.00164 & 0.00040 & 0.00081 \\ 
 & 3.343 && 0.21843 & 0.00252 & 0.00037 & 0.00087 & & 0.21127 & 0.00247 & 0.00034 & 0.00081 \\ 
 & 4.381 && 0.18342 & 0.00662 & 0.00101 & 0.00061 & & 0.17740 & 0.00641 & 0.00095 & 0.00057 \\ 
 & 5.410 && 0.15970 & 0.01568 & 0.00197 & 0.00046 & & 0.15446 & 0.01517 & 0.00184 & 0.00043 \\ 
 & 6.852 && 0.18703 & 0.03171 & 0.00640 & 0.00064 & & 0.18090 & 0.03067 & 0.00599 & 0.00060 \\ 
\hline 
 & 0.247 && 0.02840 & 0.00024 & 0.00071 & 0.00002 & & 0.02735 & 0.00024 & 0.00066 & 0.00002 \\ 
 & 0.348 && 0.04699 & 0.00025 & 0.00018 & 0.00005 & & 0.04524 & 0.00027 & 0.00017 & 0.00005 \\ 
 & 0.448 && 0.06236 & 0.00028 & 0.00015 & 0.00009 & & 0.06005 & 0.00031 & 0.00014 & 0.00009 \\ 
 & 0.548 && 0.07757 & 0.00031 & 0.00015 & 0.00014 & & 0.07469 & 0.00035 & 0.00014 & 0.00013 \\ 
 & 0.648 && 0.09141 & 0.00035 & 0.00015 & 0.00020 & & 0.08802 & 0.00040 & 0.00014 & 0.00018 \\ 
 & 0.748 && 0.10354 & 0.00039 & 0.00016 & 0.00025 & & 0.09969 & 0.00045 & 0.00015 & 0.00024 \\ 
 & 0.848 && 0.11530 & 0.00044 & 0.00019 & 0.00032 & & 0.11102 & 0.00051 & 0.00017 & 0.00029 \\ 
 & 0.948 && 0.12668 & 0.00050 & 0.00016 & 0.00038 & & 0.12198 & 0.00057 & 0.00015 & 0.00035 \\ 
 & 1.092 && 0.14106 & 0.00044 & 0.00015 & 0.00047 & & 0.13583 & 0.00054 & 0.00014 & 0.00044 \\ 
40--50\%
 & 1.292 && 0.15967 & 0.00056 & 0.00019 & 0.00061 & & 0.15374 & 0.00066 & 0.00017 & 0.00056 \\ 
$v_2$\{BBC\}
 & 1.492 && 0.17584 & 0.00075 & 0.00025 & 0.00074 & & 0.16932 & 0.00083 & 0.00023 & 0.00068 \\ 
 & 1.692 && 0.19082 & 0.00104 & 0.00031 & 0.00087 & & 0.18373 & 0.00110 & 0.00029 & 0.00080 \\ 
 & 1.892 && 0.20216 & 0.00141 & 0.00031 & 0.00097 & & 0.19466 & 0.00144 & 0.00029 & 0.00090 \\ 
 & 2.200 && 0.21274 & 0.00138 & 0.00031 & 0.00108 & & 0.20485 & 0.00142 & 0.00029 & 0.00100 \\ 
 & 2.703 && 0.22348 & 0.00256 & 0.00039 & 0.00119 & & 0.21518 & 0.00252 & 0.00036 & 0.00110 \\ 
 & 3.343 && 0.22044 & 0.00387 & 0.00067 & 0.00116 & & 0.21226 & 0.00376 & 0.00063 & 0.00107 \\ 
 & 4.381 && 0.18665 & 0.00994 & 0.00094 & 0.00083 & & 0.17973 & 0.00958 & 0.00087 & 0.00077 \\ 
 & 5.410 && 0.16716 & 0.02325 & 0.00178 & 0.00067 & & 0.16095 & 0.02239 & 0.00165 & 0.00062 \\ 
 & 6.852 && 0.15951 & 0.04732 & 0.00616 & 0.00060 & & 0.15359 & 0.04556 & 0.00571 & 0.00056 \\ 
\hline 
 & 0.247 && 0.02767 & 0.00043 & 0.00056 & 0.00003 & & 0.02604 & 0.00042 & 0.00050 & 0.00003 \\ 
 & 0.348 && 0.04569 & 0.00046 & 0.00019 & 0.00008 & & 0.04300 & 0.00046 & 0.00017 & 0.00007 \\ 
 & 0.448 && 0.06193 & 0.00050 & 0.00018 & 0.00014 & & 0.05828 & 0.00052 & 0.00016 & 0.00013 \\ 
 & 0.548 && 0.07654 & 0.00056 & 0.00014 & 0.00022 & & 0.07203 & 0.00060 & 0.00013 & 0.00019 \\ 
 & 0.648 && 0.08963 & 0.00064 & 0.00013 & 0.00030 & & 0.08435 & 0.00068 & 0.00012 & 0.00027 \\ 
 & 0.748 && 0.10358 & 0.00072 & 0.00014 & 0.00040 & & 0.09747 & 0.00077 & 0.00012 & 0.00036 \\ 
 & 0.848 && 0.11362 & 0.00082 & 0.00020 & 0.00048 & & 0.10692 & 0.00087 & 0.00018 & 0.00043 \\ 
 & 0.948 && 0.12637 & 0.00093 & 0.00011 & 0.00060 & & 0.11892 & 0.00099 & 0.00010 & 0.00053 \\ 
 & 1.092 && 0.14117 & 0.00082 & 0.00014 & 0.00075 & & 0.13284 & 0.00091 & 0.00012 & 0.00066 \\ 
50--60\%
 & 1.292 && 0.15953 & 0.00105 & 0.00020 & 0.00095 & & 0.15013 & 0.00114 & 0.00017 & 0.00085 \\ 
$v_2$\{BBC\}
 & 1.492 && 0.17233 & 0.00141 & 0.00028 & 0.00111 & & 0.16217 & 0.00146 & 0.00024 & 0.00099 \\ 
 & 1.692 && 0.18714 & 0.00196 & 0.00029 & 0.00131 & & 0.17611 & 0.00196 & 0.00026 & 0.00116 \\ 
 & 1.892 && 0.19757 & 0.00266 & 0.00054 & 0.00146 & & 0.18592 & 0.00260 & 0.00047 & 0.00130 \\ 
 & 2.200 && 0.20146 & 0.00260 & 0.00054 & 0.00152 & & 0.18959 & 0.00255 & 0.00048 & 0.00135 \\ 
 & 2.703 && 0.21521 & 0.00480 & 0.00066 & 0.00174 & & 0.20252 & 0.00458 & 0.00059 & 0.00154 \\ 
 & 3.343 && 0.19757 & 0.00712 & 0.00083 & 0.00146 & & 0.18593 & 0.00674 & 0.00074 & 0.00130 \\ 
 & 4.381 && 0.16368 & 0.01791 & 0.00363 & 0.00100 & & 0.15403 & 0.01686 & 0.00321 & 0.00089 \\ 
 & 5.410 && 0.11745 & 0.04124 & 0.00292 & 0.00052 & & 0.11053 & 0.03881 & 0.00259 & 0.00046 \\ 
 & 6.852 && -- & -- & -- & -- & & -- & -- & -- & -- \\ 
\end{tabular} \end{ruledtabular} \endgroup
\end{table*}

\begin{table*}[htbp]
\caption{\label{tab:table_v2zdc_smd_cent0-30}
$v_2$\{ZDC-SMD\} from S-N and ZDC-BBC-CNT subevents as a function of \pt in centralities
0--10\%, 10--20\%, and 20--30\%.
}
\begingroup \squeezetable
\begin{ruledtabular} \begin{tabular}{c|cccccccccccc}
Centrality
 & & & \multicolumn{4}{c}{S-N subevents} & & \multicolumn{4}{c}{ZDC-BBC-CNT subevents} \\ 
$v_2$\{\}
 &$\pt$ (\gevc) & & $v_2$ & $\sigma_{\rm stat}$ & $\sigma_{\rm syst}^B$ & $\sigma_{\rm syst}^C$ 
& & $v_2$ & $\sigma_{\rm stat}$ & $\sigma_{\rm syst}^B$ & $\sigma_{\rm syst}^C$ \\ 
\hline
 & 0.247 && 0.01342 & 0.00114 & 0.00013 & 0.00005 & & 0.01723 & 0.00158 & 0.00022 & 0.00004 \\ 
 & 0.348 && 0.01488 & 0.00131 & 0.00002 & 0.00007 & & 0.01929 & 0.00183 & 0.00003 & 0.00005 \\ 
 & 0.448 && 0.01231 & 0.00133 & 0.00002 & 0.00005 & & 0.01688 & 0.00205 & 0.00003 & 0.00004 \\ 
 & 0.548 && 0.02085 & 0.00170 & 0.00004 & 0.00013 & & 0.02643 & 0.00230 & 0.00006 & 0.00010 \\ 
 & 0.648 && 0.01557 & 0.00166 & 0.00002 & 0.00007 & & 0.02132 & 0.00256 & 0.00003 & 0.00006 \\ 
 & 0.748 && 0.02236 & 0.00203 & 0.00003 & 0.00015 & & 0.02928 & 0.00289 & 0.00005 & 0.00012 \\ 
 & 0.848 && 0.02656 & 0.00233 & 0.00003 & 0.00021 & & 0.03444 & 0.00326 & 0.00005 & 0.00017 \\ 
 & 0.948 && 0.03014 & 0.00265 & 0.00004 & 0.00027 & & 0.03909 & 0.00371 & 0.00006 & 0.00021 \\ 
 & 1.092 && 0.04275 & 0.00268 & 0.00006 & 0.00055 & & 0.04922 & 0.00319 & 0.00008 & 0.00034 \\ 
0--10\%
 & 1.292 && 0.03826 & 0.00304 & 0.00004 & 0.00044 & & 0.04801 & 0.00405 & 0.00006 & 0.00032 \\ 
$v_2$\{ZDC-SMD\}
 & 1.492 && 0.03859 & 0.00367 & 0.00003 & 0.00045 & & 0.05124 & 0.00534 & 0.00005 & 0.00037 \\ 
 & 1.692 && 0.04492 & 0.00476 & 0.00005 & 0.00061 & & 0.06137 & 0.00730 & 0.00009 & 0.00053 \\ 
 & 1.892 && 0.06318 & 0.00654 & 0.00007 & 0.00120 & & 0.08583 & 0.00992 & 0.00014 & 0.00103 \\ 
 & 2.200 && 0.06910 & 0.00672 & 0.00006 & 0.00143 & & 0.09233 & 0.00989 & 0.00011 & 0.00119 \\ 
 & 2.703 && 0.07798 & 0.01123 & 0.00010 & 0.00182 & & 0.11270 & 0.01925 & 0.00021 & 0.00178 \\ 
 & 3.343 && 0.07481 & 0.01667 & 0.00018 & 0.00168 & & 0.11230 & 0.03125 & 0.00041 & 0.00177 \\ 
 & 4.381 && -- & -- & -- & -- & & -- & -- & -- & -- \\ 
 & 5.410 && -- & -- & -- & -- & & -- & -- & -- & -- \\ 
 & 6.852 && -- & -- & -- & -- & & -- & -- & -- & -- \\ 
\hline
 & 0.247 && 0.02194 & 0.00061 & 0.00011 & 0.00006 & & 0.02145 & 0.00067 & 0.00011 & 0.00003 \\ 
 & 0.348 && 0.02987 & 0.00070 & 0.00008 & 0.00011 & & 0.02924 & 0.00074 & 0.00008 & 0.00005 \\ 
 & 0.448 && 0.03696 & 0.00078 & 0.00010 & 0.00017 & & 0.03621 & 0.00083 & 0.00010 & 0.00008 \\ 
 & 0.548 && 0.04342 & 0.00088 & 0.00015 & 0.00023 & & 0.04255 & 0.00092 & 0.00014 & 0.00011 \\ 
 & 0.648 && 0.05052 & 0.00098 & 0.00013 & 0.00031 & & 0.04951 & 0.00103 & 0.00013 & 0.00016 \\ 
 & 0.748 && 0.05556 & 0.00110 & 0.00013 & 0.00037 & & 0.05445 & 0.00115 & 0.00012 & 0.00019 \\ 
 & 0.848 && 0.06572 & 0.00125 & 0.00014 & 0.00052 & & 0.06442 & 0.00130 & 0.00014 & 0.00026 \\ 
 & 0.948 && 0.07064 & 0.00141 & 0.00014 & 0.00060 & & 0.06923 & 0.00148 & 0.00014 & 0.00030 \\ 
 & 1.092 && 0.07773 & 0.00122 & 0.00015 & 0.00073 & & 0.07626 & 0.00126 & 0.00014 & 0.00037 \\ 
10--20\%
 & 1.292 && 0.09169 & 0.00155 & 0.00016 & 0.00102 & & 0.08993 & 0.00162 & 0.00015 & 0.00051 \\ 
$v_2$\{ZDC-SMD\}
 & 1.492 && 0.10236 & 0.00204 & 0.00019 & 0.00127 & & 0.10031 & 0.00214 & 0.00019 & 0.00064 \\ 
 & 1.692 && 0.11847 & 0.00275 & 0.00020 & 0.00170 & & 0.11598 & 0.00293 & 0.00019 & 0.00085 \\ 
 & 1.892 && 0.13255 & 0.00365 & 0.00029 & 0.00212 & & 0.12960 & 0.00397 & 0.00028 & 0.00107 \\ 
 & 2.200 && 0.13748 & 0.00363 & 0.00020 & 0.00229 & & 0.13446 & 0.00393 & 0.00019 & 0.00115 \\ 
 & 2.703 && 0.15166 & 0.00640 & 0.00022 & 0.00278 & & 0.14772 & 0.00754 & 0.00021 & 0.00139 \\ 
 & 3.343 && 0.14679 & 0.00945 & 0.00028 & 0.00261 & & 0.14255 & 0.01196 & 0.00026 & 0.00129 \\ 
 & 4.381 && 0.14874 & 0.02444 & 0.00064 & 0.00268 & & 0.14410 & 0.03301 & 0.00060 & 0.00132 \\ 
 & 5.410 && 0.02580 & 0.05846 & 0.00004 & 0.00008 & & 0.02498 & 0.08004 & 0.00004 & 0.00004 \\ 
 & 6.852 && -- & -- & -- & -- & & -- & -- & -- & -- \\ 
\hline 
 & 0.247 && 0.02479 & 0.00045 & 0.00035 & 0.00005 & & 0.02523 & 0.00056 & 0.00037 & 0.00002 \\ 
 & 0.348 && 0.03843 & 0.00052 & 0.00013 & 0.00011 & & 0.03893 & 0.00061 & 0.00014 & 0.00005 \\ 
 & 0.448 && 0.04673 & 0.00058 & 0.00013 & 0.00017 & & 0.04726 & 0.00067 & 0.00013 & 0.00008 \\ 
 & 0.548 && 0.05726 & 0.00065 & 0.00016 & 0.00025 & & 0.05785 & 0.00075 & 0.00016 & 0.00012 \\ 
 & 0.648 && 0.06796 & 0.00073 & 0.00017 & 0.00036 & & 0.06860 & 0.00084 & 0.00018 & 0.00016 \\ 
 & 0.748 && 0.07649 & 0.00082 & 0.00015 & 0.00045 & & 0.07721 & 0.00094 & 0.00016 & 0.00021 \\ 
 & 0.848 && 0.08664 & 0.00093 & 0.00018 & 0.00058 & & 0.08745 & 0.00106 & 0.00018 & 0.00027 \\ 
 & 0.948 && 0.09430 & 0.00105 & 0.00018 & 0.00069 & & 0.09523 & 0.00120 & 0.00018 & 0.00032 \\ 
 & 1.092 && 0.10554 & 0.00093 & 0.00019 & 0.00086 & & 0.10622 & 0.00103 & 0.00020 & 0.00040 \\ 
20--30\%
 & 1.292 && 0.12012 & 0.00118 & 0.00020 & 0.00112 & & 0.12107 & 0.00132 & 0.00020 & 0.00051 \\ 
$v_2$\{ZDC-SMD\}
 & 1.492 && 0.13329 & 0.00153 & 0.00025 & 0.00138 & & 0.13466 & 0.00176 & 0.00025 & 0.00064 \\ 
 & 1.692 && 0.14589 & 0.00202 & 0.00026 & 0.00165 & & 0.14785 & 0.00242 & 0.00026 & 0.00077 \\ 
 & 1.892 && 0.16194 & 0.00265 & 0.00028 & 0.00204 & & 0.16454 & 0.00327 & 0.00029 & 0.00095 \\ 
 & 2.200 && 0.17353 & 0.00265 & 0.00028 & 0.00234 & & 0.17613 & 0.00322 & 0.00029 & 0.00109 \\ 
 & 2.703 && 0.18631 & 0.00458 & 0.00038 & 0.00269 & & 0.19024 & 0.00610 & 0.00039 & 0.00127 \\ 
 & 3.343 && 0.18180 & 0.00683 & 0.00042 & 0.00257 & & 0.18612 & 0.00952 & 0.00044 & 0.00121 \\ 
 & 4.381 && 0.17827 & 0.01783 & 0.00084 & 0.00247 & & 0.18283 & 0.02570 & 0.00088 & 0.00117 \\ 
 & 5.410 && 0.16731 & 0.04246 & 0.00184 & 0.00217 & & 0.17163 & 0.06153 & 0.00194 & 0.00103 \\ 
 & 6.852 && -- & -- & -- & -- & & -- & -- & -- & -- \\ 
\end{tabular} \end{ruledtabular} \endgroup
\end{table*}

\begin{table*}[htbp]
\caption{\label{tab:table_v2zdc_smd_cent30-60}
$v_2$\{ZDC-SMD\} from S-N and ZDC-BBC-CNT subevents as a function of \pt in centralities
30--40\%, 40--50\%, and 50--60\%.
}
\begingroup \squeezetable
\begin{ruledtabular} \begin{tabular}{c|cccccccccccc}
Centrality
 & & & \multicolumn{4}{c}{S-N subevents} & & \multicolumn{4}{c}{ZDC-BBC-CNT subevents} \\ 
$v_2$\{\}
 &$\pt$ (\gevc) & & $v_2$ & $\sigma_{\rm stat}$ & $\sigma_{\rm syst}^B$ & $\sigma_{\rm syst}^C$ 
& & $v_2$ & $\sigma_{\rm stat}$ & $\sigma_{\rm syst}^B$ & $\sigma_{\rm syst}^C$ \\ 
\hline
 & 0.247 && 0.02694 & 0.00045 & 0.00062 & 0.00004 & & 0.02819 & 0.00061 & 0.00068 & 0.00002 \\ 
 & 0.348 && 0.04133 & 0.00050 & 0.00014 & 0.00010 & & 0.04305 & 0.00065 & 0.00015 & 0.00004 \\ 
 & 0.448 && 0.05500 & 0.00057 & 0.00015 & 0.00018 & & 0.05713 & 0.00071 & 0.00016 & 0.00007 \\ 
 & 0.548 && 0.06605 & 0.00064 & 0.00013 & 0.00026 & & 0.06852 & 0.00079 & 0.00014 & 0.00010 \\ 
 & 0.648 && 0.07744 & 0.00073 & 0.00013 & 0.00035 & & 0.08028 & 0.00089 & 0.00014 & 0.00013 \\ 
 & 0.748 && 0.08648 & 0.00082 & 0.00016 & 0.00044 & & 0.08966 & 0.00099 & 0.00017 & 0.00016 \\ 
 & 0.848 && 0.09719 & 0.00092 & 0.00016 & 0.00055 & & 0.10077 & 0.00112 & 0.00017 & 0.00021 \\ 
 & 0.948 && 0.10647 & 0.00104 & 0.00014 & 0.00066 & & 0.11046 & 0.00128 & 0.00016 & 0.00025 \\ 
 & 1.092 && 0.12033 & 0.00093 & 0.00018 & 0.00085 & & 0.12430 & 0.00110 & 0.00020 & 0.00031 \\ 
30--40\%
 & 1.292 && 0.13425 & 0.00117 & 0.00015 & 0.00106 & & 0.13898 & 0.00141 & 0.00017 & 0.00039 \\ 
$v_2$\{ZDC-SMD\}
 & 1.492 && 0.15041 & 0.00152 & 0.00025 & 0.00133 & & 0.15615 & 0.00188 & 0.00027 & 0.00050 \\ 
 & 1.692 && 0.16789 & 0.00203 & 0.00025 & 0.00165 & & 0.17486 & 0.00260 & 0.00027 & 0.00062 \\ 
 & 1.892 && 0.18310 & 0.00266 & 0.00031 & 0.00196 & & 0.19124 & 0.00353 & 0.00033 & 0.00074 \\ 
 & 2.200 && 0.18792 & 0.00263 & 0.00029 & 0.00207 & & 0.19616 & 0.00346 & 0.00031 & 0.00078 \\ 
 & 2.703 && 0.19298 & 0.00458 & 0.00033 & 0.00218 & & 0.20250 & 0.00649 & 0.00036 & 0.00083 \\ 
 & 3.343 && 0.19902 & 0.00685 & 0.00031 & 0.00232 & & 0.20918 & 0.00995 & 0.00034 & 0.00089 \\ 
 & 4.381 && 0.15951 & 0.01765 & 0.00077 & 0.00149 & & 0.16787 & 0.02619 & 0.00085 & 0.00057 \\ 
 & 5.410 && 0.03318 & 0.04176 & 0.00008 & 0.00006 & & 0.03492 & 0.06213 & 0.00009 & 0.00002 \\ 
 & 6.852 && -- & -- & -- & -- & & -- & -- & -- & -- \\ 
\hline 
 & 0.247 && 0.02601 & 0.00054 & 0.00060 & 0.00004 & & 0.02771 & 0.00077 & 0.00068 & 0.00001 \\ 
 & 0.348 && 0.04210 & 0.00060 & 0.00014 & 0.00010 & & 0.04474 & 0.00081 & 0.00016 & 0.00003 \\ 
 & 0.448 && 0.05541 & 0.00067 & 0.00012 & 0.00017 & & 0.05880 & 0.00089 & 0.00013 & 0.00005 \\ 
 & 0.548 && 0.06853 & 0.00076 & 0.00012 & 0.00026 & & 0.07264 & 0.00099 & 0.00013 & 0.00008 \\ 
 & 0.648 && 0.08077 & 0.00086 & 0.00011 & 0.00036 & & 0.08558 & 0.00111 & 0.00013 & 0.00012 \\ 
 & 0.748 && 0.09316 & 0.00097 & 0.00013 & 0.00048 & & 0.09868 & 0.00125 & 0.00015 & 0.00015 \\ 
 & 0.848 && 0.10257 & 0.00110 & 0.00015 & 0.00059 & & 0.10868 & 0.00141 & 0.00017 & 0.00019 \\ 
 & 0.948 && 0.11494 & 0.00125 & 0.00013 & 0.00074 & & 0.12181 & 0.00161 & 0.00015 & 0.00023 \\ 
 & 1.092 && 0.12842 & 0.00112 & 0.00012 & 0.00092 & & 0.13572 & 0.00139 & 0.00014 & 0.00029 \\ 
40--50\%
 & 1.292 && 0.14455 & 0.00141 & 0.00015 & 0.00116 & & 0.15299 & 0.00179 & 0.00017 & 0.00037 \\ 
$v_2$\{ZDC-SMD\}
 & 1.492 && 0.15539 & 0.00183 & 0.00020 & 0.00134 & & 0.16483 & 0.00240 & 0.00022 & 0.00043 \\ 
 & 1.692 && 0.16641 & 0.00245 & 0.00023 & 0.00154 & & 0.17691 & 0.00333 & 0.00027 & 0.00049 \\ 
 & 1.892 && 0.18706 & 0.00325 & 0.00027 & 0.00195 & & 0.19913 & 0.00453 & 0.00030 & 0.00063 \\ 
 & 2.200 && 0.19007 & 0.00319 & 0.00025 & 0.00201 & & 0.20228 & 0.00443 & 0.00028 & 0.00065 \\ 
 & 2.703 && 0.19675 & 0.00563 & 0.00030 & 0.00215 & & 0.20991 & 0.00824 & 0.00034 & 0.00069 \\ 
 & 3.343 && 0.17518 & 0.00833 & 0.00043 & 0.00171 & & 0.18706 & 0.01244 & 0.00049 & 0.00055 \\ 
 & 4.381 && 0.15207 & 0.02120 & 0.00062 & 0.00129 & & 0.16245 & 0.03198 & 0.00071 & 0.00042 \\ 
 & 5.410 && 0.23778 & 0.04958 & 0.00360 & 0.00315 & & 0.25402 & 0.07485 & 0.00410 & 0.00102 \\ 
 & 6.852 && -- & -- & -- & -- & & -- & -- & -- & -- \\ 
\hline 
 & 0.247 && 0.02164 & 0.00071 & 0.00034 & 0.00004 & & 0.02529 & 0.00114 & 0.00047 & 0.00002 \\ 
 & 0.348 && 0.03766 & 0.00077 & 0.00013 & 0.00011 & & 0.04384 & 0.00120 & 0.00017 & 0.00007 \\ 
 & 0.448 && 0.05159 & 0.00087 & 0.00013 & 0.00021 & & 0.05986 & 0.00132 & 0.00017 & 0.00013 \\ 
 & 0.548 && 0.06277 & 0.00098 & 0.00010 & 0.00031 & & 0.07273 & 0.00148 & 0.00013 & 0.00020 \\ 
 & 0.648 && 0.07471 & 0.00111 & 0.00009 & 0.00044 & & 0.08647 & 0.00166 & 0.00012 & 0.00028 \\ 
 & 0.748 && 0.08320 & 0.00125 & 0.00009 & 0.00054 & & 0.09633 & 0.00188 & 0.00012 & 0.00035 \\ 
 & 0.848 && 0.09675 & 0.00143 & 0.00015 & 0.00074 & & 0.11196 & 0.00214 & 0.00020 & 0.00047 \\ 
 & 0.948 && 0.10720 & 0.00163 & 0.00008 & 0.00090 & & 0.12413 & 0.00244 & 0.00011 & 0.00058 \\ 
 & 1.092 && 0.11901 & 0.00146 & 0.00010 & 0.00111 & & 0.13707 & 0.00212 & 0.00013 & 0.00070 \\ 
50--60\%
 & 1.292 && 0.12717 & 0.00184 & 0.00013 & 0.00127 & & 0.14709 & 0.00274 & 0.00017 & 0.00081 \\ 
$v_2$\{ZDC-SMD\}
 & 1.492 && 0.14188 & 0.00243 & 0.00019 & 0.00158 & & 0.16469 & 0.00370 & 0.00025 & 0.00101 \\ 
 & 1.692 && 0.15811 & 0.00331 & 0.00021 & 0.00196 & & 0.18411 & 0.00516 & 0.00028 & 0.00127 \\ 
 & 1.892 && 0.15997 & 0.00439 & 0.00035 & 0.00201 & & 0.18679 & 0.00701 & 0.00048 & 0.00131 \\ 
 & 2.200 && 0.16724 & 0.00431 & 0.00037 & 0.00220 & & 0.19518 & 0.00684 & 0.00051 & 0.00142 \\ 
 & 2.703 && 0.18027 & 0.00776 & 0.00047 & 0.00255 & & 0.21100 & 0.01265 & 0.00064 & 0.00166 \\ 
 & 3.343 && 0.13888 & 0.01139 & 0.00041 & 0.00152 & & 0.16274 & 0.01878 & 0.00056 & 0.00099 \\ 
 & 4.381 && 0.12204 & 0.02867 & 0.00202 & 0.00117 & & 0.14306 & 0.04745 & 0.00277 & 0.00077 \\ 
 & 5.410 && -- & -- & -- & -- & & -- & -- & -- & -- \\ 
 & 6.852 && -- & -- & -- & -- & & -- & -- & -- & -- \\ 
\end{tabular} \end{ruledtabular} \endgroup
\end{table*}

\clearpage


\begin{thebibliography}{51}

\expandafter\ifx\csname natexlab\endcsname\relax\def\natexlab#1{#1}\fi
\expandafter\ifx\csname bibnamefont\endcsname\relax
  \def\bibnamefont#1{#1}\fi
\expandafter\ifx\csname bibfnamefont\endcsname\relax
  \def\bibfnamefont#1{#1}\fi
\expandafter\ifx\csname citenamefont\endcsname\relax
  \def\citenamefont#1{#1}\fi
\expandafter\ifx\csname url\endcsname\relax
  \def\url#1{\texttt{#1}}\fi
\expandafter\ifx\csname urlprefix\endcsname\relax\def\urlprefix{URL }\fi
\providecommand{\bibinfo}[2]{#2}
\providecommand{\eprint}[2][]{\url{#2}}

\bibitem[{\citenamefont{Arsene et~al.}(2005)}]{Arsene:2004fa}
\bibinfo{author}{\bibfnamefont{I.}~\bibnamefont{Arsene}} \bibnamefont{et~al.}
  (\bibinfo{collaboration}{BRAHMS}), \bibinfo{journal}{Nucl. Phys.}
  \textbf{\bibinfo{volume}{A757}}, \bibinfo{pages}{1} (\bibinfo{year}{2005}).

\bibitem[{\citenamefont{Adcox et~al.}(2005)}]{Adcox:2004mh}
\bibinfo{author}{\bibfnamefont{K.}~\bibnamefont{Adcox}} \bibnamefont{et~al.}
  (\bibinfo{collaboration}{PHENIX}), \bibinfo{journal}{Nucl. Phys.}
  \textbf{\bibinfo{volume}{A757}}, \bibinfo{pages}{184} (\bibinfo{year}{2005}).

\bibitem[{\citenamefont{Back et~al.}(2005)}]{Back:2004je}
\bibinfo{author}{\bibfnamefont{B.~B.} \bibnamefont{Back}} \bibnamefont{et~al.}
  (\bibinfo{collaboration}{PHOBOS}), \bibinfo{journal}{Nucl. Phys.}
  \textbf{\bibinfo{volume}{A757}}, \bibinfo{pages}{28} (\bibinfo{year}{2005}).

\bibitem[{\citenamefont{Adams et~al.}(2005{\natexlab{a}})}]{Adams:2005dq}
\bibinfo{author}{\bibfnamefont{J.}~\bibnamefont{Adams}} \bibnamefont{et~al.}
  (\bibinfo{collaboration}{STAR}), \bibinfo{journal}{Nucl. Phys.}
  \textbf{\bibinfo{volume}{A757}}, \bibinfo{pages}{102}
  (\bibinfo{year}{2005}{\natexlab{a}}).

\bibitem[{\citenamefont{Gyulassy and McLerran}(2005)}]{Gyulassy:2004zy}
\bibinfo{author}{\bibfnamefont{M.}~\bibnamefont{Gyulassy}} \bibnamefont{and}
  \bibinfo{author}{\bibfnamefont{L.}~\bibnamefont{McLerran}},
  \bibinfo{journal}{Nucl. Phys.} \textbf{\bibinfo{volume}{A750}},
  \bibinfo{pages}{30} (\bibinfo{year}{2005}).

\bibitem[{\citenamefont{Muller}()}]{Muller:2004kk}
\bibinfo{author}{\bibfnamefont{B.}~\bibnamefont{Muller}},
  \bibinfo{note}{nucl-th/0404015}.

\bibitem[{\citenamefont{Shuryak}(2005)}]{Shuryak:2004cy}
\bibinfo{author}{\bibfnamefont{E.~V.} \bibnamefont{Shuryak}},
  \bibinfo{journal}{Nucl. Phys.} \textbf{\bibinfo{volume}{A750}},
  \bibinfo{pages}{64} (\bibinfo{year}{2005}).

\bibitem[{\citenamefont{Ollitrault}(1992)}]{Ollitrault:1992bk}
\bibinfo{author}{\bibfnamefont{J.-Y.} \bibnamefont{Ollitrault}},
  \bibinfo{journal}{Phys. Rev. D} \textbf{\bibinfo{volume}{46}},
  \bibinfo{pages}{229} (\bibinfo{year}{1992}).

\bibitem[{\citenamefont{Kolb et~al.}(2001)\citenamefont{Kolb, Heinz, Huovinen,
  Eskola, and Tuominen}}]{Kolb:2001qz}
\bibinfo{author}{\bibfnamefont{P.~F.} \bibnamefont{Kolb}},
  \bibinfo{author}{\bibfnamefont{U.~W.} \bibnamefont{Heinz}},
  \bibinfo{author}{\bibfnamefont{P.}~\bibnamefont{Huovinen}},
  \bibinfo{author}{\bibfnamefont{K.~J.} \bibnamefont{Eskola}},
  \bibnamefont{and} \bibinfo{author}{\bibfnamefont{K.}~\bibnamefont{Tuominen}},
  \bibinfo{journal}{Nucl. Phys.} \textbf{\bibinfo{volume}{A696}},
  \bibinfo{pages}{197} (\bibinfo{year}{2001}).

\bibitem[{\citenamefont{Hirano and Nara}(2004)}]{Hirano:2004en}
\bibinfo{author}{\bibfnamefont{T.}~\bibnamefont{Hirano}} \bibnamefont{and}
  \bibinfo{author}{\bibfnamefont{Y.}~\bibnamefont{Nara}},
  \bibinfo{journal}{Nucl. Phys.} \textbf{\bibinfo{volume}{A743}},
  \bibinfo{pages}{305} (\bibinfo{year}{2004}).

\bibitem[{\citenamefont{Ajitanand}(2003)}]{Ajitanand:2002qd}
\bibinfo{author}{\bibfnamefont{N.~N.} \bibnamefont{Ajitanand}}
  (\bibinfo{collaboration}{PHENIX}), \bibinfo{journal}{Nucl. Phys.}
  \textbf{\bibinfo{volume}{A715}}, \bibinfo{pages}{765} (\bibinfo{year}{2003}).

\bibitem[{\citenamefont{Chiu}(2003)}]{Chiu:2002ma}
\bibinfo{author}{\bibfnamefont{M.}~\bibnamefont{Chiu}}
  (\bibinfo{collaboration}{PHENIX}), \bibinfo{journal}{Nucl. Phys.}
  \textbf{\bibinfo{volume}{A715}}, \bibinfo{pages}{761} (\bibinfo{year}{2003}).

\bibitem[{\citenamefont{Adler et~al.}(2003{\natexlab{a}})}]{Adler:2002tq}
\bibinfo{author}{\bibfnamefont{C.}~\bibnamefont{Adler}} \bibnamefont{et~al.}
  (\bibinfo{collaboration}{STAR}), \bibinfo{journal}{Phys. Rev. Lett.}
  \textbf{\bibinfo{volume}{90}}, \bibinfo{pages}{082302}
  (\bibinfo{year}{2003}{\natexlab{a}}).

\bibitem[{\citenamefont{Adler et~al.}(2005)}]{Adler:2004zd}
\bibinfo{author}{\bibfnamefont{S.~S.} \bibnamefont{Adler}} \bibnamefont{et~al.}
  (\bibinfo{collaboration}{PHENIX}), \bibinfo{journal}{Phys. Rev. C}
  \textbf{\bibinfo{volume}{71}}, \bibinfo{pages}{051902}
  (\bibinfo{year}{2005}).

\bibitem[{\citenamefont{Adams et~al.}(2005{\natexlab{b}})}]{Adams:2005ph}
\bibinfo{author}{\bibfnamefont{J.}~\bibnamefont{Adams}} \bibnamefont{et~al.}
  (\bibinfo{collaboration}{STAR}), \bibinfo{journal}{Phys. Rev. Lett.}
  \textbf{\bibinfo{volume}{95}}, \bibinfo{pages}{152301}
  (\bibinfo{year}{2005}{\natexlab{b}}).

\bibitem[{\citenamefont{Adler et~al.}(2006{\natexlab{a}})}]{Adler:2005ee}
\bibinfo{author}{\bibfnamefont{S.~S.} \bibnamefont{Adler}} \bibnamefont{et~al.}
  (\bibinfo{collaboration}{PHENIX}), \bibinfo{journal}{Phys. Rev. Lett.}
  \textbf{\bibinfo{volume}{97}}, \bibinfo{pages}{052301}
  (\bibinfo{year}{2006}{\natexlab{a}}).

\bibitem[{\citenamefont{Adare et~al.}(2008)}]{Adare:2008cqb}
\bibinfo{author}{\bibfnamefont{A.}~\bibnamefont{Adare}} \bibnamefont{et~al.}
  (\bibinfo{collaboration}{PHENIX}), \bibinfo{journal}{Phys. Rev. C}
  \textbf{\bibinfo{volume}{78}}, \bibinfo{pages}{014901}
  (\bibinfo{year}{2008}).

\bibitem[{\citenamefont{Bleicher and Stoecker}(2002)}]{Bleicher:2000sx}
\bibinfo{author}{\bibfnamefont{M.}~\bibnamefont{Bleicher}} \bibnamefont{and}
  \bibinfo{author}{\bibfnamefont{H.}~\bibnamefont{Stoecker}},
  \bibinfo{journal}{Phys. Lett.} \textbf{\bibinfo{volume}{B526}},
  \bibinfo{pages}{309} (\bibinfo{year}{2002}).

\bibitem[{\citenamefont{Adler et~al.}(2003{\natexlab{b}})}]{Adler:2003kt}
\bibinfo{author}{\bibfnamefont{S.~S.} \bibnamefont{Adler}} \bibnamefont{et~al.}
  (\bibinfo{collaboration}{PHENIX}), \bibinfo{journal}{Phys. Rev. Lett.}
  \textbf{\bibinfo{volume}{91}}, \bibinfo{pages}{182301}
  (\bibinfo{year}{2003}{\natexlab{b}}).

\bibitem[{\citenamefont{Xu et~al.}(2008)\citenamefont{Xu, Greiner, and
  Stocker}}]{Xu:2008dv}
\bibinfo{author}{\bibfnamefont{Z.}~\bibnamefont{Xu}},
  \bibinfo{author}{\bibfnamefont{C.}~\bibnamefont{Greiner}}, \bibnamefont{and}
  \bibinfo{author}{\bibfnamefont{H.}~\bibnamefont{Stocker}},
  \bibinfo{journal}{J. Phys.} \textbf{\bibinfo{volume}{G35}},
  \bibinfo{pages}{104016} (\bibinfo{year}{2008}).

\bibitem[{\citenamefont{Adams et~al.}(2004{\natexlab{a}})}]{Adams:2003am}
\bibinfo{author}{\bibfnamefont{J.}~\bibnamefont{Adams}} \bibnamefont{et~al.}
  (\bibinfo{collaboration}{STAR}), \bibinfo{journal}{Phys. Rev. Lett.}
  \textbf{\bibinfo{volume}{92}}, \bibinfo{pages}{052302}
  (\bibinfo{year}{2004}{\natexlab{a}}).

\bibitem[{\citenamefont{Adare et~al.}(2007)}]{Adare:2006ti}
\bibinfo{author}{\bibfnamefont{A.}~\bibnamefont{Adare}} \bibnamefont{et~al.}
  (\bibinfo{collaboration}{PHENIX}), \bibinfo{journal}{Phys. Rev. Lett.}
  \textbf{\bibinfo{volume}{98}}, \bibinfo{pages}{162301}
  (\bibinfo{year}{2007}).

\bibitem[{\citenamefont{Huang}(2008)}]{Huang:2008vd}
\bibinfo{author}{\bibfnamefont{S.}~\bibnamefont{Huang}}
  (\bibinfo{collaboration}{PHENIX}), \bibinfo{journal}{J. Phys.}
  \textbf{\bibinfo{volume}{G35}}, \bibinfo{pages}{104105}
  (\bibinfo{year}{2008}).

\bibitem[{\citenamefont{Moln\'ar and Voloshin}(2003)}]{PhysRevLett.91.092301}
\bibinfo{author}{\bibfnamefont{D.}~\bibnamefont{Moln\'ar}} \bibnamefont{and}
  \bibinfo{author}{\bibfnamefont{S.~A.} \bibnamefont{Voloshin}},
  \bibinfo{journal}{Phys. Rev. Lett.} \textbf{\bibinfo{volume}{91}},
  \bibinfo{pages}{092301} (\bibinfo{year}{2003}).

\bibitem[{\citenamefont{Fries et~al.}(2003)\citenamefont{Fries, M\"uller,
  Nonaka, and Bass}}]{PhysRevC.68.044902}
\bibinfo{author}{\bibfnamefont{R.~J.} \bibnamefont{Fries}},
  \bibinfo{author}{\bibfnamefont{B.}~\bibnamefont{M\"uller}},
  \bibinfo{author}{\bibfnamefont{C.}~\bibnamefont{Nonaka}}, \bibnamefont{and}
  \bibinfo{author}{\bibfnamefont{S.~A.} \bibnamefont{Bass}},
  \bibinfo{journal}{Phys. Rev. C} \textbf{\bibinfo{volume}{68}},
  \bibinfo{pages}{044902} (\bibinfo{year}{2003}).

\bibitem[{\citenamefont{Greco et~al.}(2003)\citenamefont{Greco, Ko, and
  L\'evai}}]{PhysRevC.68.034904}
\bibinfo{author}{\bibfnamefont{V.}~\bibnamefont{Greco}},
  \bibinfo{author}{\bibfnamefont{C.~M.} \bibnamefont{Ko}}, \bibnamefont{and}
  \bibinfo{author}{\bibfnamefont{P.}~\bibnamefont{L\'evai}},
  \bibinfo{journal}{Phys. Rev. C} \textbf{\bibinfo{volume}{68}},
  \bibinfo{pages}{034904} (\bibinfo{year}{2003}).

\bibitem[{\citenamefont{Shor}(1985)}]{Shor:1984ui}
\bibinfo{author}{\bibfnamefont{A.}~\bibnamefont{Shor}}, \bibinfo{journal}{Phys.
  Rev. Lett.} \textbf{\bibinfo{volume}{54}}, \bibinfo{pages}{1122}
  (\bibinfo{year}{1985}).

\bibitem[{\citenamefont{Poskanzer and Voloshin}(1998)}]{Poskanzer:1998yz}
\bibinfo{author}{\bibfnamefont{A.~M.} \bibnamefont{Poskanzer}}
  \bibnamefont{and} \bibinfo{author}{\bibfnamefont{S.~A.}
  \bibnamefont{Voloshin}}, \bibinfo{journal}{Phys. Rev. C}
  \textbf{\bibinfo{volume}{58}}, \bibinfo{pages}{1671} (\bibinfo{year}{1998}).

\bibitem[{\citenamefont{Jia}(2007)}]{Jia:2006sb}
\bibinfo{author}{\bibfnamefont{J.}~\bibnamefont{Jia}}
  (\bibinfo{collaboration}{PHENIX}), \bibinfo{journal}{Nucl. Phys.}
  \textbf{\bibinfo{volume}{A783}}, \bibinfo{pages}{501} (\bibinfo{year}{2007}).

\bibitem[{\citenamefont{Alver et~al.}()}]{Alver:2009id}
\bibinfo{author}{\bibfnamefont{B.}~\bibnamefont{Alver}} \bibnamefont{et~al.},
  \bibinfo{note}{arXiv:0903.2811 [nucl-ex]}.

\bibitem[{\citenamefont{Sorensen}(2008)}]{Sorensen:2008zk}
\bibinfo{author}{\bibfnamefont{P.}~\bibnamefont{Sorensen}}
  (\bibinfo{collaboration}{STAR}), \bibinfo{journal}{J. Phys.}
  \textbf{\bibinfo{volume}{G35}}, \bibinfo{pages}{104102}
  (\bibinfo{year}{2008}).

\bibitem[{\citenamefont{Ollitrault et~al.}()\citenamefont{Ollitrault,
  Poskanzer, and Voloshin}}]{Ollitrault:2009ie}
\bibinfo{author}{\bibfnamefont{J.-Y.} \bibnamefont{Ollitrault}},
  \bibinfo{author}{\bibfnamefont{A.~M.} \bibnamefont{Poskanzer}},
  \bibnamefont{and} \bibinfo{author}{\bibfnamefont{S.~A.}
  \bibnamefont{Voloshin}}, \bibinfo{note}{arXiv:0904.2315 [nucl-ex]}.

\bibitem[{\citenamefont{Adcox et~al.}(2003)}]{Adcox:2003zm}
\bibinfo{author}{\bibfnamefont{K.}~\bibnamefont{Adcox}} \bibnamefont{et~al.}
  (\bibinfo{collaboration}{PHENIX}), \bibinfo{journal}{Nucl. Instrum. Meth.}
  \textbf{\bibinfo{volume}{A499}}, \bibinfo{pages}{469} (\bibinfo{year}{2003}).

\bibitem[{\citenamefont{Allen et~al.}(2003)}]{Allen:2003zt}
\bibinfo{author}{\bibfnamefont{M.}~\bibnamefont{Allen}} \bibnamefont{et~al.}
  (\bibinfo{collaboration}{PHENIX}), \bibinfo{journal}{Nucl. Instrum. Meth.}
  \textbf{\bibinfo{volume}{A499}}, \bibinfo{pages}{549} (\bibinfo{year}{2003}).

\bibitem[{\citenamefont{Adler et~al.}(2003{\natexlab{c}})}]{Adler:2003sp}
\bibinfo{author}{\bibfnamefont{C.}~\bibnamefont{Adler}} \bibnamefont{et~al.},
  \bibinfo{journal}{Nucl. Instrum. Meth.} \textbf{\bibinfo{volume}{A499}},
  \bibinfo{pages}{433} (\bibinfo{year}{2003}{\natexlab{c}}).

\bibitem[{\citenamefont{Miller et~al.}(2007)\citenamefont{Miller, Reygers,
  Sanders, and Steinberg}}]{Miller:2007ri}
\bibinfo{author}{\bibfnamefont{M.~L.} \bibnamefont{Miller}},
  \bibinfo{author}{\bibfnamefont{K.}~\bibnamefont{Reygers}},
  \bibinfo{author}{\bibfnamefont{S.~J.} \bibnamefont{Sanders}},
  \bibnamefont{and}
  \bibinfo{author}{\bibfnamefont{P.}~\bibnamefont{Steinberg}},
  \bibinfo{journal}{Ann. Rev. Nucl. Part. Sci.} \textbf{\bibinfo{volume}{57}},
  \bibinfo{pages}{205} (\bibinfo{year}{2007}).

\bibitem[{\citenamefont{Glauber and Matthiae}(1970)}]{Glauber:1970jm}
\bibinfo{author}{\bibfnamefont{R.~J.} \bibnamefont{Glauber}} \bibnamefont{and}
  \bibinfo{author}{\bibfnamefont{G.}~\bibnamefont{Matthiae}},
  \bibinfo{journal}{Nucl. Phys.} \textbf{\bibinfo{volume}{B21}},
  \bibinfo{pages}{135} (\bibinfo{year}{1970}).

\bibitem[{\citenamefont{Adcox et~al.}(2001)}]{Adcox:2000sp}
\bibinfo{author}{\bibfnamefont{K.}~\bibnamefont{Adcox}} \bibnamefont{et~al.}
  (\bibinfo{collaboration}{PHENIX}), \bibinfo{journal}{Phys. Rev. Lett.}
  \textbf{\bibinfo{volume}{86}}, \bibinfo{pages}{3500} (\bibinfo{year}{2001}).

\bibitem[{\citenamefont{Adler et~al.}(2004)}]{Adler:2003au}
\bibinfo{author}{\bibfnamefont{S.~S.} \bibnamefont{Adler}} \bibnamefont{et~al.}
  (\bibinfo{collaboration}{PHENIX}), \bibinfo{journal}{Phys. Rev. C}
  \textbf{\bibinfo{volume}{69}}, \bibinfo{pages}{034910}
  (\bibinfo{year}{2004}).

\bibitem[{\citenamefont{Mitchell et~al.}(2002)}]{Mitchell:2002wu}
\bibinfo{author}{\bibfnamefont{J.~T.} \bibnamefont{Mitchell}}
  \bibnamefont{et~al.} (\bibinfo{collaboration}{PHENIX}),
  \bibinfo{journal}{Nucl. Instrum. Meth.} \textbf{\bibinfo{volume}{A482}},
  \bibinfo{pages}{491} (\bibinfo{year}{2002}).

\bibitem[{\citenamefont{Adler et~al.}(2006{\natexlab{b}})}]{Adler:2005ad}
\bibinfo{author}{\bibfnamefont{S.~S.} \bibnamefont{Adler}} \bibnamefont{et~al.}
  (\bibinfo{collaboration}{PHENIX}), \bibinfo{journal}{Phys. Rev. C}
  \textbf{\bibinfo{volume}{73}}, \bibinfo{pages}{054903}
  (\bibinfo{year}{2006}{\natexlab{b}}).

\bibitem[{\citenamefont{Borghini
  et~al.}(2001{\natexlab{a}})\citenamefont{Borghini, Dinh, and
  Ollitrault}}]{Borghini:2001vi}
\bibinfo{author}{\bibfnamefont{N.}~\bibnamefont{Borghini}},
  \bibinfo{author}{\bibfnamefont{P.~M.} \bibnamefont{Dinh}}, \bibnamefont{and}
  \bibinfo{author}{\bibfnamefont{J.-Y.} \bibnamefont{Ollitrault}},
  \bibinfo{journal}{Phys. Rev. C} \textbf{\bibinfo{volume}{64}},
  \bibinfo{pages}{054901} (\bibinfo{year}{2001}{\natexlab{a}}).

\bibitem[{\citenamefont{Borghini
  et~al.}(2001{\natexlab{b}})\citenamefont{Borghini, Dinh, and
  Ollitrault}}]{Borghini:2001zr}
\bibinfo{author}{\bibfnamefont{N.}~\bibnamefont{Borghini}},
  \bibinfo{author}{\bibfnamefont{P.~M.} \bibnamefont{Dinh}}, \bibnamefont{and}
  \bibinfo{author}{\bibfnamefont{J.-Y.} \bibnamefont{Ollitrault}}
  (\bibinfo{year}{2001}{\natexlab{b}}).

\bibitem[{\citenamefont{Alt et~al.}(2003)}]{Alt:2003ab}
\bibinfo{author}{\bibfnamefont{C.}~\bibnamefont{Alt}} \bibnamefont{et~al.}
  (\bibinfo{collaboration}{NA49}), \bibinfo{journal}{Phys. Rev. C}
  \textbf{\bibinfo{volume}{68}}, \bibinfo{pages}{034903}
  (\bibinfo{year}{2003}).

\bibitem[{\citenamefont{Adler et~al.}(2002)}]{Adler:2002pu}
\bibinfo{author}{\bibfnamefont{C.}~\bibnamefont{Adler}} \bibnamefont{et~al.}
  (\bibinfo{collaboration}{STAR}), \bibinfo{journal}{Phys. Rev. C}
  \textbf{\bibinfo{volume}{66}}, \bibinfo{pages}{034904}
  (\bibinfo{year}{2002}).

\bibitem[{\citenamefont{Selyuzhenkov and Voloshin}(2008)}]{Selyuzhenkov:2007zi}
\bibinfo{author}{\bibfnamefont{I.}~\bibnamefont{Selyuzhenkov}}
  \bibnamefont{and} \bibinfo{author}{\bibfnamefont{S.}~\bibnamefont{Voloshin}},
  \bibinfo{journal}{Phys. Rev. C} \textbf{\bibinfo{volume}{77}},
  \bibinfo{pages}{034904} (\bibinfo{year}{2008}).

\bibitem[{\citenamefont{Adams et~al.}(2005{\natexlab{c}})}]{Adams:2004bi}
\bibinfo{author}{\bibfnamefont{J.}~\bibnamefont{Adams}} \bibnamefont{et~al.}
  (\bibinfo{collaboration}{STAR}), \bibinfo{journal}{Phys. Rev. C}
  \textbf{\bibinfo{volume}{72}}, \bibinfo{pages}{014904}
  (\bibinfo{year}{2005}{\natexlab{c}}).

\bibitem[{\citenamefont{Alver et~al.}(2007)}]{Alver:2006wh}
\bibinfo{author}{\bibfnamefont{B.}~\bibnamefont{Alver}} \bibnamefont{et~al.}
  (\bibinfo{collaboration}{PHOBOS}), \bibinfo{journal}{Phys. Rev. Lett.}
  \textbf{\bibinfo{volume}{98}}, \bibinfo{pages}{242302}
  (\bibinfo{year}{2007}).

\bibitem[{\citenamefont{Adams et~al.}(2004{\natexlab{b}})}]{Adams:2004wz}
\bibinfo{author}{\bibfnamefont{J.}~\bibnamefont{Adams}} \bibnamefont{et~al.}
  (\bibinfo{collaboration}{STAR}), \bibinfo{journal}{Phys. Rev. Lett.}
  \textbf{\bibinfo{volume}{93}}, \bibinfo{pages}{252301}
  (\bibinfo{year}{2004}{\natexlab{b}}).

\bibitem[{\citenamefont{Bhalerao and Ollitrault}(2006)}]{Bhalerao:2006tp}
\bibinfo{author}{\bibfnamefont{R.~S.} \bibnamefont{Bhalerao}} \bibnamefont{and}
  \bibinfo{author}{\bibfnamefont{J.-Y.} \bibnamefont{Ollitrault}},
  \bibinfo{journal}{Phys. Lett.} \textbf{\bibinfo{volume}{B641}},
  \bibinfo{pages}{260} (\bibinfo{year}{2006}).

\bibitem[{\citenamefont{Miller and Snellings}()}]{Miller:2003kd}
\bibinfo{author}{\bibfnamefont{M.}~\bibnamefont{Miller}} \bibnamefont{and}
  \bibinfo{author}{\bibfnamefont{R.}~\bibnamefont{Snellings}},
  \bibinfo{note}{nucl-ex/0312008}.

\end{thebibliography}

\end{document}